\documentclass[%
 amsmath,amssymb,
 aps, 
 reprint,
prd,
superscriptaddress,
floatfix,
nofootinbib,
]{revtex4-1}

\usepackage{latexsym}
\usepackage{amssymb} 
\usepackage{amsbsy} 
\usepackage{amsmath} 
\usepackage{amsthm} 
\usepackage{amsfonts} 
\usepackage{epsfig} 
\usepackage{graphics,color}
\usepackage{multirow} 
\usepackage{xcolor} 
\usepackage{soul} 
\usepackage{mathtools} 
\usepackage[caption=false]{subfig}
\usepackage[%
colorlinks=true,    
citecolor=blue,     
linkcolor=black,    
urlcolor=blue,      
]{hyperref}
\usepackage{tabularx} 

\newcommand{\be}{\begin{equation}}
\newcommand{\ee}{\end{equation}}
\newcommand{\bea}{\begin{eqnarray}}
\newcommand{\eea}{\end{eqnarray}}

\newcommand{\nn}{\nonumber}

\graphicspath{ {./images/} } 

\newcommand{\Norm}[1]{\left\lVert #1 \right\lVert}
\newcommand{\bvec}[1]{\boldsymbol{#1}}
\newcommand{\Tr}{{\rm Tr}}

\newcommand{\sexpv}[2][]{\mathbb{E}_{#1} \left[ #2 \right]}

\begin{document}

\title{Spectral phase transitions and trainability in neural network learning dynamics}

\author{Chanju Park}
\email{chanju.b.park@gmail.com}
\affiliation{Centre for Quantum Fields and Gravity, Department of Physics, Swansea University, Swansea SA2 8PP, United Kingdom}

\author{Dario Bocchi}
\email{dario.bocchi@uniroma1.it}
\affiliation{Physics Department, University of Rome Sapienza, Piazzale Aldo Moro 5, 00185 Rome, Italy}
\affiliation{Institute of Nanotechnology, CNR-NANOTEC, Piazzale Aldo Moro 5, 00185 Rome, Italy}

\author{Francesco D'Amico}
\email{francesco.damico@uniroma1.it}
\affiliation{Physics Department, University of Rome Sapienza, Piazzale Aldo Moro 5, 00185 Rome, Italy}
\affiliation{Institute of Nanotechnology, CNR-NANOTEC, Piazzale Aldo Moro 5, 00185 Rome, Italy}

\author{Biagio Lucini}
\email{b.lucini@qmul.ac.uk}
\affiliation{School of Mathematical Sciences, Queen Mary University of London, London E1 4NS, United Kingdom}

\author{Gert Aarts}
\email{g.aarts@swansea.ac.uk}
\affiliation{Centre for Quantum Fields and Gravity, Department of Physics, Swansea University, Swansea SA2 8PP, United Kingdom}

\date{\today}

\begin{abstract}
    The emergence of low-dimensional structures in the spectra of neural network weight matrices is a common empirical feature of trained models, but the dynamical origin of this phenomenon during learning remains an open problem.
    We formulate neural network training as the stochastic evolution of an initially random matrix ensemble, driven by stochastic gradient descent (SGD) updates that reshape the spectral bulk while amplifying signal strength.
    This induces a Baik-Ben Arous-P\'ech\'e (BBP) transition during training, where isolated eigenvalues detach from the random bulk distribution, providing a dynamical framework for representation formation in high-dimensional learning dynamics.
    We demonstrate this in a solvable linear teacher-student model, where spectral evolution is analytically tractable and a phase diagram of trainability governed by the step size (or learning rate) and initial weight variance is obtained, and subsequently extend our formalism beyond the linear regime to nonlinear and stochastic settings.
    Numerical simulations in realistic settings support this picture, showing robust emergence of spectral alignment during training.
    Our results suggest that spectral analysis may provide a unified perspective of stochastic learning dynamics, linking trainability, optimisation hyperparameters, spectral phase transitions, and representation learning in neural networks.
\end{abstract}

{
\let\clearpage\relax
\maketitle
}

\setlength{\parindent}{2em}

\section{Introduction}
\label{sec:1.introduction}
    Phase transitions provide a powerful framework for understanding the emergence of collective behaviour in complex systems, where macroscopic order may emerge from the microscopic interactions of a large number of constituents.
    In the context of the learning dynamics of neural networks, the interplay between high-dimensional non-convex loss landscapes and stochastic optimisation algorithms can give rise to emergent phenomena, including transitions associated with trainability, signal propagation, and the formation of low-dimensional spectral structure in trained models
    \cite{sompolinskyChaosRandomNeural1988, mahoneyTraditionalHeavyTailed2019, martinImplicitSelfRegularizationDeep2021, dandiRandomMatrixTheory2024, huLoRALowRankAdaptation2021}.

    Empirical studies have revealed that the trainability of a neural network exhibits distinct dynamical regimes, with sharp changes in performance as a function of optimisation hyperparameters such as the step size (or learning rate) and initial weight variance \cite{granziolLearningRatesFunction, sclocchiDissectingEffectsSGD2023, kalraPhaseDiagramEarly2023, aartsStochasticWeightMatrix2025, parkPhaseDiagramEigenvalue2025}.
    This perspective has led to the hypothesis that learning dynamics may exhibit collective phenomena analogous to phase transitions, where the system undergoes an abrupt change in dynamical behaviour as it crosses critical thresholds in its parameter space.

    Random matrix theory has emerged as an important framework for understanding high-dimensional learning systems.
    Early work established its role in the spectral analysis of random neural networks \cite{louartRandomMatrixApproach2017, penningtonNonlinearRandomMatrix2017}, while subsequent studies revealed nontrivial spectral structure in trained models, including heavy-tailed spectra \cite{mahoneyTraditionalHeavyTailed2019, martinImplicitSelfRegularizationDeep2021, dandiRandomMatrixTheory2024}, implicit self-regularisation, and emergent scaling behaviour \cite{defilippisScalingLawsSpectra2025, atanasovScalingRenormalizationHighdimensional2025, damicoImplicitBiasProduces2025}.
    Collectively, these results suggest a close connection between optimisation, representation learning, and spectral phase transitions.
    However, a unified dynamical theory linking these phenomena remains absent.
    %

    \begin{figure*}[tp!]
        \centering
        \includegraphics[trim={0.5cm 0cm 0.5cm 0cm,},clip,width=\linewidth]{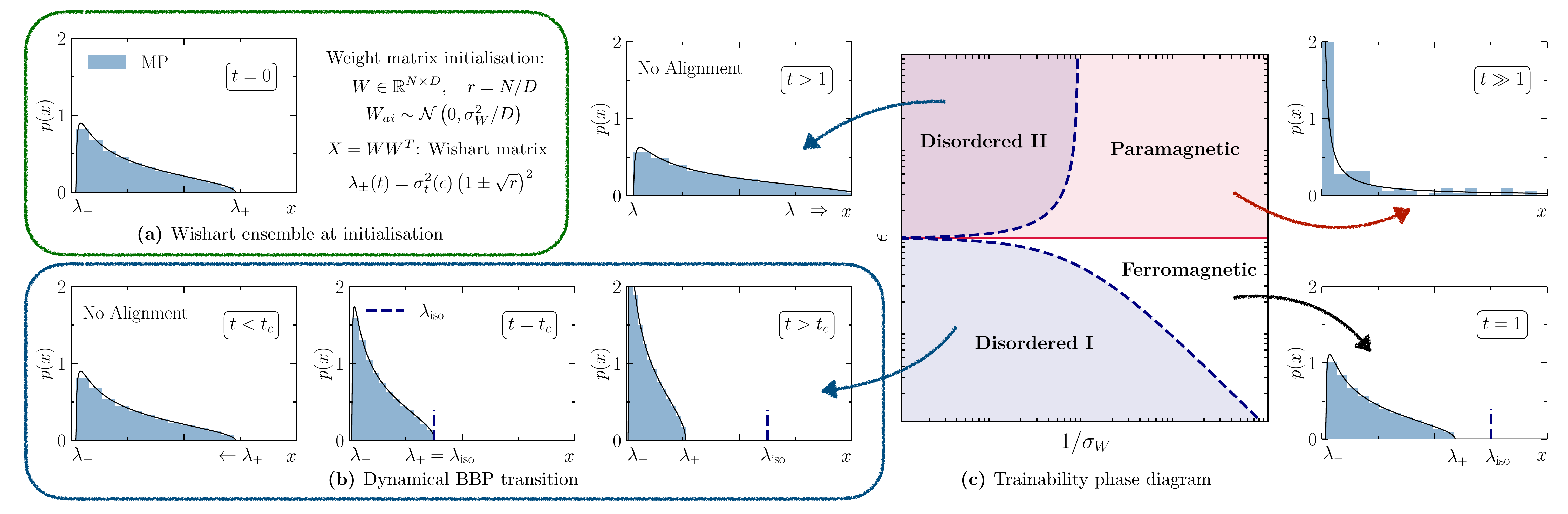}
        \caption{Summary and outline of the paper. 
        ({\bf a}) Rectangular $N\times D$ weight matrices $W$ with aspect ratio $r=N/D$ are initialised from a normal distribution with a variance proportional to $\sigma_{t=0}^2 = \sigma_W^2$. 
        The covariance matrix $X=WW^T$ is a Wishart matrix at initialisation and eigenvalues $x$ of $X$ follow the Marchenko-Pastur (MP) distribution with compact support, $\lambda_-<x<\lambda_+$. 
        ({\bf b}) Gradient descent contains finite-rank signals propagating from the data. 
        During training, the compact support of the bulk distribution is reduced due to the time dependence of the variance $\sigma_t^2$ while the signal strength is enhanced, resulting in a dynamical Baik-Ben Arous-P\'ech\'e (BBP) transition at a critical time $t=t_c$. 
        ({\bf c}) The occurrence of this phenomenon depends on the initial variance 
        $\sigma_W^2$ and the step size (or learning rate) $\epsilon$ and can be indicated in a trainability phase diagram.
        Relying on concepts from random matrix theory and disordered systems, the dynamical BBP transition takes place after a single update in the ferromagnetic phase (lower right). 
        In disordered phase I (lower left), the bulk shrinks and the dynamical BBP transition occurs at a later $t=t_c$. 
        In contrast, in disordered phase II (upper left), the bulk expands and isolated eigenvalues do not emerge, while in the paramagnetic regime (upper right), the dynamics are either unstable or spurious isolated eigenvalues appear, not aligned to the signal, and a power law spectrum develops eventually.
        In Sec.~\ref{sec:2.solvable_model}, the full time-dependent scenario is derived analytically for a linear teacher-student model with gradient descent.
        The analysis is then extended to nonlinear networks in Sec.~\ref{sec:3.extensions}, to stochastic gradient descent in Sec.~\ref{sec:4.stochastic}, and to realistic data sets in Sec.~\ref{sec:5.realistic}.
        }
        \label{fig:introdunction}
    \end{figure*}
    %
    In this work, we propose that a dynamical spectral phase transition provides a common mechanism underlying trainability and representation learning in neural networks.
    We argue that stochastic gradient descent (SGD) generates a competition between a finite-rank gradient signal \cite{baHighdimensionalAsymptoticsFeature2022, dandiHowTwoLayerNeural2025} and the disorder associated with the initially random weight matrix, continuously reshaping the spectrum throughout training.
    The emergence of informative representations is then realised by isolated eigenvalues detaching from the random spectral bulk of a weight matrix, with the corresponding eigenvectors acquiring nonzero alignment with the learning signal.

    This spectral phase transition, namely the appearance of an isolated and informative eigenmode outside the continuous spectral bulk, is known as the Baik-Ben Arous-Péché (BBP) transition \cite{baikPhaseTransitionLargest2005}.
    Originally observed in disordered magnetic systems \cite{edwards1976eigenvalue} and studied in unsupervised learning and high-dimensional inference \cite{watkin1994optimal, hoyleLimitingFormSample2003, paul2007asymptotics, montanari2015limitation, perry2018optimality, adomaitytePCARecoveryThresholds2025, bocchiDiscontinuousBBPTransitions2026}, BBP transitions have become a central paradigm for understanding information recovery in noisy high-dimensional systems.
    In neural networks, BBP spectral transitions have also been observed, with isolated modes emerging in weight covariances \cite{wangNonlinearSpikedCovariance2024} and in Hessian spectra \cite{annesiOverparametrizationBendsLandscape2025, bonnaireRoleTimeDependentHessian2025}.
    Very recently, dynamical spectral alignment during learning has been studied as well \cite{arousSpectralAlignmentStochastic2025, arousLocalGeometryHighdimensional2026, lauditiSpectralDynamicsDeep2026}.
    Together, these results support that BBP-type transitions may provide a unifying framework for understanding representation formation and trainability in high-dimensional learning systems.

    To establish this picture, we first demonstrate the dynamical BBP transition during learning in a solvable linear teacher-student model with a rank-1 teacher matrix in Sec.~\ref{sec:2.solvable_model}.
    A graphical overview of the paper is shown in Fig.~\ref{fig:introdunction}.
    We employ this solvable model to establish the conceptual framework of dynamical BBP transitions and define the trainability phase diagram in terms of step size and initial weight variance.
    The phase diagram reveals regimes in which the BBP transition is transient, occurring only over a finite range of learning rates for a given initial variance.

    Although the rank-1 teacher-student setting serves as a minimal example, the mechanism extends naturally to general cases, where the teacher matrix or ground state value is not known a priori. 
    The spectral alignment in this case is studied through the self-overlap, which acts as an effective order parameter of the BBP-type spectral phase transition, allowing the same framework to characterise the emergence and stability of the spectral properties.
    In Sec.~\ref{sec:3.extensions}, we then show that the same mechanism persists in nonlinear and multilayer networks, where the phase boundaries retain the qualitative structure predicted by the linear theory. 

    \vspace{0.5cm}
    We further investigate the effects of stochasticity and finite datasets in Sec.~\ref{sec:4.stochastic}.
    In particular, we demonstrate that finite sample fluctuations scale the transient BBP behaviour, controlled by the load parameter $\alpha = P/D$, where $P$ and $D$ denote the dataset size and input dimension, respectively.
    In the stationary regime of stochastic gradient descent, we show that the weight covariance satisfies a fluctuation-dissipation relation determined by the Hessian and gradient fluctuation around the ground state, linking late-time spectral structure to the local geometry of the loss landscape.

    Finally, in Sec.~\ref{sec:5.realistic} we demonstrate that the proposed framework remains predictive in realistic learning settings, where the informative signals span low-dimensional subspaces \cite{henaffLocalLowdimensionalityNatural, goldtModelingInfluenceData2020, leviUnderlyingScalingLaws2024}.
    Using deep neural networks trained on real datasets, we show that the empirical phase diagram of final test loss and the associated spectral alignment persist beyond the solvable model, providing empirical evidence that dynamical BBP transitions offer a useful organising principle for understanding trainability and representation learning in neural networks.

    Technical derivations, additional numerical results, and analyses of deeper architecture are presented in the Appendices.

\section{Solvable model of a dynamical BBP transition}
\label{sec:2.solvable_model}

    The gradient descent method and its variants are the most widely used optimisation algorithms for training neural networks.
    Given a dataset, the task of learning is to find the optimal values of weight matrices $W$ in neural networks that minimise a loss function, defined as the distance between a model output and the target values.
    Denoting the per-sample loss function as $\ell (W, x)$, where $x$ is a sample from a dataset, a single gradient descent step with step size $\epsilon$ is
    \begin{align}
        W_{ai}' = W_{ai} - \epsilon K_{ai},
        \qquad
        K_{ai} \equiv \sexpv[x]{\frac{\partial \ell(W,x)}{\partial W_{ai}}},
    \end{align}
    where the drift $K_{ai}$ is a gradient matrix averaged over the data set.

    In practice, the weight matrix $W$ at initialisation is randomly sampled from an i.i.d.\ distribution with a variance proportional to $\sigma_W^2$, and its singular values asymptotically follow the Marchenko-Pastur distribution in the large matrix size limit, defined on a compact support.
    The informative component of the learning signal is encoded in the drift matrix $K$, which often contains a finite-rank signal on top of a noisy bulk \cite{baHighdimensionalAsymptoticsFeature2022, dandiHowTwoLayerNeural2025}.
    From the perspective of random matrix theory, the gradient update is a finite-rank perturbation to the initial random matrix, where the signal strength is determined by the step size and norm of the drift matrix.
    Depending on the initial variance $\sigma_W^2$ and the signal strength, isolated eigenvalues may emerge from the spectrum of the updated weight matrix $W'$, whose associated eigenvectors have a nonzero overlap with the signal direction.
    This is precisely the dynamical manifestation of a BBP transition during optimisation, whereby an isolated mode aligned with the training signal emerges from the random spectral bulk.

    To demonstrate the mechanism of dynamical BBP transitions in a gradient update, consider a linear model in a teacher-student setting as a fully solvable case.
    Conceptually, the teacher weight matrix represents the ground state value that the student model aims to learn from the data.
    Subsequent analysis can be applied in general neural network architectures; the teacher-student setup is introduced here as a theoretical demonstration.

    Define the teacher model $y$ and student model $\hat{y}$ as
    \begin{align}
        y_a = \sum_{i=1}^{D} W_{ai}^{\ast} x_{i} ,
        \qquad
        \hat{y}_a = \sum_{i=1}^{D} W_{ai} x_{i} ,
    \end{align}
    where $\boldsymbol{x} \in \mathbb{R}^{D}$ is the input data vector, $W^{\ast} \in \mathbb{R}^{N\times D}$ is the teacher weight matrix, and $W \in \mathbb{R}^{N\times D}$ is the student weight matrix.
    In the teacher-student setting considered here, the BBP transition corresponds to the emergence of eigendirections in the student matrix aligned with the teacher signal.

    For a quadratic loss function 
    \begin{align}
    \begin{aligned}
        \ell \left( W, x \right) &= \frac{1}{2} \sum_{a=1}^{N}
        \left( y_a (x)- \hat{y}_a \left(W, x\right) \right)^2 \\
        &= \frac{1}{2} \sum_{a=1}^{N}
        \left( \sum_{i=1}^{D} \left( W_{ai}^{\ast} - W_{ai} \right) x_{i} \right)^2,
        \end{aligned}
    \end{align}
    the gradient of the loss function with respect to the student weight matrix is given as
    \begin{align}
        \frac{\partial \ell}{\partial W_{ai}}
        = - \sum_{j=1}^{D} \left( W_{aj}^{\ast} - W_{aj} \right) x_{j} x_{i}.
    \end{align}

    Assuming we have an infinite dataset, the full batch gradient is obtained as an expectation value of the gradient over the true data distribution, which we assume to be an uncorrelated normalised Gaussian distribution (extensions to a finite dataset, stochastic optimisation, and a non-Gaussian data distribution are considered in Sec.~\ref{sec:4.stochastic} and Appendix~\ref{appendix:I.non-gaussian}).
    The expectation value of the gradient becomes
    \begin{align} \begin{split} \label{eq:grad_loss}
        \sexpv[x]{\frac{\partial \ell}{\partial W_{ai}}}
        &= - \sum_{j=1}^{D} \left( W_{aj}^{\ast} - W_{aj} \right)
        \sexpv[x]{x_{j} x_{i}} \\
        &= - \left( W_{ai}^{\ast} - W_{ai} \right).
    \end{split} \end{align}
    A single gradient descent step with step size $\epsilon$ is 
    \begin{align} \label{eq:linear_update_rule}
        W_{ai}' = \left( 1 - \epsilon \right) W_{ai} + \epsilon W^{\ast}_{ai}.
    \end{align}

    As a minimal analytically tractable setting, we consider a rank-1 teacher matrix, constructed as
    \begin{align} \label{eq:teacher_w}
        W^{\ast} = \bvec{u}_{} \bvec{v}^T,
    \end{align}
    where $\bvec{u} \in \mathbb{R}^N$ and $\bvec{v} \in \mathbb{R}^{D}$ are arbitrary vectors on the unit sphere, $\Norm{\bvec{u}} = \Norm{\bvec{v}}=1$, not known a priori in general.
    The eigenmode $\bvec{w}$ resonant to the signal direction is separated from the student matrix
    \begin{align}
        W = W_{\perp} + \bvec{w_{}} \bvec{v}^T,
    \end{align}
    where each component is defined as
    \begin{align}
        W_{\perp} = W \left( I - \bvec v_{} \bvec v^T\right),
        \qquad
        \bvec{w} = W_{} \bvec v.
    \end{align}
    The random bulk and the signal are separated in the update equation,
    \begin{align} \begin{split} \label{eq:masked_update}
        W' &= (1 - \epsilon ) W_{\perp} + \left((1 - \epsilon)_{} \bvec w + \epsilon_{} \bvec u \right) \bvec v^T\\
        &= (1 - \epsilon ) W_{\perp} + \bvec s_{} \bvec v^T,
    \end{split} \end{align}
    where we have defined the masked rank-1 signal vector 
    \be
    \bvec s = (1 - \epsilon) \bvec w + \epsilon \bvec u.
    \ee
    This is a typical scenario of a finite rank perturbation to a rectangular random matrix \cite{benaych-georgesSingularValuesVectors2012, fornerBBPPhaseTransition2025, pottersFirstCourseRandom}.

    Generalisation to the rank-$k$ case can be incorporated following standard methods \cite{pottersFirstCourseRandom, rosHighdimensionalRandomLandscapes2025}. 
    This finite-rank signal setting captures an important feature often observed in overparameterised learning, namely that task-relevant information can be concentrated in a low-dimensional subspace of a much larger parameter space \cite{dandiHowTwoLayerNeural2025, huLoRALowRankAdaptation2021, kaushikUniversalWeightSubspace2025, sclocchiProbingLatentHierarchical2024, yooGeometricStructureLayer2026, huangSharpDescriptionLocal2026}.
    Although the solvable theory focuses on a rank-1 signal, Sec.~\ref{sec:5.realistic} provides empirical evidence that the same mechanism extends to realistic settings, where aligned signal subspaces emerge from a random spectral bulk.
    For the special case where the rank of the perturbation matrix also scales with the size of the network, one can utilise the extensive rank BBP transition formalism for rectangular random matrices \cite{fornerBBPPhaseTransition2025}.

    To study the spectral properties of the weight matrix, we consider the weight covariance matrix
    \begin{align}
        X_{ab} \equiv \sum_{i=1}^{D} W_{ai} W_{bi}.
    \end{align}
    After a single update step,
    \begin{align} \label{eq:BBP_cov}
        X' = (1 - \epsilon)^2 X_{\perp} + \bvec s_{} \bvec s^T.
    \end{align}
    Assuming that the student weight matrix elements are initialised from a centred Gaussian distribution with variance $\sigma_W^2/D$, $W_{ai} \sim \mathcal{N} \left(0, \sigma_W^2 / D\right)$, which is a standard practice in neural network initialisation \cite{glorotUnderstandingDifficultyTraining2010, heDelvingDeepRectifiers2015}, the matrix $X_{\perp}$ is a Wishart matrix with variance $\sigma_W^2$ and aspect ratio $r=N/D$.
    Introducing the normalised student weight covariance matrix $\hat X_{\perp}$ and the normalised signal vector $\bvec{\hat s}$ as
    \be
    \hat X_{\perp} = X_{\perp} / \sigma_W^2, 
    \qquad
    \bvec{\hat s} = \bvec{s} / \Norm{\bvec{s}},
    \ee
    the single step update Eq.~(\ref{eq:BBP_cov}) is rewritten as
    \begin{align}
        X' = \sigma^2 (\epsilon) \hat X_{\perp} + \theta^2 (\epsilon) \bvec{\hat s} \bvec{\hat s}^T,
    \end{align}
    where the effective variance $\sigma^2 (\epsilon)$ of the bulk and the effective signal strength $\theta^2 (\epsilon)$ are extracted, as
    \begin{align}
    \begin{aligned}
        \sigma^2 (\epsilon) &= (1 - \epsilon)^2 \sigma_W^2,
        \\
        \theta^2 (\epsilon) &= \Norm{\bvec{s}}^2 = (1-\epsilon)^2 r \sigma_W^2 + \epsilon^2.
        \end{aligned}
    \end{align}
    This model contains all the essential elements for the mechanism of BBP transitions in gradient descent optimisation, including the role of the hyperparameters.
    The competition between disorder in the system and the signal strength, scaled by factors $\sigma^2(\epsilon)$ and $\theta^2 (\epsilon)$, determines the BBP transition in the updated weight covariance matrix.
    Unlike as in standard BBP settings, however, the step size $\epsilon$ simultaneously controls the spike amplitude and the random bulk, which makes the BBP transition dynamical and induces transient behaviour.

\subsection{Optimal step size and critical step size}
\label{sec:linear_dynamic}
    In the case of a linear model with full batch update, the time-dependent covariance matrix can be calculated analytically.
    The weight covariance matrix at iteration time $t$ is written in terms of the initial bulk covariance matrix, as
    \begin{align} \begin{split} \label{eq:linear_multistep}
        X^{(t)} &= \sigma_t^2 (\epsilon) \hat X_{\perp}^{(0)} + \theta_t^2 (\epsilon) \bvec{\hat s}_{t} {\bvec{\hat s}_{t}}^T,
    \end{split} \end{align}
    which can be obtained by applying Eq.~(\ref{eq:BBP_cov}) recursively. 
    The time dependent signal vector is
    \begin{align}
        \bvec{s}_t &= (1-\epsilon)^{t} \bvec{w} + \left[1 - \left(1 - \epsilon\right)^t\right] \bvec{u},
    \end{align}
    and the effective variance and signal strength read
    \begin{align} \begin{split}
        \sigma_t^2 (\epsilon) &= (1 - \epsilon)^{2t} \sigma_W^2,
        \\
        \theta_t^2 (\epsilon) &= (1 - \epsilon)^{2t} r \sigma_W^2 + \left[1 - (1 - \epsilon)^t \right]^2.
    \end{split} \end{align}
    Considering this time dependence, the initial randomness shrinks monotonically for step sizes smaller than one, $\epsilon < 1$, as $t$ increases.
    Then, in the long-time limit,
    \begin{align}
        \lim_{t\to \infty} \sigma_{t}^{2} (\epsilon) \to 0,
        \quad
        \lim_{t \to \infty} \theta_t^2 (\epsilon) \to 1,
        \quad
        \lim_{t \to \infty} \bvec{\hat s}_t \to \bvec{u},
    \end{align}
    and the student weight covariance matrix converges to the target value $W^{\ast} {W^{\ast}}^T = \bvec{u} \bvec{u}^T$ exponentially, 
    \begin{align}
        X^{(t)} - \bvec{u}_{} \bvec{u}^T\sim C e^{- \gamma (\epsilon)_{} t},
    \end{align}
    with convergence rate $\gamma (\epsilon)$,
    \begin{align}
        \gamma(\epsilon) = - \log \left(1 - \epsilon \right) > 0.
    \end{align}

    When $1 < \epsilon < 2$, the optimisation still converges, but with oscillatory behaviour, and if $\epsilon > 2$, the optimisation diverges.
   If the step size is equal to one, the ground state is immediately reached after a single update and the random bulk is completely suppressed.
    We denote this special value by $\epsilon_o$ and refer to it as the optimal step size.
    In the linear model, $\epsilon_o=1$ separates the monotone convergence regime ($\epsilon<\epsilon_o$) from oscillatory or unstable dynamics ($\epsilon>\epsilon_o$).
    This observation motivates a more general definition of the optimal step size as the value that minimises the effective bulk variance,
    \begin{align} \label{eq:optimal_stepsize}
        \left. \frac{ d \sigma_t^{2}(\epsilon)}{d \epsilon}\right|_{\epsilon = \epsilon_o} = 0.
    \end{align}
     This is interpreted as a spectral reference scale characterising the most efficient suppression of the random bulk, rather than as a general statement about optimal loss or generalisation.
    In general, $\epsilon_o$ is time dependent.
    However, as the early stage dynamics of the student matrix is exponential, the spectral properties of the weight matrix are largely determined by the initial conditions and the first few updates.
    As a result, the optimal step size computed for the initial step remains highly predictive.
    This dynamical behaviour is further shown in Sec.~\ref{sec:dyanmical_bbp_transition}.

    The critical step size $\epsilon_c$, marking the onset of the BBP transition and the emergence of an isolated signal mode, is determined by the bulk spectrum of the weight covariance matrix $X$.
    The spectral density of $X$ follows the Marchenko-Pastur distribution with variance $\sigma_t^2 (\epsilon)$ and aspect ratio $r=N/D$, 
    \begin{align}
        \rho_{{\rm MP}_r}(z) = \frac{1}{2 \pi r \sigma_t^2 (\epsilon) z} \sqrt{\left(\lambda_{+} - z\right) \left(z - \lambda_{-}\right)},
    \end{align}
    for $\lambda_- < z < \lambda_+$ and zero elsewhere, and 
    \begin{align}
        \lambda_{\pm} = \sigma_t^2 (\epsilon) \left( 1 \pm \sqrt{r}\right)^2
    \end{align}
    are the left $(-)$ and right $(+)$ edge of the bulk distribution.
    Then, $\epsilon_c$ is determined by the BBP threshold equation (see Appendix~\ref{appendix:A.derivation_linear}),
    \begin{equation} \label{eq:bbp_condition}
        \theta_t^2 (\epsilon_c) = \frac{1}{g_X(\lambda_{+})},
    \end{equation}
    where $g_X(z)$ is the resolvent of the weight covariance matrix $X$.
    For the Marchenko-Pastur distribution, the BBP threshold equation simplifies to
    \begin{align} \label{eq:eps_c_linear}
        \theta_t^2(\epsilon_c) = \sigma^2_t (\epsilon_c) \left( \sqrt{r} + r \right),
    \end{align}
    and the critical step size $\epsilon_c$ is defined as the solution to Eq.~(\ref{eq:eps_c_linear}).
    Typically, Eq.~(\ref{eq:eps_c_linear}) is quadratic in $\epsilon_c$, and there exist two solutions that define lower and upper critical boundaries.
    Along the lower branch, $\epsilon_c < \epsilon_o$, the random bulk shrinks during training, eventually allowing an isolated signal mode to detach from the spectrum through a dynamical BBP transition, as discussed in Sec.~\ref{sec:dyanmical_bbp_transition}.
    By contrast, for the upper branch with $\epsilon_c > \epsilon_o$, the bulk expands as training progresses, keeping the signal embedded within the random spectrum and preventing the formation of an isolated aligned mode.

    Below, we define the trainability phase diagram of optimisation dynamics, where different phases are separated by the optimal step size $\epsilon_o$ and critical step size~$\epsilon_c$.

    \begin{figure}[tp!]
        \centering
        \includegraphics[trim={0.5cm 0.8cm 0.5cm 0.5cm,},clip,width=\columnwidth]{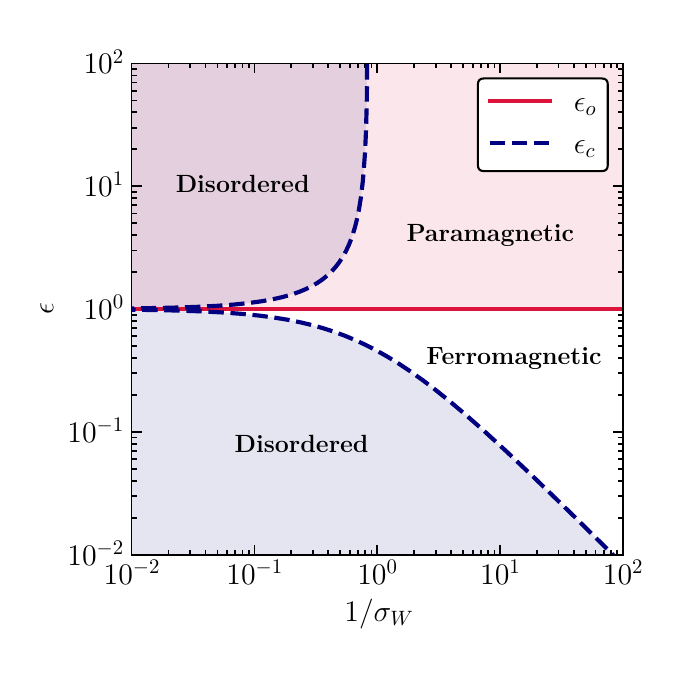}
        \caption{Trainability phase diagram of the linear teacher-student model predicted by random matrix. Three phases separate the phase diagram into four dynamical regimes. The phase boundaries are defined by the optimal step size $\epsilon_o$ (red solid line) and the critical step size $\epsilon_c$ (blue dashed line). Above $\epsilon_o$ corresponds to a non-monotonic large-step regime. In the linear model, optimisation remains convergent but oscillatory for $1<\epsilon<2$, while it diverges for $\epsilon>2$. We refer to this large-step regime as paramagnetic. The disordered phases are classified by $\epsilon_c$, where the signal remains submerged within the random bulk, and no spectral alignment occurs after a single update. There are two disordered phases. The disordered phase reappearing with the larger step size originates from the fact that the step size also rescales the bulk along the signal strength. This leads to the transient behaviour of the BBP transition, further discussed in Sec.~\ref{sec:proportional_limit}. Between these boundaries lies the ferromagnetic phase, where a BBP transition produces an isolated eigenvalue aligned with the learning signal. This region lies within the monotonic stable-convergence regime and corresponds to the emergence of spectrally visible informative structure. The phase diagram is shown for the aspect ratio $r=0.5$.}
        \label{fig:phase_boundary_linear}
    \end{figure}

\subsection{Trainability phase diagram after a single update}
    The trainability phase diagram is spanned by the step size $\epsilon$ and $1/\sigma_W$, where $\sigma_W^2$ is the initial randomness of the weight matrix. 
    The initial variance $\sigma_W^2$ sets the scale of the initial random spectral bulk, while $\epsilon$ controls the magnitude of the gradient induced perturbation and the location of the bulk edge.
    Varying these parameters tunes the balance between signal and noise in the learning dynamics.
    The trainability phase diagram for the linear teacher-student model after a single optimisation step ($t=1$) is shown in Fig.~\ref{fig:phase_boundary_linear}.

    Two characteristic step sizes, $\epsilon_o$ and $\epsilon_c$, partition the phase diagram into distinct learning regimes.
    Above $\epsilon_o$, optimisation becomes unstable, defining the paramagnetic regime.
    Below $\epsilon_c$, the learning signal remains embedded within the random bulk, corresponding to a disordered phase.
    Between these two boundaries, a BBP transition produces an isolated mode aligned to the signal, giving rise to a ferromagnetic phase.

    We have adopted the nomenclature of disordered systems, but note that while the boundary $\epsilon_c$ is associated with a spectral phase transition, $\epsilon_o$ is an optimisation threshold.
    We therefore refer to the regions separated by $\epsilon_c$ as the ferromagnetic and disordered phases, while the unstable region above $\epsilon_o$ is denoted the paramagnetic regime.

    For linear models with full batch gradient after a single update, the optimal boundary is located at $\epsilon_o =1 $, and the BBP boundary $\epsilon_c$ is given by the solution to Eq.~(\ref{eq:eps_c_linear}),
    \begin{align} \label{eq:boundaries_linear}
        \epsilon_o = 1,
        \qquad
        \epsilon_c^{\pm} = \left(1 \pm \frac{1}{r^{\frac{1}{4}}\sigma_W} \right)^{-1}.
    \end{align}
    The two boundaries separate the four regimes of training dynamics.
    Above the optimal boundary $(\epsilon > \epsilon_o)$, the optimisation is unstable, oscillatory or divergent, corresponding to the paramagnetic regime observed in Ref.~\cite{parkPhaseDiagramEigenvalue2025}.
    In the paramagnetic regime, informative eigenvalues may emerge from the bulk after one gradient descent step due to the strong perturbation.
    However, the initial nonzero overlap in this regime decreases as the optimisation proceeds, as shown later in this section, causing this alignment to be lost during the dynamics.

    Below the lower BBP transition boundary ($\epsilon < \epsilon_c^+$), the system is in a disordered phase. The signal strength from the gradient is relatively weak compared to the initial randomness and the signal is submerged in the random bulk.
    In this phase, none of the eigenvalues is aligned to the signal direction, slowing down the optimisation and information retrieval.
    A second disordered phase exists in the large step size and large initial randomness region ($\epsilon > \epsilon_c^{-}$), embedded in the paramagnetic regime in the upper left corner of Fig.~\ref{fig:phase_boundary_linear}, corresponding to the upper branch of critical step size, see Eq.~(\ref{eq:boundaries_linear}).
    This phase is related to the transient BBP transition \cite{coeurdouxRandomMatrixTheory2026a}, and will be further discussed in Sec.~\ref{sec:proportional_limit}.

    The gap between the optimal boundary and the critical BBP boundary ($\epsilon_c^+ < \epsilon < \epsilon_o$), appearing for large $1/\sigma_W$ and smaller $\epsilon$, is the favourable region to initialise the hyperparameters. This is the ferromagnetic phase, where the convergence is guaranteed at a faster speed with respect to other regimes \cite{parkPhaseDiagramEigenvalue2025}.
    For large initial weight variance ($\sigma^2_W \gg 1$), the critical step size asymptotically approaches the optimal step size, and the ferromagnetic gap closes.
    This indicates that an optimal set of hyperparameters does not exist in this limit,
    \begin{align}
        \epsilon_c^{\pm} \ \to \ \epsilon_o, \quad {\rm for} \ \sigma_W^2 \gg 1 .
    \end{align}
    The dependence of the phase diagram on the aspect ratio is shown in Appendix~\ref{appendix:B.alpha_r}.

    \begin{figure}[tp!]
        \centering
        \includegraphics[trim={0.5cm 0.5cm 0.5cm 0.2cm,},clip,width=\columnwidth]
        {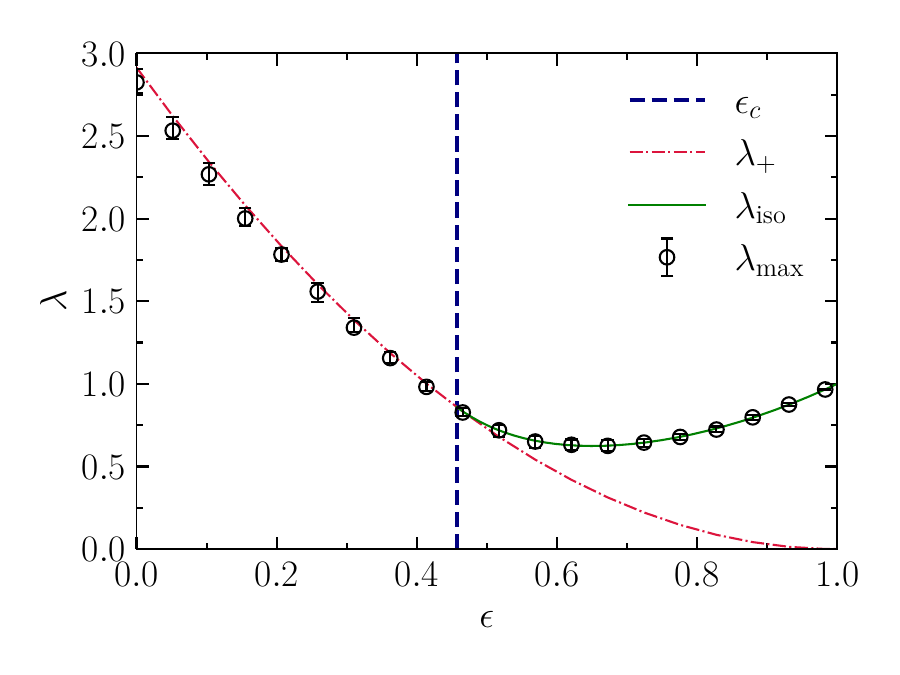}
        \caption{Location of the maximum eigenvalue $\lambda_{\rm max}$ with respect to the step size at fixed initial randomness $\sigma^2_W = 1$. The maximum eigenvalue of the weight covariance matrix after a single update $X^{(1)}$ (black dots) follows the right edge of the Marchenko-Pastur bulk $\lambda_{+}$ (red dash-dotted line) below the critical step size. An isolated eigenvalue emerges when the step size crosses the critical step size $\epsilon_c$ (blue dashed line), and follows the random matrix theory prediction $\lambda_{\rm iso}$ (green solid line). We only show the lower critical branch $\epsilon_c = \epsilon_c^+$ here, as the upper critical branch is in the paramagnetic regime for linear models, and is irrelevant. For figures in Sec.~\ref{sec:order_parameters_and_observables}, the dimension of the weight matrix used in the numerical experiments is $N=100$, with aspect ratio $r=N/D=0.5$, unless indicated otherwise. }
        \label{fig:lambda_flow_linear}
    \end{figure}
\subsection{Order parameters and observables}
\label{sec:order_parameters_and_observables}

    To numerically confirm the presence of the different regions in the trainability phase diagram, we construct a set of observables, namely the location of the maximum eigenvalue $\lambda_{\rm max}$ of the weight covariance matrix, the square overlap $q^2$ between the signal and the top eigenvector, and the self-overlap $q^{\alpha \beta}$.
    Their definitions are given below.
    Detailed derivations of the corresponding theoretical predictions are provided in Appendix A, while comparisons with numerical results are presented here.

    As a realisation of the rank-1 teacher matrix used in the simulation, we choose two unit vectors
    \begin{align} \label{eq:teacher_vectors_definitions}
        \bvec{u} &= \frac{\textbf{u}}{\Norm{\textbf{u}}}\in \mathbb{R}^{N},
        \quad
        \text{u}_a \sim \mathcal{N} (0, 1),\\
        \bvec{v} &= \frac{\textbf{v}}{\Norm{\textbf{v}}}\in \mathbb{R}^{D},
        \quad
        \text{v}_i \sim \mathcal{N} (0, 1),
    \end{align}
    to construct the rank-1 teacher matrix
    \begin{align}
        W_{ai}^{\ast} = u_a v_i.
    \end{align}
    The dimension of the weight matrices used in the simulations is $N=100$ with aspect ratio $r=0.5$.

    Between the lower and upper critical branches, ($\epsilon_c^{+} < \epsilon < \epsilon_c^{-}$), the maximum eigenvalue $\lambda_{\rm max}$ of the updated weight covariance matrix $X'$ detaches from the bulk spectrum and emerges as an isolated eigenvalue, whose corresponding eigenvector aligns with the signal direction.
    Fig.~\ref{fig:lambda_flow_linear} shows the location of $\lambda_{\rm max}$ after a single update, obtained using numerical simulations.
    The maximum eigenvalue $\lambda_{\rm max}$ follows the right edge of the Marchenko-Pastur bulk $\lambda_{+}$ below the critical step size $\epsilon_{c}$, then detaches from the bulk and follows the predicted location of the isolated eigenvalue $\lambda_{\rm iso}$ above the critical value.
    The theoretical prediction of the isolated eigenvalue is given as a function of effective bulk variance $\sigma_t^2 (\epsilon)$ and signal strength $\theta_t^2 (\epsilon)$ as
    \begin{align} \label{eq:iso_linear}
        \lambda_{\rm iso} (\epsilon) = \theta_t^2 (\epsilon)\left[1 + \frac{ \sigma_t^2 (\epsilon)}{\theta_t^2 (\epsilon) - r\sigma_t^2(\epsilon)}\right],
    \end{align}
    for all $\epsilon_c^+ < \epsilon < \epsilon_c^-$.

    \begin{figure}[tp!]
        \centering
        \includegraphics[trim={0.5cm 0.6cm 0.5cm 0.5cm,},clip,width=\columnwidth]{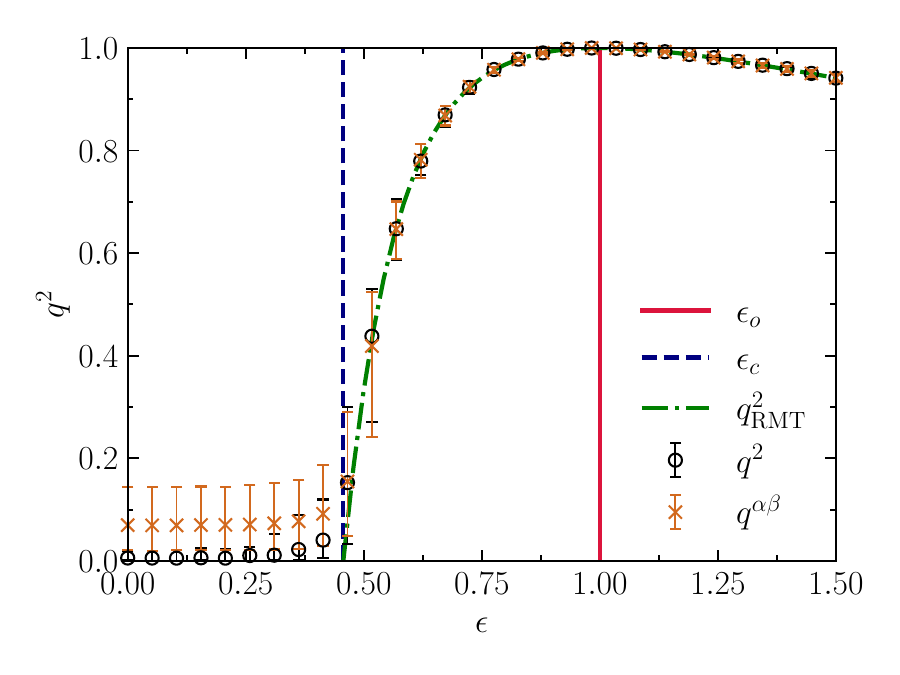}
        \caption{Squared overlap $q^2$ and self-overlap $q^{\alpha \beta}$ at different step size $\epsilon$ for fixed initial randomness $\sigma_W^2=1$.  The overlap is small below the critical step size and achieves perfect alignment, $q^2 = 1$ at $\epsilon = \epsilon_o$, then the overlap deteriorates as the step size exceeds the optimal step size $\epsilon_o$. The gap between critical step size $\epsilon_c$ and the optimal step size $\epsilon_o$ is the ferromagnetic phase. The computed overlaps match the theoretical value $q_{\rm RMT}^2$. Finite-size scaling of the overlap $q^2$ is presented in Appendix~\ref{appendix:C.finite_size}.}
        \label{fig:q2_overlap_linear}
    \end{figure}
    The effective order parameter of the BBP transition is the squared overlap between the signal vector $\bvec{u}$ and the eigenvector $\bvec{w}_{\rm max}$ corresponding to the maximum eigenvalue of $X^{(t)}$.
    The squared overlap $q^2$ is defined as
    \begin{align}
        q^2 \equiv \sexpv[\bvec{w}]{(\bvec{w}_{\rm max} \cdot \bvec{u})^2},
    \end{align}
    where the expectation value is obtained by averaging over different realisations of the student weight matrices.
    The overlap is nonzero only if the step size is larger than the critical value,
    \begin{align}\begin{split}
        q^2 \neq 0, &\qquad \epsilon_c^+ < \epsilon < \epsilon_c^- , \\ 
        q^2=0, &\qquad \text{otherwise} .
    \end{split}\end{align}
    The value of the overlap can be calculated analytically (see Appendix~\ref{appendix:A.derivation_linear}),
    \begin{align} \label{eq:q2_linear}
        q^2(\epsilon,\sigma_W;t)
        = 1 - \frac{r \sigma_t^2 (\epsilon)}{\theta_t^2 (\epsilon)}
        \left[1 + \frac{\sigma_t^2 (\epsilon)}{\theta_t^2 (\epsilon) - r \sigma_t^2 (\epsilon) } \right],
    \end{align}
    for all $\epsilon_c^+ < \epsilon < \epsilon_c^-$.
    Outside this range, the maximum eigenvector remains uncorrelated to the signal direction, implying that $q^2=0$ in the large matrix limit.

    \begin{figure}[tp!]
        \centering        
        \includegraphics[trim={0.2cm 0.3cm 0.2cm 0cm,},clip,width=\columnwidth]{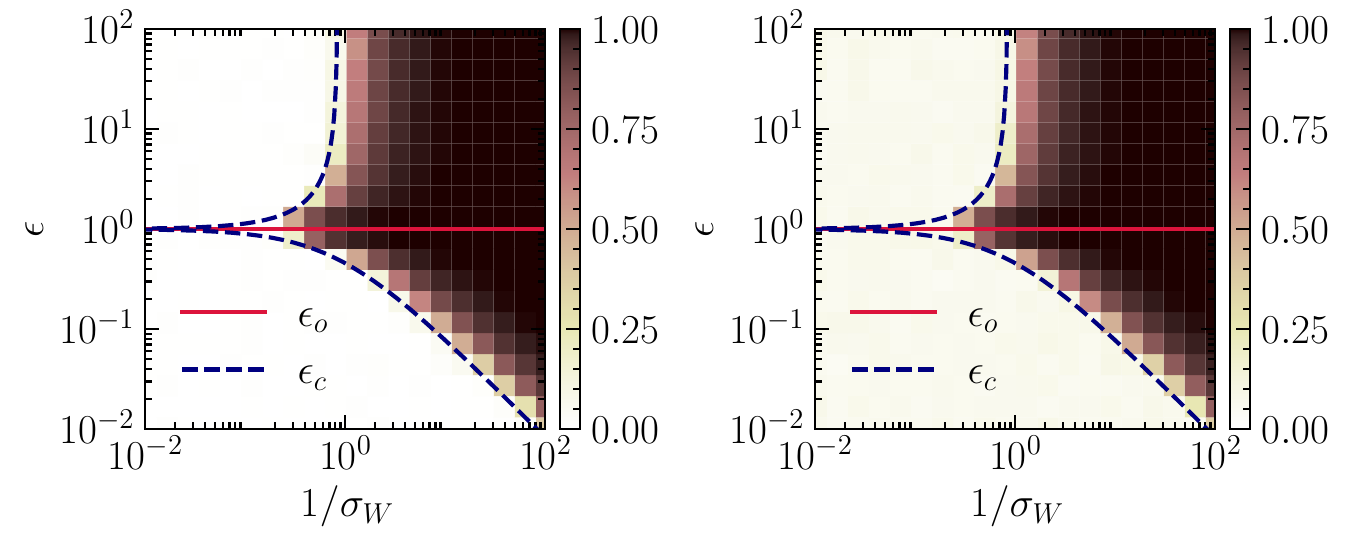}
        \caption{(Left) squared overlap $q^2$ and (right) self-overlap $q^{\alpha \beta}$. The self-overlap $q^{\alpha \beta}$ effectively captures the BBP transition without the need to know the teacher matrix. This is useful in realistic neural network training scenario where the notion of a teacher matrix is absent.}
        \label{fig:top_eig_diff_exp_label}
    \end{figure}
    \begin{figure*}[tp!]
        \centering
        \includegraphics[trim={1.2cm 1.2cm 1.2cm 1cm,},clip,width=0.32\linewidth]{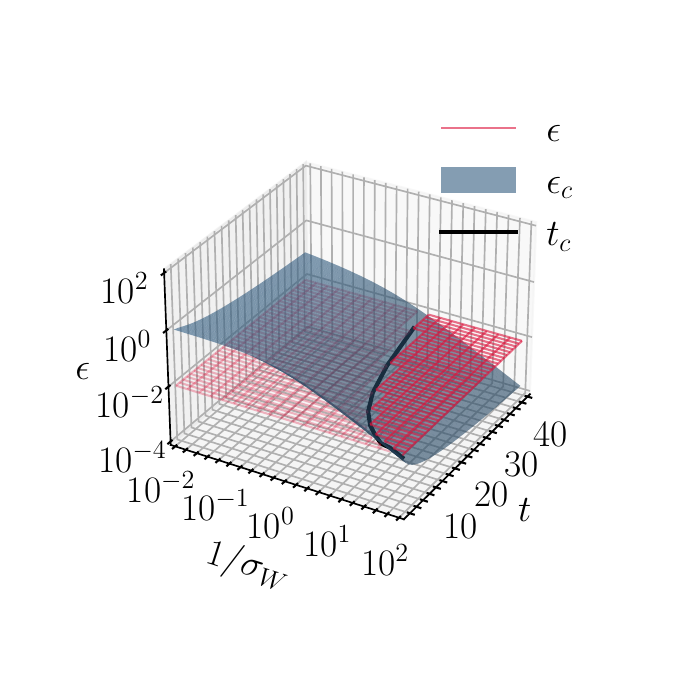}
        \includegraphics[trim={1.2cm 1.2cm 1.2cm 1cm,},clip,width=0.32\linewidth]{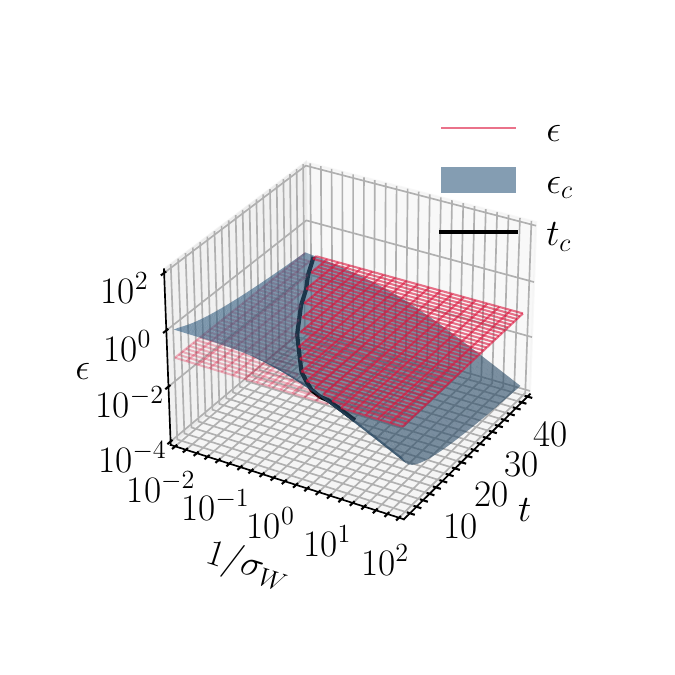}
        \includegraphics[trim={1.2cm 1.2cm 1.2cm 1cm,},clip,width=0.32\linewidth]{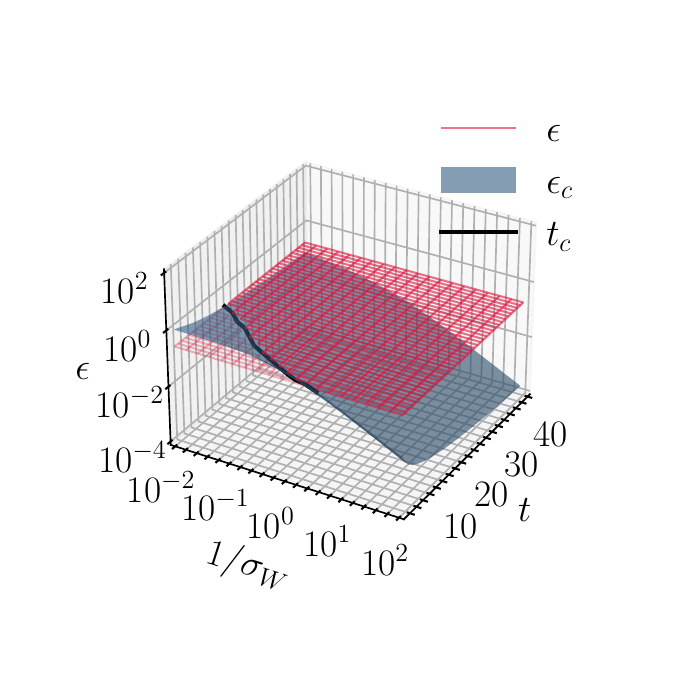}
        \caption{BBP transition surface for different step size $\epsilon = (0.01, 0.1, 0.25)$ from left to right. The step size used in the optimisation is marked with a red horizontal mesh, and the critical step size surface is shown by a blue curved plane. Even if the step size is smaller than the critical value at initialisation, the critical step size decreases in time and eventually becomes smaller than the step size, inducing a dynamical BBP transition at critical time $t_c$. Only the lower branch of the critical step size $\epsilon_c$ is shown here, as the upper branch is above the optimal boundary and becomes irrelevant.}
        \label{fig:dynamical_bbp_linear}
    \end{figure*}
    The numerical evaluation of the squared overlap $q^2$ for fixed variance $\sigma^2_W=1$ is shown in Fig.~\ref{fig:q2_overlap_linear}, and the values for the complete phase diagram are shown in the left panel of Fig.~\ref{fig:top_eig_diff_exp_label}.
    Outside of the critical step size boundary, there is no overlap as the signal is submerged in the random bulk, and the overlap increases as the step size becomes larger than the critical value.
    If the step size becomes bigger than the optimal step size $\epsilon_o$, the overlap decreases again, leading to suboptimal training.
    Finite-size scaling behaviour of the overlap $q^2$ is shown in Appendix~\ref{appendix:C.finite_size}.

    In the general case where the teacher (or ground state) matrix is not known, the self-overlap between the maximum eigenvector of different initial random seeds can be used as a complementary effective order parameter.
    The self-overlap is defined as
    \begin{align}
        q^{\alpha \beta} \equiv \sexpv[\bvec{w}]{\left| \bvec{w}_{\rm max}^{\alpha} \cdot \bvec{w}_{\rm max}^{\beta} \right|},
    \end{align}
    where $\bvec{w}_{\rm max}^{\alpha}$ is the eigenvector corresponding to the largest eigenvalue direction of $X^{(1)}$, with a seed index $\alpha$.
    Then the self-overlap is computed by taking the average over all pairs of seeds $\alpha$ and $\beta$.
    The self-overlap is small if there is no alignment, while the overlap becomes larger when the largest eigenvectors are globally aligned to the signal direction across different samples,
    \begin{align} \begin{split}
        q^{\alpha \beta} \neq 0, &\qquad \epsilon_c^+ < \epsilon < \epsilon_c^- , \\ 
        q^{\alpha \beta}=0, &\qquad \text{otherwise} .
    \end{split}\end{align}
    The equivalence between $q^2$ and $q^{\alpha \beta}$ is shown in Appendix~\ref{appendix:D.equivalence}.

    The numerical computation of the self-overlap is shown in Fig.~\ref{fig:q2_overlap_linear} and the right panel of Fig.~\ref{fig:top_eig_diff_exp_label}.
    If the step size is smaller than the critical value, the eigenvectors of the bulk are randomly distributed. On the other hand, if the step size is larger than the critical value, the isolated eigenvalue emerges, and the largest eigenvectors of different samples all align to the same signal direction.

    Although the overlap parameters can be nonzero in the paramagnetic regime, the associated isolated eigenvector is not stable under continued training as shown in the following section.

\subsection{Dynamical BBP transition}
\label{sec:dyanmical_bbp_transition}
    As the initial random bulk is scaled by a factor $(1 - \epsilon)^{2t}$, the bulk spectrum shrinks when the step size $\epsilon$ is smaller than the optimal step size $\epsilon_o$.
    Subsequently, the critical step size $\epsilon_c (t)$ decreases as the optimisation proceeds, and at the critical time $t_c$, the critical step size becomes smaller than the step size, inducing the dynamical BBP transition.
    The critical time $t_c$ is obtained by setting $\epsilon = \epsilon_c \left(t_c \right)$ in Eq.~(\ref{eq:eps_c_linear}), yielding
    \begin{align}
        t_c = - \frac{\ln\left(1 + r^{\frac{1}{4}}\sigma_W\right)}{\ln \left(1 - \epsilon \right)}
        \qquad 
        \forall \quad 0 < \epsilon < 1 .
    \end{align}
    The time-dependent critical step size surface is shown in Fig.~\ref{fig:dynamical_bbp_linear}.
    The phase diagram shown in Fig.~\ref{fig:phase_boundary_linear} is the cross section of Fig.~\ref{fig:dynamical_bbp_linear} at time $t=1$.

    For the step size below the optimal boundary $(\epsilon < \epsilon_o = 1)$, the critical time $t_c$ is always positive and finite, implying that the optimisation always converges to the correct ground state.
    This is an expected behaviour, reflecting the fact that the gradient descent optimisation is just an interpolation between the initial and the target covariance matrices as shown in Eq.~(\ref{eq:linear_multistep}).
    This is an another manifestation of the fact that linear models do not possess spurious local minima \cite{venturiSpuriousValleysOnehiddenlayer2019}, implying that gradient descent converges to the global minimum whenever optimisation remains stable.

    \begin{figure}[tp!]
        \centering
        \includegraphics[trim={0.2cm 0.3cm 0.2cm 0cm,},clip,width=\linewidth]{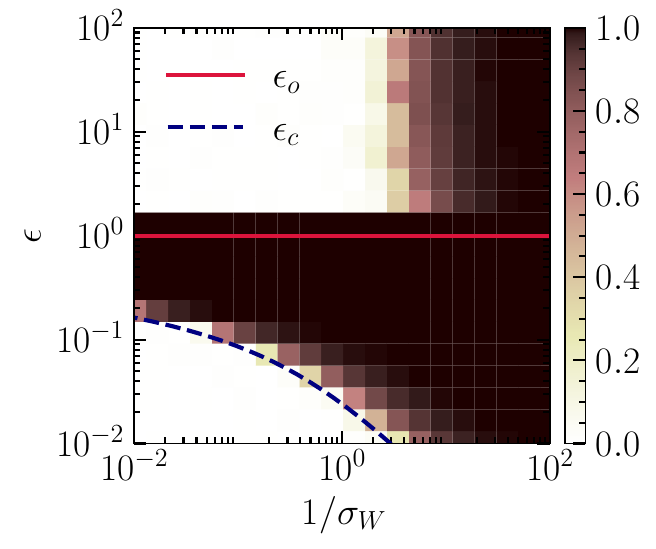}
        \caption{The squared overlap $q^2$ after 25 iterations. Above the optimal step size, the overlap dilutes as the optimisation proceeds. The critical step size $\epsilon_c$ becomes smaller compared to the initial critical boundary as the initial random bulk is reduced. Note that the dashed line indicating $\epsilon_c$ has shifted down compared to the $t=1$ boundary shown in Fig.~\ref{fig:top_eig_diff_exp_label}.}
        \label{fig:q2_last}
    \end{figure}
    The squared overlap after 25 iterations is shown in Fig.~\ref{fig:q2_last}.
    The spurious overlaps in the paramagnetic region decrease as the training proceeds, and the BBP phase boundary shifts to lower values of $\epsilon$ as the random bulk decreases.
    Further time-dependent phase diagrams are shown in Appendix~\ref{appendix:E.q2_t}.

\section{Extensions beyond the solvable linear model}
\label{sec:3.extensions}
    For realistic neural networks, nonlinear activations and multiple layers are essential features that make the training dynamics nontrivial. 
    At the same time, these features also make the analytic tractability of the training dynamics lost in general.
    Here we present an analytical study of the spectral properties of the weight matrix after one iteration, and show numerical results beyond one iteration.

    As in the linear case, the updated weight matrix after a single update can be written in terms of the initial weight matrix and the signal, multiplied by the effective variance and the signal strength.  
    The effect of the nonlinearity and deeper structure is absorbed in the expressions of the effective variance and the signal strength.
    After a single update, the rescaling factors can be numerically integrated with the Gaussian measure, utilising the Gaussian equivalence \cite{leeDeepNeuralNetworks2018, goldtGaussianEquivalenceGenerative2022}.
    In the following sections, we study the effect of each additional feature on the phase diagram.

\subsection{Nonlinearity}
\label{sec:bbp_non_linear}
    Consider the teacher-student setting with a nonlinear activation function $\phi$.
    The teacher model $y$ and student model $\hat y$ are now defined as 
    \begin{align} \begin{split}
        y_{a} &= \phi \left( \sum_{i=1}^{D} W_{ai}^{\ast} x_i \right) = \phi \left( h_a^{\ast} \right) , 
        \\
        \hat y_{a} &= \phi \left( \sum_{i=1}^{D} W_{ai} x_i \right) = \phi \left( h_a \right) ,
    \end{split} \end{align}
    where $h_a^{\ast}$ and $h_a$ denote pre-activations of the teacher and student network respectively.
    The gradient of the squared error loss function with respect to the student weight matrix follows from the chain rule,
    \begin{align} \begin{split} \label{eq:non-linear_grad}
        \frac{\partial \ell}{\partial W_{ai}}
        &= - \left( y_a - \hat y_a \right) \phi'(h_a) x_i \\
        &= - \left( \phi \left(h_{a}^{\ast} \right) - \phi \left( h_a \right) \right) \phi'\left(h_a\right)x_i .
    \end{split} \end{align}

    To calculate the expectation value of the gradient over the Gaussian input distribution with unit variance, we cannot simply average over $x_i$ as in the linear case, because pre-activations $\bvec{h}^{\ast}$ and $\bvec{h}$ are functions of $\bvec{x}$, which appears in the argument of the nonlinear activation $\phi$.
    Instead, we employ Wick-Isserlis's theorem (or Stein's lemma, partial integration with respect to a Gaussian measure) to obtain the expectation value of the gradient matrix,
    \begin{align} \label{eq:nonline_grad}
        \sexpv[x]{\frac{\partial \ell}{\partial W_{ai}}}
        &= - \sexpv[x]{\left( \phi\left(h_a^{\ast}\right) - \phi \left(h_a\right) \right) \phi'\left(h_a\right)_{} x_i} \nn \\
        &= - \sexpv[x]{\frac{\partial}{\partial x_i} \left(\left( \phi \left( h_a^{\ast} \right) - \phi \left( h_a \right)\right)\phi'\left(h_a\right) \right)} \nn \\
        &= - \left(\mu_{1,a} W_{ai}^{\ast} - \mu_{2,a} W_{ai} \right) ,
    \end{align}
    where, after gathering the terms by $W^{\ast}$ and $W$, the factors $\bvec{\mu}_{1} \in \mathbb{R}^{N}$ and $\bvec{\mu}_{2} \in \mathbb{R}^{N}$ are 
    \begin{align}
    \begin{aligned}
        \mu_{1,a} &= \sexpv[x]{\phi'\left( h_{a}^{\ast} \right) \phi'\left(h_a\right)}, \\
        \mu_{2,a} &= \sexpv[x]{\left(\phi' \left(h_a \right) \right)^2 + \left( \phi \left(h_a\right) -\phi\left( h_a^{\ast}\right) \right) \phi'' \left( h_a \right)}.
         \label{eq:non_lin_rescale}
    \end{aligned}
    \end{align} 

    For given activation function $\phi$, $\bvec{\mu}_1$ and $\bvec{\mu}_2$ can be computed using a Gaussian measure.
    The preactivation $h_a$ and $h_a^{\ast}$ are decomposed into
    \begin{align}\begin{split}
        h_a &= \sum_{j=1}^{D} W_{aj} x_j = \sum_{j=1}^{D} \left({W_{\perp}}_{aj} + w_{a} v_j \right) x_j \\
        &= {h_{\perp}}_a + w_a z, \\
        h_a^{\ast} &= \sum_{j=1}^{D} u_{a} v_j x_j = u_a z,
    \end{split}\end{align}
    where $\bvec{h}_{\perp} = W_{\perp} \bvec{x}$ and $z = \sum_{i=1}^{D} v_i x_i \sim \mathcal{N}\left(0,1\right)$.
    $\bvec{u} \in \mathbb{R}^{N}$ is the unit row signal vector composing the teacher matrix (see Eq.~(\ref{eq:teacher_w})).
    At initialisation, the student matrix is sampled from centred Gaussian so we assume $\bvec{w} \sim 0$.
    Then, $\bvec{\mu}_1$ and $\bvec{\mu}_2$ are integrated over two uncorrelated Gaussian measures,
    \begin{align} \label{eq:nonequilibrium_rescaling_decomposition}
        \mu_{1,a} &= \sexpv[\bvec{h},z]{\phi'(u_a z) \phi'({h_{\perp}}_{a})}, \\
        \mu_{2,a} &= \sexpv[\bvec{h},z]{\left(\phi'({h_{\perp}}_a)\right)^2 + \left(\phi({h_{\perp}}_{a}) - \phi(u_a z) \right) \phi''({h_{\perp}}_{a})},\nn
    \end{align}
    where $\bvec{h}_{\perp}$ and $z$ are uncorrelated Gaussian random variables with variance $\sigma_W^2$ and one, respectively.

    The nonlinearity rescales the coefficients of the single update equation to
    \begin{align} \label{eq:ts_decomposition}
        W'_{ai} = \left(1 - \mu_{2,a} \epsilon \right)W_{ai} + \mu_{1,a} \epsilon W_{ai}^{\ast}, 
    \end{align}
    which should be compared with Eq.~(\ref{eq:linear_update_rule}) for the linear case.
    Separating the bulk and signal segments of the weight covariance matrix, the effective bulk variance and signal strength become
    \begin{align} \begin{split}
        \sigma_{ab}^2 (\epsilon) &= \left(1 - \mu_{2,a} \epsilon\right) \left(1 - \mu_{2,b} \epsilon\right)_{} \sigma_W^2,
        \\
        \theta_{ab}^2 (\epsilon) &= \left(1 - \mu_{2,a} \epsilon\right) \left(1 - \mu_{2,b} \epsilon\right)_{} r_{} \sigma_W^2 + \mu_{1,a} \mu_{1, b} \epsilon^2.
    \end{split} \end{align}
    Since $\mu_{1,a}$ and $\mu_{2,a}$ rescale the effective bulk variance and signal strength, we refer to them as rescaling factors.

    The optimal step size and critical step size now depend on the matrix indices,
    \begin{align}
        \epsilon_{o, ab}=\frac{\mu_{2, a} + \mu_{2, b}}{2 \mu_{2,a} \mu_{2,b}},
    \end{align}
    and critical step size is a solution to 
    \begin{align}
        \theta_{ab}^2 (\epsilon_{c, ab}) = \sigma_{ab}^{2}(\epsilon_{c, ab}) \left(\sqrt{r} + r \right).
    \end{align}

    However, since the weight matrices are initialised homogeneously, the components containing ${h_{\perp}}_a$ self-average in the large $N,D$ limit and do not depend on the matrix index.
    Moreover, if we assume that the unit signal vector $\bvec{u}$ does not condense into a specific direction so that $\sexpv[a]{u_a} \sim 1/\sqrt{N}$, we can assume $u_a \to 0$ in the large matrix limit.
    Then, we can safely approximate $\bvec{\mu}_1$ and $\bvec{\mu}_2$ to a self-averaged value $\bar \mu_1$ and $\bar \mu_2$, 
    \begin{align}
        \bar \mu \equiv \lim_{N \to \infty} \frac{1}{N} \sum_{a=1}^N \mu_a (u_a = 0).
    \end{align}
    With this self-averaged approximation, the optimal step size and the critical step size after the first iteration become
    \begin{align}
        \epsilon_{o} \simeq \frac{1}{\bar \mu_2},
        \qquad
        \epsilon_{c}^{\pm} \simeq \left(\bar \mu_2 \pm \bar \mu_1 \frac{1}{r^{\frac{1}{4}}\sigma_W} \right)^{-1}.
    \end{align}
    The only case where the self-averaging approximation fails is when the signal vector $\bvec{u}$ contains a condensation aligned to one of the canonical basis vectors, $\bvec{e}_i$.
    This case is discussed in detail in Appendix~\ref{appendix:F.variance_and_signal_nonlinear}.
    In the following, we apply this setup to the hyperbolic tangent and ReLU activation functions.  

\subsubsection{Hyperbolic tangent}
    \begin{figure}[tp!]
        \centering
        \includegraphics[trim={0.2cm 0cm 0.2cm 0cm,},clip,width=\linewidth]{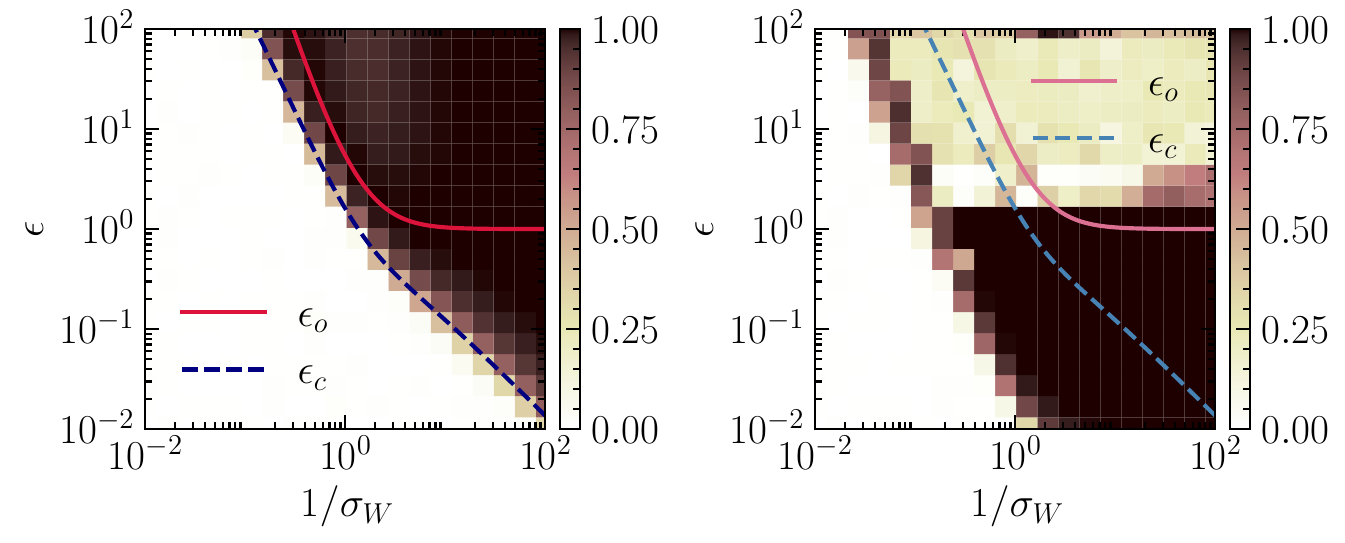}
        \caption{Trainability phase diagram for hyperbolic tangent activation. The colour map shows the squared overlap $q^2$ after a first step (left) and after 75 iterations (right). Even though the phase diagram is derived for the first iteration, the two boundaries are still informative, separating different phases of training effectively. This indicates that the trainability of the optimisation is sensitive to the initialisation. The spurious overlap in the paramagnetic regime reduces as the training proceeds.}
        \label{fig:top_eig_tanh}
    \end{figure}
    For the hyperbolic tangent activation function, $\phi(x)=\tanh(x)$, we have
    \begin{align} \begin{split}
        \tanh'(x) &= 1 - \tanh^2 (x), \\
        \tanh''(x) &= 2 \tanh^3(x) - 2 \tanh(x) .
    \end{split} \end{align}
    The trainability phase diagram after a single update is shown in the left panel of Fig.~\ref{fig:top_eig_tanh}.
    The hyperbolic tangent function is bounded from above and below, and a large portion of the gradient reaches the gradient vanishing region for large values of $\sigma_W^2$.
    In this region, it requires a larger step size to escape the plateau and trigger the BBP transition, leading to a higher critical step size $\epsilon_c$ compared to the linear case. 
    We observe excellent agreement between the theoretical predictions for $\epsilon_c$ and $\epsilon_o$ and the numerical results.

    In the right panel of Fig.~\ref{fig:top_eig_tanh}, the phase boundaries calculated for the first step are overlaid on top of the overlap evaluated after 75 iterations.
    We observe that the boundaries computed for the first iteration effectively separate the different phases well, even after many iterations.
    After 75 iterations, the spurious overlap in the paramagnetic regime diminishes, and the ferromagnetic phase extends to the smaller value of step size.
    The mechanism of this fading overlap and expanding ferromagnetic phase can be explained in the same way as in the linear case, where the suppression of the initial random bulk lowers the critical step size leading to the dynamical BBP transition.
    The phase diagrams for intermediate steps are shown in Appendix~\ref{appendix:E.q2_t}, where the gradual fading of the spurious overlaps in paramagnetic regime is more apparent.

\subsubsection{ReLU}
    \begin{figure}[tp!]
        \centering
        \includegraphics[trim={0.2cm 0cm 0.2cm 0cm,},clip,width=\linewidth]{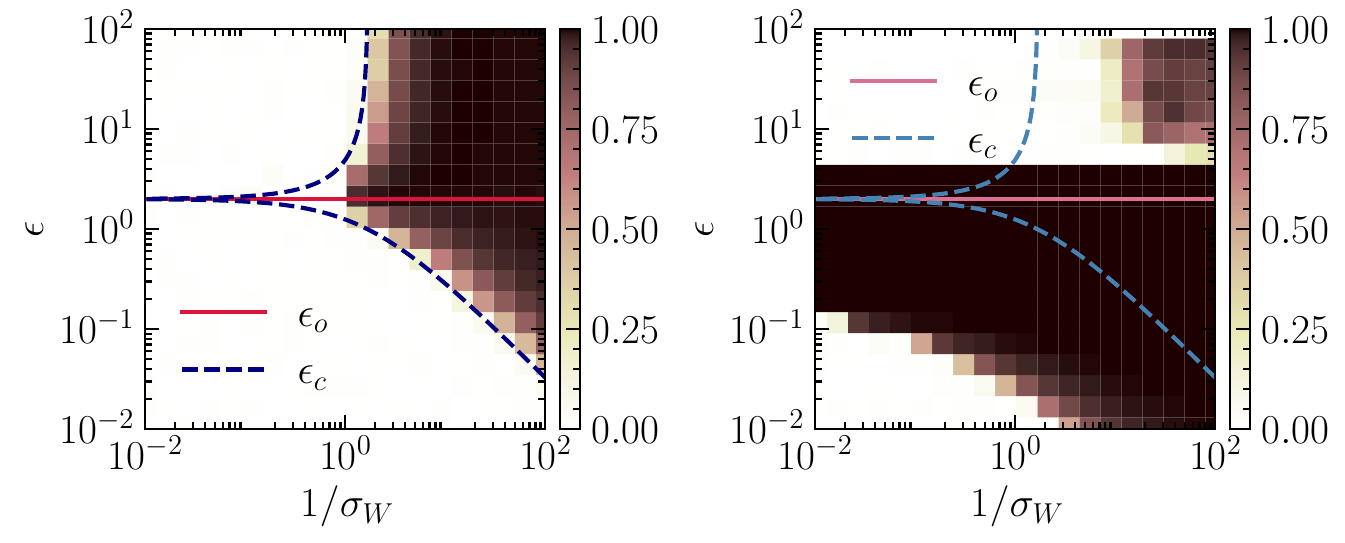}
        \caption{Trainability phase diagram for ReLU activation. The colour map shows the squared overlap $q^2$ after a first step (left) and after 75 iterations (right). The initial overlap above the optimal boundary $\epsilon_o$ diminishes as the training proceeds. The dynamics of the BBP boundary exhibits a qualitative behaviour similar to that of the linear model, see Fig.~\ref{fig:q2_last}, consistent with the locally linear nature of the ReLU activation.}
        \label{fig:top_eig_relu}
    \end{figure}
    For the ReLU activation function, ${\rm ReLU}(x) = x_{} \Theta(x)$, we have 
    \begin{align} 
        {\rm ReLU}'(x) = \Theta(x), \quad
        {\rm ReLU}''(x) = \delta(x),
    \end{align}
    where $\Theta (x)$ is the Heaviside function.
    In this case the functions can be integrated analytically, yielding,
    \begin{align} \label{eq:relu_rescaling_sa}
        \bar \mu_{1} = \frac{1}{4},
        \qquad
        \bar \mu_{2} = \frac{1}{2}.
    \end{align}
    Following the same steps as above, we determine the 
    trainability phase diagram for ReLU activation, shown in Fig.~\ref{fig:top_eig_relu}.
    The BBP phase boundary is similar to the linear case, slightly shifted to the higher value.
    Time-dependent behaviours of $q^2$ for intermediate time steps are shown in Appendix~\ref{appendix:E.q2_t}.

\subsection{Multiple layers with nonlinearity}
\label{sec:bbp_multiple_layers}
    \begin{figure}[tp!]
        \centering
        \includegraphics[trim={0.8cm 1cm 0.8cm 1cm,},clip,width=\linewidth]{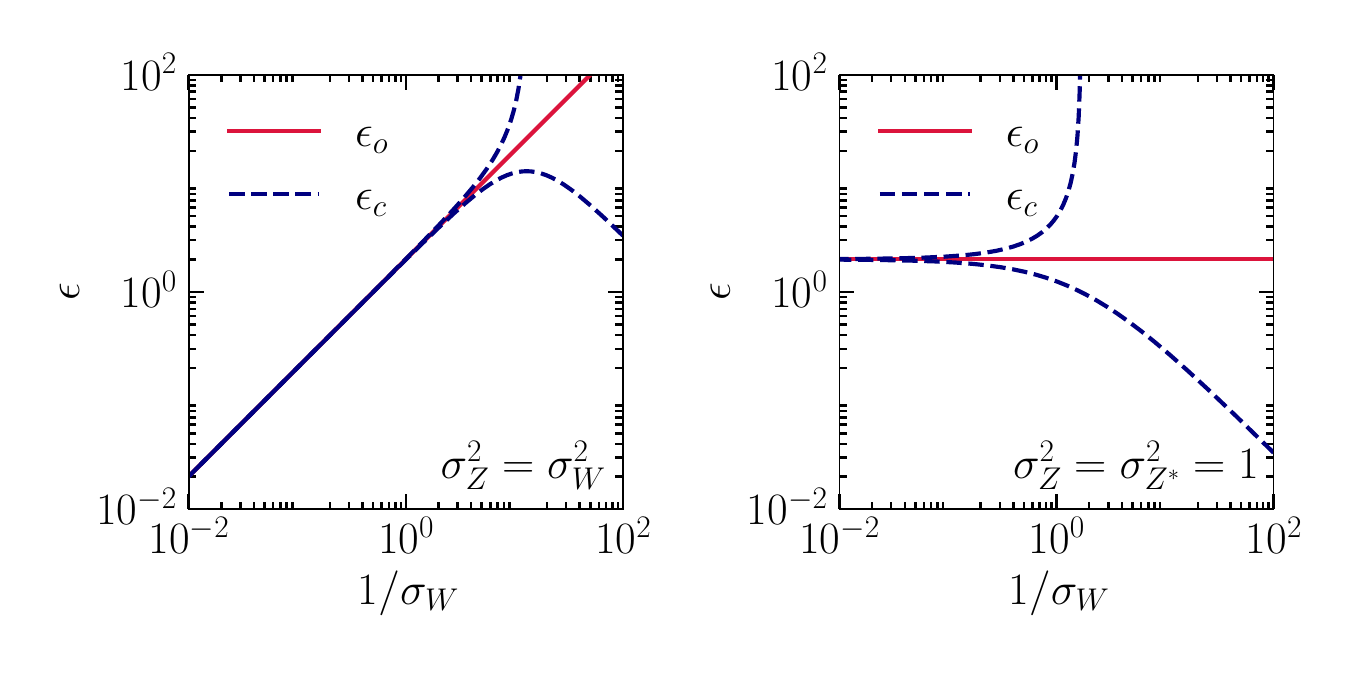}
        \caption{Trainability phase diagram for a single hidden layer ReLU
        network with different teacher-student scenarios. The left panel shows the phase diagram for the case where the initial variances of all of the layers are scaled equally. The teacher-student overlap $\mathcal{J}$ is assumed to be small ($\mathcal{O}\left( 1 / N \right)$ with $N=100$), which corresponds to the situation at initialisation. The right panel shows the phase diagram for the case where the variance of the student readout layer matches the teacher layer and the overlap $\mathcal{J}$ is large.}
        \label{fig:multi_nonlin_phase_boundary}
    \end{figure}
    Adding more layers to the model further modifies the rescaling factors $\mu_1$ and $\mu_2$.
    Consider a model with multiple layers ($l=1,\ldots, L$) and a nonlinear activation $\phi$,
    \begin{align}
        \hat y_{\alpha} = W_{\alpha a}^{L} \phi_{a}^{L},
        \quad
        \phi_{a}^{l} \equiv \phi^{l} \left(W_{ai}^{l-1} \phi_{i}^{l-1} \right),
        \quad
        \phi_{i}^{0} = x_i ,
    \end{align}
    where all indices are contracted except the ones matching the left-hand side, unless otherwise stated.
    The gradient with respect to the weight matrix in the $l$-th layer is
    \begin{align} \label{eq:multi_nonlin}
        \frac{\partial \ell}{\partial W_{ai}^{l}}
        = \left( J_{ab}^{l+1}\phi_{b}^{L}
        - H_{a}^{l+1} \right) {\phi_{a}^{l+1}}' \phi_{i}^{l},
    \end{align}
    where
    \begin{align}
    \begin{aligned}
    & {\phi_{a}^{l+1}}'(x)  = \frac{\partial}{\partial x}\phi_{a}^{l+1}(x), \\
     &   J_{ab}^{l}  = W_{ci}^{L} {W_{bi}^{L}} T_{ca}^{L,l},
        \qquad
        H_{a}^{l} = y_{\alpha} W_{\alpha b}^{L} T_{ba}^{L,l},
    \end{aligned}
    \end{align}
    and $T^{L,l}$ are the transfer matrices which propagate the derivatives from the layer $L$ to the layer $l$, defined as
    \begin{align} \label{eq:transfer_matrix}
        T_{ba}^{L,l} = T_{bi}^{L, L-1} T_{ij}^{L-1, L-2} \, \cdots \, T_{ka}^{l+1, l},
        \quad
        T_{ai}^{l, l-1} \equiv {\phi_{a}^{l}}' W_{ai}^{l-1}.
    \end{align}

    By comparing Eq.~(\ref{eq:multi_nonlin}) to the case of a single-layer, see Eq.~(\ref{eq:non-linear_grad}), it is immediately apparent that the teacher-student decomposition in terms of $W^{l}$ and $W^{\ast l}$, as in Eq.~(\ref{eq:ts_decomposition}), is no longer preserved.
    This decomposition can still be done for the first layer $W^{0}$.
    For the first layer, the rescaling factors $\mu_1$ and $\mu_2$ become
    \begin{align}  \label{eq:multi_non_lin_rescale}
        \mu_{1, ac} &= \sexpv[x]{\mathcal{J}_{db}T_{bc}^{\ast L,2}T_{da}^{L,1}{\phi_{c}^{\ast 1}}'} , \\
        \mu_{2, ac} &= \sexpv[x]{g_{ac} - \mathcal{J}_{db}\phi_{b}^{\ast L}
        \mathcal{M}_{adc} {\phi_{c}^{1}}' - \delta_{ac}
        H_{b}T_{bc}^{L,1} {\phi_{c}^{1}}''},\nn
    \end{align}
    where $\mathcal{J}_{ab} = W_{\alpha a}^{L} W_{\alpha b}^{\ast L}$ is the
    overlap between the final layers of the teacher and student networks.
    The full derivation and the definitions of functions $\mathcal{M}$ and $g$
    are shown in Appendix~\ref{appendix:G.mult_nonlin}.

    For the single hidden layer case, the teacher and student networks are 
    \begin{align}
        y_{\alpha} = Z_{\alpha a}^{\ast} \phi \left( W_{ai}^{\ast} x_i \right),
        \qquad
        \hat y_{\alpha} = Z_{\alpha a} \phi \left( W_{ai} x_i \right),
    \end{align}
    and the rescaling factors are given as
    \begin{align} \begin{split}
        \mu_{1, ab} &= \sexpv[x]{ \mathcal{J}_{ab} {\phi_{b}^{\ast}}' \phi_{a}'}, \\
        \mu_{2, ab} &= \sexpv[x]{J_{ab}\phi_{a}'\phi_{b}' + \left(\delta_{ab} J_{bc} \phi_c
        - \mathcal{J}_{ab}\phi_b^{\ast} \right) \phi_{a}''}\,.
    \end{split} \end{align}
    The expression reduces to the case without a hidden layer, see Eq.~(\ref{eq:non_lin_rescale}), if $J_{ab} = \mathcal{J}_{ab} = \delta_{ab}$.

    With the additional layer, the definition of the teacher matrix becomes
    ambiguous, as the teacher matrix can be decomposed into multiple layers in
    different ways.
    Here, we show three possible scenarios with ReLU activation to schematically understand the effect of additional layers.

    First, at initialisation, the student matrices $Z$ and $W$ are not aligned
    with the teacher matrices $Z^{\ast}$ and $W^{\ast}$, and the
    teacher-student overlap $\mathcal{J}$ is small.
    If the readout layer matrix $Z$ is initialised with a variance
    $\sigma_Z^2$ proportional to the variance of the first layer weight matrix
    $\sigma_W^2$, the initial gradient gets suppressed by the additional factor
    of $\sigma_Z^2$ compared to the single-layer case, and the disordered phase below the BBP transition line becomes larger in the smaller $\sigma_W$
    range, as shown in the left panel of Fig.~\ref{fig:multi_nonlin_phase_boundary}.

    Then, in the hypothetical case where the readout layer of the student network is already aligned with the teacher matrix, $Z \simeq Z^{\ast}$, the trainability phase diagram depends on the statistical properties of the teacher matrix $Z^{\ast}$, and the optimal step size and the critical step size are rescaled by the variance of the teacher matrix $\sigma_{Z^{\ast}}^2$.
    For a Gaussian teacher matrix with variance $\sigma_{Z^{\ast}}^2 =
    1$, the trainability phase diagram is similar to the single-layer case, as
    shown in the right panel of the Fig.~\ref{fig:multi_nonlin_phase_boundary}.

    While an explicit analytical treatment of deeper layers is challenging, the first-layer analysis already captures the fundamental signal-bulk competition responsible for the BBP transition.
    Numerical evidence shown in Sec.~\ref{sec:5.realistic} suggests that the resulting spectral phenomenology persists more generally in multilayer networks.

\section{Stochastic optimisation and effect of finite data size}
\label{sec:4.stochastic}
    Relaxing the assumption of an infinite dataset, a finite sample size introduces stochasticity into the gradient descent dynamics.
    The expectation value of the gradient over the data distribution in Eq.~(\ref{eq:grad_loss}) is replaced by an empirical average over a finite amount of data.
    In general, the additional stochastic noise carries a nontrivial correlation structure, complicating the analysis of the training dynamics by modifying the bulk structure and the effective spike.

    However, assuming that the dataset is sampled independently from the true data distribution, the stochastic measure of the gradient term decomposes into a deterministic drift and a stochastic noise using the central limit theorem \cite{chaudhariStochasticGradientDescent2018, mandtContinuousTimeLimitStochastic, mandtStochasticGradientDescent2018,
    yaidaFluctuationdissipationRelationsStochastic2019,
    aartsStochasticWeightMatrix2025}.
    Up to the leading order, the stochastic update equation for gradient descent is written as
    \begin{align} \label{eq:sgd_update}
        W_{ai}' = W_{ai} - \epsilon K_{ai} + \sqrt{\epsilon T}\left(\frac{1}{\sqrt{n}} + \sqrt{1 - \frac{1}{n}} \right) \tilde \eta_{ai},
    \end{align}
    where $n = P/|\mathcal{B}|$ is the number of batches, and $T = \epsilon / |\mathcal{B}|$ is the effective temperature, which controls the stochasticity in the training dynamics \cite{chaudhariStochasticGradientDescent2018, mandtContinuousTimeLimitStochastic, mandtStochasticGradientDescent2018,
    yaidaFluctuationdissipationRelationsStochastic2019,
    aartsStochasticWeightMatrix2025, goyalAccurateLargeMinibatch2018, smithDontDecayLearning2018}.
    The deterministic drift term is
    \begin{align}
        K_{ai} \equiv \sexpv[x]{\frac{\partial \ell}{\partial W_{ai}}} ,
    \end{align}
    and the noise term satisfies 
    \begin{align} \begin{split} \label{eq:properties_of_noise_term}
        \sexpv[x]{\tilde \eta_{ai}} &= 0  ,
        \\
        \sexpv[x]{\tilde \eta_{ai} \tilde \eta_{bj}}
        &= \sexpv[x]{\frac{\partial \ell}{\partial W_{ai}} \frac{\partial \ell}{\partial W_{bj}}} - K_{ai} K_{bj}  .
    \end{split} \end{align}

    The update rule (\ref{eq:sgd_update}) effectively represents the dynamics of stochastic gradient descent, where different scaling regimes are recovered by tuning four scaling ratios,
    \begin{align}
        n = \frac{P}{|\mathcal{B}|},
        \quad
        r = \frac{N}{D},
        \quad
        T = \frac{\epsilon}{|\mathcal{B}|},
        \quad
        \alpha = \frac{P}{D},
    \end{align}
    where $P$, $|\mathcal{B}|$, $N$, $D$, and $\epsilon$ represent the size of the dataset, size of the minibatch, network width, dimension of the data, and the step size, respectively.

    The ratio between the total dataset size and the batch size, $n$, is the number of batches in one epoch of training, where each epoch reiterates on the same $P$ number of data.
    The full batch gradient flow update corresponds to the case of fixed $n=1$ and $P \to \infty$. In this case, the noise term is removed from Eq.~(\ref{eq:sgd_update}) and the update rule reduces to Eq.~(\ref{eq:linear_update_rule}).
    The aspect ratio $r$ is the ratio determining the shape of the random matrix bulk spectrum.
    In the $\epsilon \to 0$ limit with fixed effective temperature $T$, 
    SGD dynamics reduces to a stochastic differential equation (SDE).

    The load parameter $\alpha$ \cite{engel2001statistical} measures the number of samples per input dimension and therefore controls the amount of information available per degree of freedom.
    At fixed $\alpha$, the empirical gradient remains a finite-sample object even in the thermodynamic limit $P, D \to \infty$, producing sample-induced fluctuations around the population drift.
    Increasing $\alpha$ suppresses these finite-data fluctuations and drives the dynamics towards the population, or infinite-data, limit.
    Equivalently, $\alpha$ plays the role of a sample-complexity parameter.
    Small $\alpha$ corresponds to an underconstrained regime where the number of data points is insufficient to determine all input directions, while large $\alpha$ corresponds to a data-rich regime where the empirical covariance concentrates around its population value.

    Summarising different scaling limits, we identify
    \begin{itemize}
        \item {\it Full batch gradient flow:} $n=1$, $\alpha \to \infty$, $T\to 0$,
            \begin{align*}
                W_{ai}' = W_{ai} - \epsilon K_{ai}.
            \end{align*}
        \item {\it Full batch proportional regime:} $n=1$, $\alpha$ fixed, $T=\epsilon/P$,
            \begin{align*}
                W_{ai}' = W_{ai} - \epsilon K_{ai} + \frac{\epsilon}{\sqrt{P}} \tilde \eta_{ai}.
            \end{align*}
        \item {\it Minibatch Stochastic Gradient Descent:} $n \to \infty$,
            \begin{align*}
                W_{ai}' &= W_{ai} - \epsilon K_{ai} + \sqrt{\epsilon T} \tilde \eta_{ai},
            \end{align*}
        \item {\it SDE limit of SGD}, $\epsilon \to 0$, $n\to \infty$, $T$ fixed,
            \begin{align*}
                \frac{d W_{ai}}{dt} \simeq - K_{ai} + \sqrt{T} \tilde \eta_{ai}.
            \end{align*}
    \end{itemize}
    The full batch gradient flow has been considered in Sec.~\ref{sec:2.solvable_model}.
    We now consider the full batch proportional regime and stochastic gradient descent regime in following sections.

\subsection{Proportional regime}
\label{sec:proportional_limit}

    We start by considering the proportional regime $\alpha=\mathcal O(1)$, where the empirical covariance of the dataset does not self-average to the identity matrix, and finite-sample fluctuations remain macroscopic even in the thermodynamic limit.
    For a finite dataset of size $P$, define a data matrix
    \begin{align} \label{eq:data_matrix}
        \mathbf{X} = \begin{pmatrix}
        \bvec{x}^1 & \bvec{x}^2 & \cdots & \bvec{x}^P \end{pmatrix} 
        \ \in \ \mathbb R^{D\times P},
    \end{align}
    with data vectors $\bvec{x} \in \mathbb{R}^{D}$. The empirical data covariance matrix is given as
    \begin{align} \label{eq:empirical_data_covariance}
        \widehat C = \sexpv[x \sim P]{\bvec{x} \bvec{x}^T}
        = \frac{1}{P} \mathbf{X} \mathbf{X}^T.
    \end{align}
    The gradient matrix averaged over the finite dataset is now written in terms of the empirical data covariance matrix as
    \begin{align}
    \begin{aligned}
        \sexpv[x \sim P]{\frac{\partial \ell}{\partial W_{ai}}} 
        = -\left( W_{aj}^{\ast} - W_{aj} \right) \widehat C_{ji} 
        \equiv -\Delta_{aj} \widehat C_{ji},
        \end{aligned}
    \end{align}
    where we have denoted $\Delta \equiv W^{\ast} - W$ for notational simplicity.

    Using the central limit theorem, the empirical covariance matrix decomposes into
    \begin{align} \begin{split}
        \widehat C_{ij} &= I_{ij} + \frac{1}{\sqrt{P}} D_{ij},
        \\
        D_{ij} &= \frac{1}{\sqrt{P}} \sum_{\mu=1}^{P}\left(x_i^{\mu} x_j^{\mu} - I_{ij} \right),
    \end{split} \end{align}
    where the fluctuation matrix $D_{ij}$ is of order $\mathcal{O} (1)$, and in the infinite sample size limit $P \to \infty$, the empirical covariance matrix reduces to identity.
    In terms of the empirical covariance matrix $\widehat C$, an update step becomes
    \begin{align}
        W_{ai}' = W_{ai} + \epsilon \Delta_{ai} + \frac{\epsilon}{\sqrt{P}} \Delta_{aj} D_{ji},
    \end{align}
    where the noise matrix $\tilde \eta_{ai} = \Delta_{aj} D_{ji}$ can be identified. 

    In terms of random matrix theory, keeping the empirical covariance matrix explicitly in the equation makes the random matrix argument cleaner.
    The gradient update for the linear teacher-student model becomes
    \begin{align} \label{eq:prop_update}
        W' = W \left( I - \epsilon \widehat C\right)
        + \epsilon W^{\ast} \widehat C,
        \quad
        W^{\ast} = \bvec{u}_{} \bvec{v}^T,
    \end{align}
    where $\Norm{\bvec{u}} = \Norm{\bvec{v}} = 1$ are the unit vectors forming the rank-1 teacher matrix $W^{\ast}$, as before.
    Define quantities masked by the empirical data covariance matrix as
    \begin{align}
        \widetilde W_{ai} = W_{aj} \left(1 - \epsilon \widehat C_{ji}\right),
        \quad
        \tilde v_i = \widehat C_{ij} v_j,
    \end{align}
    and the normalised signal direction,
    \begin{align}
        \bvec{\hat v} = \frac{\bvec{\tilde v}}{\Norm{\bvec{\tilde v}}}.
    \end{align}
    The weight matrix is separated into the bulk and the masked signal direction,
    \begin{align}
        \widetilde W = \widetilde W_{\perp} + \bvec{\tilde w}_{} \bvec{\hat v}^T,
    \end{align}
    where
    \begin{align}
        \widetilde W_{\perp} = \widetilde W \left( I - \bvec{\hat v} \bvec{\hat v}^T \right),
        \quad
        \bvec{\tilde w} = \tilde W \bvec{\hat v}.
    \end{align}
    The single-step update equation is written as
    \begin{align}
    \begin{aligned}
        W' &= \widetilde W_{\perp} + \left( \bvec{\tilde w} + \epsilon_{} \vartheta_{} \bvec{u} \right)_{} \bvec{\hat v}^T\\
        &= \widetilde W_{\perp} + \bvec{s} \bvec{\hat v}^T,
        \end{aligned}
    \end{align}
    where we have denoted $\Norm{\bvec{\tilde v}} = \vartheta$, and defined the effective signal vector $\bvec{s} = \bvec{\tilde w} + \epsilon_{} \vartheta_{} \bvec{u}$.
    The weight covariance matrix after an update becomes
    \begin{align}
        X' = \widetilde W_{\perp} \widetilde W_{\perp}^T + \bvec{s}_{} \bvec{s}^T,
    \end{align}
    where the first term describes the bulk and the second term is the effective rank-1 spike.

    The expression is similar to the infinite data limit discussed in Sec.~\ref{sec:2.solvable_model}, with the difference that the empirical covariance matrix is a Wishart matrix with aspect ratio $1/\alpha$.
    As a consequence, finite-data effects simultaneously deform the random bulk spectrum and modify the BBP outlier condition.
    The proportional regime dynamics therefore still correspond to a BBP-type transition, but now taking place on top of a deformed Wishart ensemble controlled by the load parameter $\alpha=P/D$.

    \begin{figure}[tp!]
        \centering
        \includegraphics[trim={0.5cm 0.5cm 0.5cm 0.5cm,},clip,width=\columnwidth]{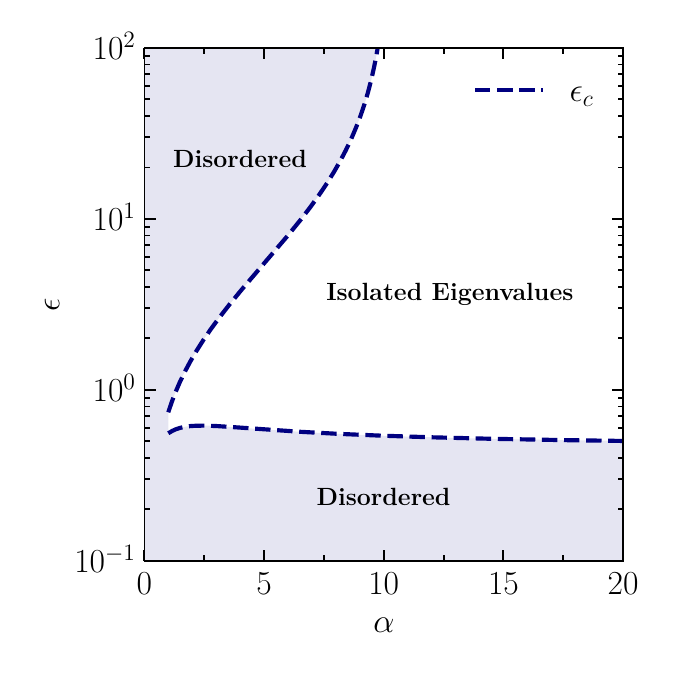}
        \caption{Dependence of the critical step size $\epsilon_c$ on the finite value of the load parameter $\alpha=P/D$. The isolated eigenvalues only exist in the region between the two shaded areas, corresponding to the ferromagnetic phase or paramagnetic regime, depending on the value of the bulk variance $\sigma_W^2$. The phase diagram is drawn for $\sigma_W^2 = 1$ and $r = 0.5$.}
        \label{fig:alpha_epsilon}
    \end{figure}

    \begin{figure*}[tp!]
        \centering
        \includegraphics[trim={0.5cm 0.7cm 0.5cm 0.7cm},clip,width=\linewidth]{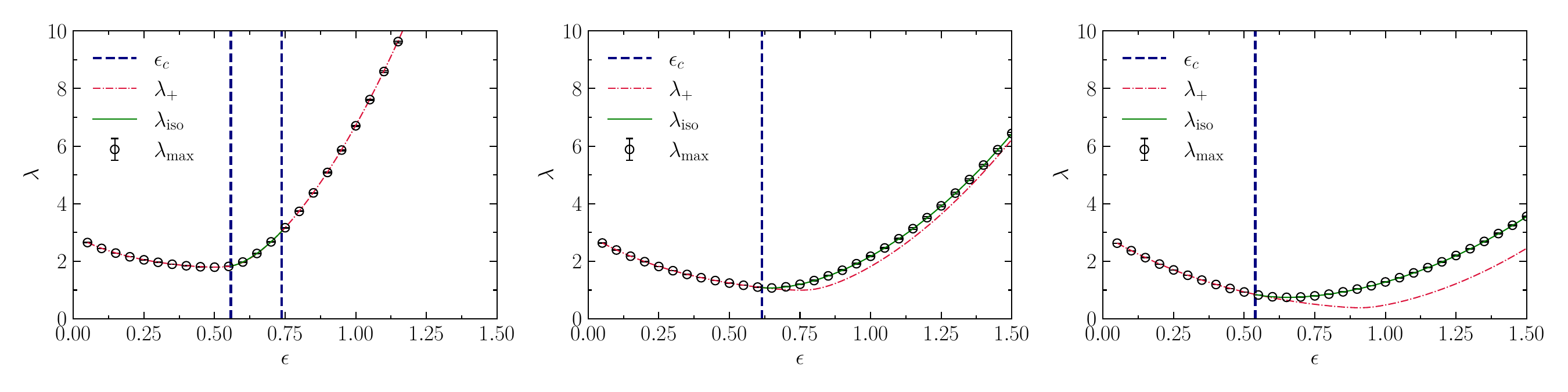}
        \caption{
        Largest eigenvalue of the covariance matrix in the proportional regime for representative values of the load parameter $\alpha=P/D$, for $\alpha = (1, 3, 10)$ from left to right.
        The red dash-dotted curve denotes the edge of the deformed random bulk $\lambda_{+}$, while the green solid curve $\lambda_{\rm iso}$ is the predicted location of the isolated eigenvalue.
        As $\alpha$ increases, the theory continuously approaches the isotropic infinite-data limit.
        For $\alpha = 1$, the isolated eigenvalue exists for a finite range of step size $\epsilon$ between two BBP transition boundaries, then gets reabsorbed into the bulk. 
        This transient BBP transition is the result of step size rescaling both the bulk and the signal strength.
        For the simulations in this section, we have used the matrices of size $N = 12800$ with aspect ratio $r=0.5$.
        }
        \label{fig:proportional_limit_eigs}
        \centering
        \includegraphics[trim={0.5cm 0.7cm 0.5cm 0.0cm},clip,width=\linewidth]{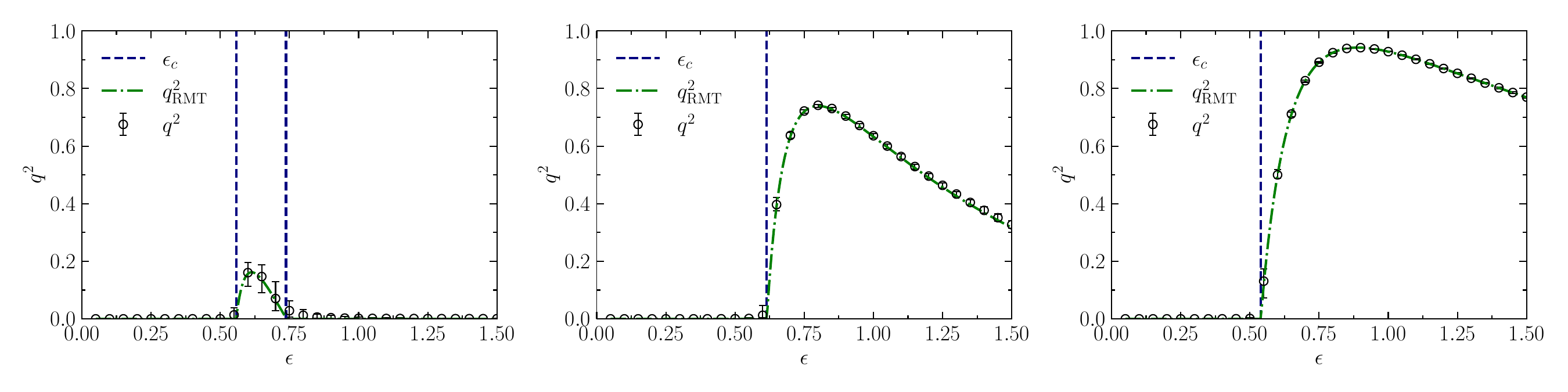}
        \caption{
        Squared overlap $q^2$ in the proportional regime for representative values of the load parameter $\alpha=P/D$, for $\alpha = (1, 3, 10)$ from left to right.
        The green dash-dotted curve denotes the theoretical overlap $q^2_{\rm RMT}$ in the large $N, D$ limit.
        As $\alpha$ increases, the theory continuously approaches the isotropic infinite-data limit.
        For $\alpha = 1$, the transient behaviour of the BBP transition becomes significant, where the maximum eigenvalue detaches from the bulk for a small range of step sizes $\epsilon$ until it gets reabsorbed into the bulk as $\epsilon$ increases.
        }
        \label{fig:proportional_limit_overlap}
    \end{figure*}
    The full derivation is presented in Appendix~\ref{appendix:finite_sample_size}.
    Here, we will only quote the final result.
    The critical step size $\epsilon_c$ is determined implicitly by the generalised BBP condition,
    \begin{align} \label{eq:prop_spike}
        \theta^2 (\epsilon_c;g) = \frac{1}{g(\lambda_{+})},
    \end{align}
    where the effective signal strength $\theta^2 (\epsilon)$ is defined as
    \begin{align}
        \theta^2 (\epsilon;g) \equiv \epsilon^2 \int \frac{x^2}{1 - r \sigma_W^2 (1 - \epsilon_{} x)^2 g} \rho_{{\rm MP}_{1/\alpha}}(x) dx,
    \end{align}
    and $\rho_{{\rm MP}_{1/\alpha}}(x) dx$ denotes the probability measure of the Marchenko-Pastur distribution with aspect ratio ${1/\alpha}$.
    Similarly, the right edge of the deformed bulk $\lambda_{+}$ is obtained as a solution of
    \begin{align}
        \left. \frac{d z (g)}{d g} \right|_{g=g(\lambda_{+})} = 0,
    \end{align}
    where the inverse resolvent is given as,
    \begin{align}
        z(g)
        &=
        \frac{1}{g}
        +
        \int
        \frac{
        \sigma_W^2(1-\epsilon x)^2
        }
        {
        1-r\sigma_W^2(1-\epsilon x)^2g
        }
        \,
        \rho_{{\rm MP}_{1/\alpha}}(x) dx.
    \end{align}

    In the limit $\alpha\to\infty$, the Marchenko-Pastur distribution collapses to $\delta(x-1)$ and the isotropic full-batch result of Sec.~\ref{sec:linear_dynamic} is recovered.
    For small $\alpha$, finite-sample fluctuations strongly deform the bulk spectrum and suppress signal extraction.
    As the load parameter $\alpha$ increases, the empirical covariance concentrates around the identity matrix and the theory smoothly approaches the infinite-data limit.

    We observe a transient BBP behaviour with respect to the step size $\epsilon$, where an isolated eigenvalue temporarily detaches from the bulk before merging back as $\epsilon$ is increased.
    While this behaviour was already shown for linear full-batch gradient flow in Sec.~\ref{sec:2.solvable_model}, here we show that finite $\alpha$ further controls the relative scaling between the signal and the random bulk, modifying the BBP boundaries.
    Similar transient spectral transitions were recently observed in evolving random matrix ensembles \cite{coeurdouxRandomMatrixTheory2026a}.

    The effect of finite load parameter $\alpha$ and the transient BBP transition is shown in Fig.~\ref{fig:alpha_epsilon} and in Appendix~\ref{appendix:B.alpha_r}.
    For small values of the load parameter $\alpha$, the gap between the lower and upper branches of the critical step size becomes narrower.
    This transient behaviour originates from the fact that the step size rescales both the bulk spectrum and the signal strength.

    Representative numerical results for the largest eigenvalue and the squared overlap are shown in Figs.~\ref{fig:proportional_limit_eigs} and \ref{fig:proportional_limit_overlap}.
    For finite $\alpha$, the largest eigenvalue remains attached to the bulk edge below the critical step size, then detaches as the step size becomes larger than the critical value.
    As $\alpha$ increases, the finite-data theory smoothly converges to the infinite-data prediction.
    The numerical evaluations are performed with $N=12800$.

\subsection{SGD and spectral density near the ground state}
\label{sec:stationary}
    In the stochastic gradient descent limit, the effective load parameter for a single batch update, number of data points per dimension $|\mathcal{B}|/D$, converges to zero, and the thermodynamic limit of random matrix theory does not provide a meaningful result for the spectral density of the weight matrix.
    The macroscopic effect of the optimisation only emerges in the stationary limit, where the microscopic update accumulates and deforms the macroscopic spectral density of the weight matrix.
    Furthermore, it becomes apparent that the spectral density of the weight matrix is related to the ground state geometry represented by the Hessian of the
    loss, and the thermal fluctuations of the weight matrix around the ground
    state.

    Starting from the SGD limit of Eq.~(\ref{eq:sgd_update}), the weight covariance matrix after a single batch update is
    \begin{align} \label{eq:sgd_cov_update}
        X_{ab} ' &= \sum_{i=1}^{D} W_{ai}' W_{bi}' \\
        &= \sum_{i=1}^{D} \left( W_{ai} - \epsilon K_{ai} \right) \left( W_{bi} - \epsilon K_{bi} \right)
        + \frac{\epsilon^2}{|\mathcal{B}|} \sum_{i=1}^{D} \tilde \eta_{ai} \tilde \eta_{bi} . \nn
    \end{align}
    Contracting the indices $i$ and $j$ in Eq.~(\ref{eq:properties_of_noise_term}), the fluctuation term is a superstatistic Gaussian distribution \cite{beckSuperstatisticsTheoryApplications2004, adomaityteClassificationHeavytailedFeaturesa} with covariance
    \begin{align}
         C_{ab} \equiv \sum_{i=1}^{D} \sexpv[x]{\tilde \eta_{ai} \tilde \eta_{bi}} &= F_{ab} - \sum_{i=1}^{D} K_{ai} K_{bi},
    \end{align}
    where
    \begin{align}
        F_{ab} = \sum_{i=1}^{D} \sexpv[x]{\frac{\partial \ell}{\partial W_{ai}} \frac{\partial \ell}{\partial W_{bi}}}
    \end{align}
    represents a generalised Fisher Information matrix describing the fluctuations of the gradient matrix.

    In the stationary limit, assuming that $W\sim W^\ast$, the dynamics of the weight matrix can be described as a
    fluctuation around it, and the gradient drift is
    approximated by the Hessian of the loss function near the minimum.
    The drift term in Eq.~(\ref{eq:sgd_cov_update}) can then be approximated in terms of the Hessian as
    \begin{align} \begin{split}
        K_{ai} &\simeq {H_{ai}}^{bj} \left(W_{bj}^{\ast} - W_{bj} \right) = {H_{ai}}^{bj} \Delta_{bj} ,
        \\
        {H_{ai}}^{bj} &\equiv \left. \sexpv[x]{\frac{\partial^2 \ell}{\partial W_{ai} \partial W_{bj}}} \right|_{W = W^{\ast}} ,
    \end{split} \end{align}
    where $W^{\ast}$ is the ground state, or the teacher matrix in the teacher-student setting.
    Here, repeated indices are summed over except the ones appearing on the left-hand side.

    Taking the expectation value over $W$ and denoting the weight fluctuation covariance matrix as
    \begin{align}
        \Sigma_{ai, bj} \equiv \sexpv[W]{\Delta_{ai} \Delta_{bj}},
        \qquad
        \Sigma_{ab} \equiv \sum_{i=1}^{M} \Sigma_{ai, bi},
    \end{align}
    the weight covariance matrix $X$ and the fluctuation covariance matrix $\Sigma$ are related by
    \begin{align}
        X_{ab} = \Sigma_{ab} + \left(W^{\ast} {W^{\ast}}^T \right)_{ab}.
        \label{eq:powerLawPlusSpikes}
    \end{align}
    Then, the update equation in terms of the weight fluctuation covariance matrix becomes
    \begin{align} \begin{split}
        \Sigma_{ab}' = \Sigma_{ab} &- \epsilon \left[ \Sigma_{ai, cj} {H_{bi}}^{cj} + {H_{ai}}^{cj} \Sigma_{cj, bi} \right] \\
        &+ \epsilon^2 {H_{ai}}^{cj} \Sigma_{cj, dk} {H_{bi}}^{dk} \\
        &+ \frac{\epsilon^2}{|\mathcal{B}|}\left[\mathcal{F}_{ab} - {H_{ai}}^{cj} \Sigma_{cj, dk} {H_{bi}}^{dk} \right],
    \end{split} \end{align}
    where $\mathcal{F} = \sexpv[W]{F}$.

    In the continuous time limit of SGD preserving the fluctuations  \cite{chaudhariStochasticGradientDescent2018, mandtContinuousTimeLimitStochastic, mandtStochasticGradientDescent2018,
    yaidaFluctuationdissipationRelationsStochastic2019,
    aartsStochasticWeightMatrix2025}, i.e., $\epsilon \to 0$, $T = \epsilon/|\mathcal{B}|$ fixed, the stochastic differential equation describing the dynamics of $\Sigma$ is
    \begin{align}
        \frac{d \Sigma}{d t} = - \left[ \Sigma H + H \Sigma \right] + T \left[ \mathcal{F} - H \Sigma H \right] ,
    \end{align}
    and at the fixed point $\Sigma$ satisfies the Lyapunov-type equation
    \begin{align} \label{eq:lyapunov}
        \Sigma H + H \Sigma = T \left[ \mathcal{F} - H \Sigma H \right] ,
    \end{align}
    where we have omitted matrix indices for aesthetic reasons.
    Eq.~(\ref{eq:lyapunov}) is the fluctuation-dissipation relation connecting the drift on the left-hand side to the fluctuation on the right-hand side.
    By solving the equation for $\Sigma$, the spectral density of the weight covariance matrix $X$ is written in terms of the ground state geometry $H$ and the fluctuation $\mathcal{F}$ around it.

\subsubsection{Low-temperature limit}
    In the limit of low temperature $(T \to 0)$, the fluctuating term on the right-hand side is negligible, and the equation only has a trivial solution,
    \begin{align}
        \Sigma H + H \Sigma = 0
        \quad \Rightarrow \quad
        \Sigma = 0.
    \end{align}
    The weight covariance converges to the correct ground state $X = W^{\ast} {W^{\ast}}^T$, which covers the case of full batch update, shown in Sec.~\ref{sec:2.solvable_model}.

\subsubsection{High-temperature limit}
    In the high-temperature limit, thermal fluctuations become significant, and the weight fluctuations satisfy
    \begin{align}
        \Sigma = H^{-1}\mathcal F H^{-1}.
    \end{align}
    Assuming a well-behaved loss function with a compact ground-state manifold, the Hessian is almost surely invertible up to a finite number of zero modes \cite{tanakaNoethersLearningDynamics2021}. 
    The spectral properties of the weight covariance matrix are determined by the interplay between the Hessian and the Fisher information matrix.
    While the relation above motivates the discussion below, the subsequent analysis is purely schematic and serves only to illustrate a possible route to heavy-tailed spectra.
    The high-temperature expression suggests that large fluctuations originate from soft directions of the loss landscape. Let
    \begin{align}
        H e_\mu = h_\mu e_\mu,
        \qquad
        h_\mu>0,
    \end{align}
    denote the Hessian eigenvalue decomposition. In the Hessian eigenbasis,
    \begin{align}
        \Sigma_{\mu\nu}
        =
        \langle e_\mu,\Sigma e_\nu\rangle
        \sim
        \frac{\mathcal F_{\mu\nu}}
             {h_\mu h_\nu}, \quad     \mathcal F_{\mu\nu}
        =
        \langle e_\mu,\mathcal F e_\nu\rangle.
    \end{align}
    In particular,
    \begin{align}
        \Sigma_{\mu\mu}
        \sim
        \frac{\mathcal F_{\mu\mu}}
             {h_\mu^2}.
    \end{align}
    Assuming that the Fisher matrix remains regular and that the off-diagonal matrix elements of $\mathcal F$ in the Hessian eigenbasis do not modify the leading scaling of the soft modes, the largest eigenvalues of the fluctuation matrix are controlled by
    \begin{align}
        \lambda_\Sigma
        \sim
        h^{-2}.
    \end{align}
    If the Hessian density behaves near the origin as
    \begin{align}
        \rho_H(h)
        \sim
        h^\beta,
        \qquad
        h\to0^+,
    \end{align}
    with $\beta>0$, the change of variables $h\sim\lambda^{-1/2}$ yields
    \begin{align}
    \begin{aligned}
        \rho_\Sigma(\lambda)
        &\sim
        \rho_H(\lambda^{-1/2})
        \left|
        \frac{dh}{d\lambda}
        \right|
        \\
        &\sim
        \lambda^{-\beta/2}
        \lambda^{-3/2},
        \end{aligned}
    \end{align}
    and therefore
    \begin{align}
        \rho_\Sigma(\lambda)
        \sim
        \lambda^{-(3+\beta)/2}.
    \end{align}
    
    Interestingly, heavy-tailed spectra are frequently reported in deep neural networks \cite{mahoneyTraditionalHeavyTailed2019,martinImplicitSelfRegularizationDeep2021}. 
    Although the derivation is highly heuristic and relies on many simplifying assumptions, it suggests a possible mechanism by which soft directions of the loss landscape and gradient fluctuations may cooperatively generate heavy-tailed weight spectra, with additional isolated spikes arising from the teacher contribution in Eq.~\eqref{eq:powerLawPlusSpikes}.
  
    \begin{figure}[tp!]
        \centering
        {\includegraphics[width=\linewidth]{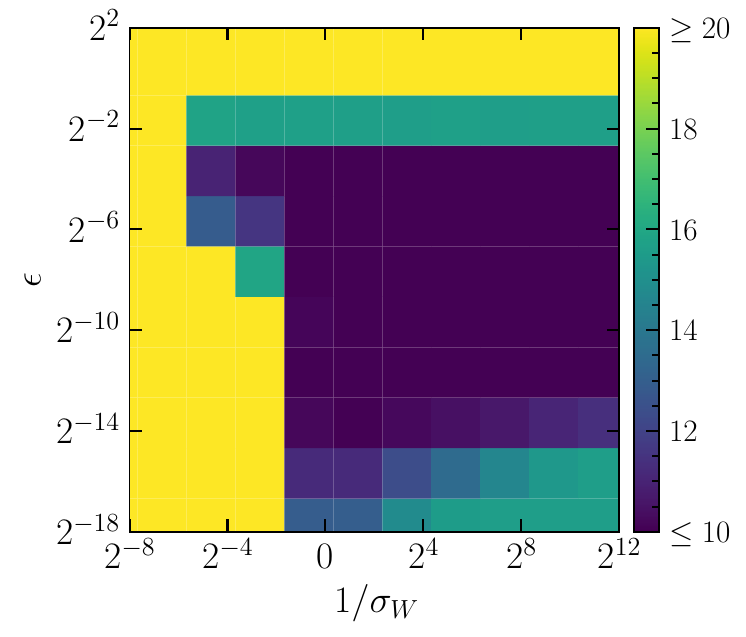}}
        \caption{Empirical trainability phase diagram for UTKFace data. The diagram shows the final test Mean Absolute Error (MAE) of the model after 300 epochs of training for different values of step size $\epsilon$ and initial variance $\sigma_W^2$. The models are only well trained for a specific set of step sizes and initial variance near the centre of the figure, with lower final loss. The diagram shows that the theoretical description of learning as a BBP transition is valid even with the realistic architecture and dataset, and the learning dynamics are classified into three qualitatively distinct spectral phases. Specific architecture and training settings are explained in Appendix~\ref{appendix:real_data}.}
        \label{fig:real_loss}
    \end{figure}
    \begin{figure*}[tp!]
        \centering
        \includegraphics[width=\linewidth]{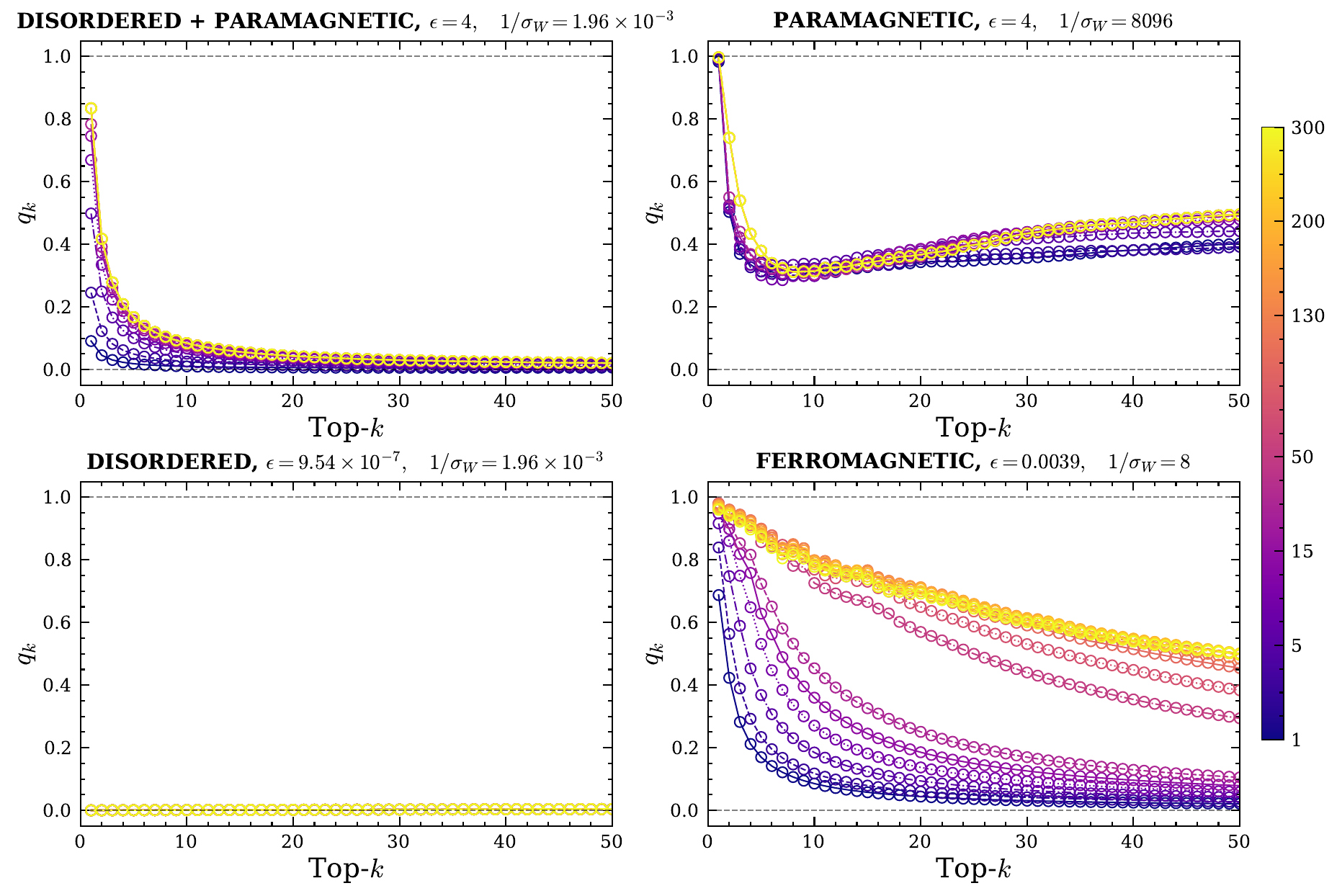}
        \caption{The top-$k$ subspace self-overlap in the (top left) mixed paramagnetic disordered phase, (top right) paramagnetic regime, (bottom left) disordered phase, and (bottom right) ferromagnetic phase. 
        The colourbar indicates the epochs of each line. The darker colour corresponds to the earlier epoch in the training and the brighter colour to the later. In the paramagnetic regime, the model reduces to performing a PCA decomposition, and only trivial alignments exist. In the disordered phase, no eigenvalue is aligned to the signal direction, and all eigenvectors are randomly distributed. In the ferromagnetic phase, the top-$k$ subspace between different initialisations acquires alignment to the same signal direction during training.}
        \label{fig:real_overlaps}
    \end{figure*}
\section{Trainability phase diagram in realistic settings}
\label{sec:5.realistic}

    In this final section, we test the predictions of the dynamical BBP framework in a realistic learning setting using the UTKFace dataset \cite{utkface}.
    The dataset consists of $\mathcal{O}\left(10^4\right)$ facial images, with the task of predicting the age associated with each image.

\subsection{Empirical trainability phase diagram}

    To construct an empirical trainability phase diagram, we consider the test loss as the observable studied in the plane spanned by the step size and the initial variance.
    In Fig.~\ref{fig:real_loss}, we report this empirical trainability phase diagram of a one-hidden-layer ReLU network trained on the UTKFace dataset.
    The darker area corresponds to the range of hyperparameters for which the test prediction loss is low. 
    The figure suggests that the performance of the trained network depends strongly on the choice of step size $\epsilon$ and initial variance $\sigma_W^2$.

    The resulting loss values separate the phase diagram into three distinct regions.
    The upper region, corresponding to large step sizes, is associated with the paramagnetic regime.
    The lower left and lower right regions correspond to disordered phases, while the central region, characterised by low final loss, corresponds to the ferromagnetic phase.
    While the phase boundaries are already visible in the final loss, they become more evident when examined through the top-$k$ subspace self-overlap and extracted features, introduced in the following subsection.

    Corresponding results for deeper architectures are reported in Appendix~\ref{appendix:deep_real}, although our analysis is limited to the spectral properties of the first layer, as discussed in Sec.~\ref{sec:bbp_multiple_layers}.
    Details of the architectures, dataset and optimisations can be found in Appendix~\ref{appendix:real_data}.

\subsection{Evaluation of self-overlap in different phases}
    We proceed to measure the overlap between different initialisations of the weight matrices in different phases.
    In realistic learning settings, the informative signal is generally not confined to a single direction but spans a low-dimensional subspace \cite{henaffLocalLowdimensionalityNatural, goldtModelingInfluenceData2020, leviUnderlyingScalingLaws2024}.
    We therefore extend the self-overlap order parameter $q^{\alpha \beta}$, that captures the spectral alignment and BBP transition, to a rank-$k$ subspace overlap $q_k$ allowing the same framework to characterise the emergence and stability of multidimensional signal structure.
    We define the top-$k$ subspace self-overlap for two realisations of the networks with seed indices $\alpha$ and $\beta$ as
    \begin{equation}
        q_k
        = \sexpv[\mathbf{W}]{\frac{1}{k}
        \left\| {\mathbf{W}_k^{\alpha}}^T \mathbf{W}^{\beta}_k \right\|_F^2},
    \end{equation}
    where $\mathbf{W}_k \in \mathbb{R}^{D \times k}$ is the subspace defined as
    \begin{align}
        \mathbf{W}_k = \begin{pmatrix}
            \bvec{w}_1 & \bvec{w}_1 & \cdots & \bvec{w}_k
        \end{pmatrix}.
    \end{align}
    Here $\bvec{w}_i \in \mathbb{R}^{D}$ is the right singular vector corresponding to the $i$-th largest singular value of the first layer weight matrix, and the overlap is averaged over all pairs of $\alpha$ and $\beta$.

    By measuring this overlap as a function of $k$, one can characterise the similarity between the $k$-dimensional subspace spanned by the leading singular vectors of different samples.
    For $k=1$, the definition recovers the original overlap, while for uncorrelated subspaces, the typical overlap scales as $q_k = k/D$.
    Deviations above this random baseline indicate the emergence of universal alignment between samples and the formation of a common lower-dimensional signal subspace.

    Fig.~\ref{fig:real_overlaps} shows the evolution of the top-$k$ subspace overlap $q_k$ in the four phases identified in the trainability phase diagram, Fig.~\ref{fig:real_loss}.
    The panels are arranged clockwise from the top left as mixed paramagnetic-disordered, paramagnetic, ferromagnetic, and disordered.
    These correspond to the top-left, top-right, centre, and bottom-left regions of the phase diagram, respectively.
    The disordered phase is characterised by $q_k \sim k/D$, consistent with random subspaces prediction for all $k$, indicating the absence of any subspace alignment between different training runs.
    In the paramagnetic regime, the leading direction exhibits alignment across samples, while higher-order directions remain uncorrelated.
    As discussed below, this dominant mode appears to reflect a global feature of the data rather than a nontrivial feature.
    In contrast, the ferromagnetic phase develops substantial overlap over an extended range of $k$, signalling the emergence of a shared $k$-dimensional signal subspace.
    The mixed phase interpolates between the paramagnetic regime and the disordered phase.

    \begin{figure}[tp!]
        \centering
        \includegraphics[width=\linewidth]{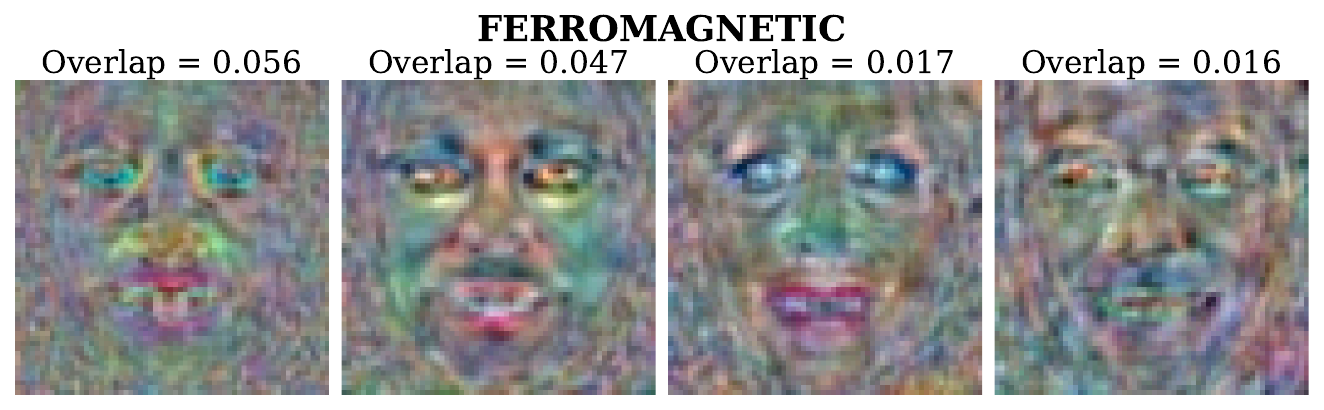}
        \includegraphics[width=\linewidth]{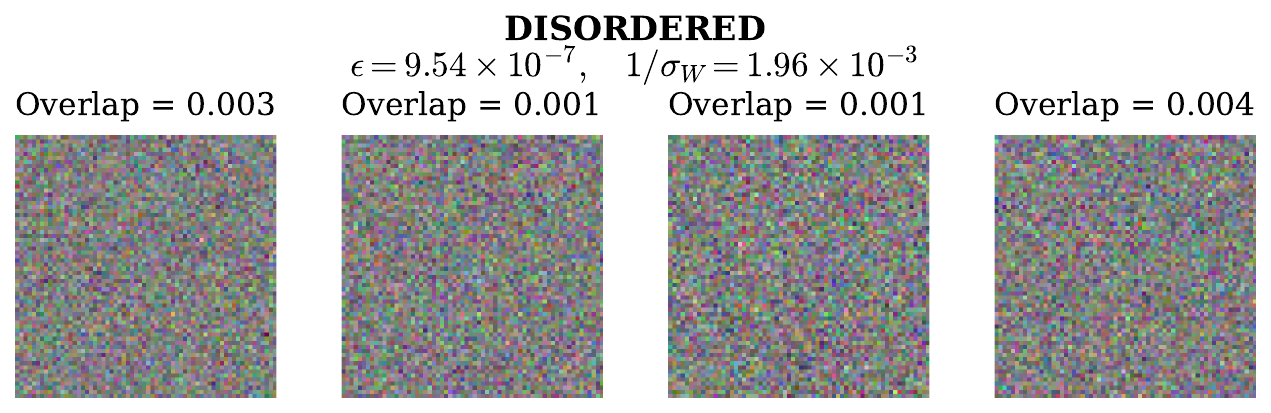}
        \includegraphics[width=\linewidth]{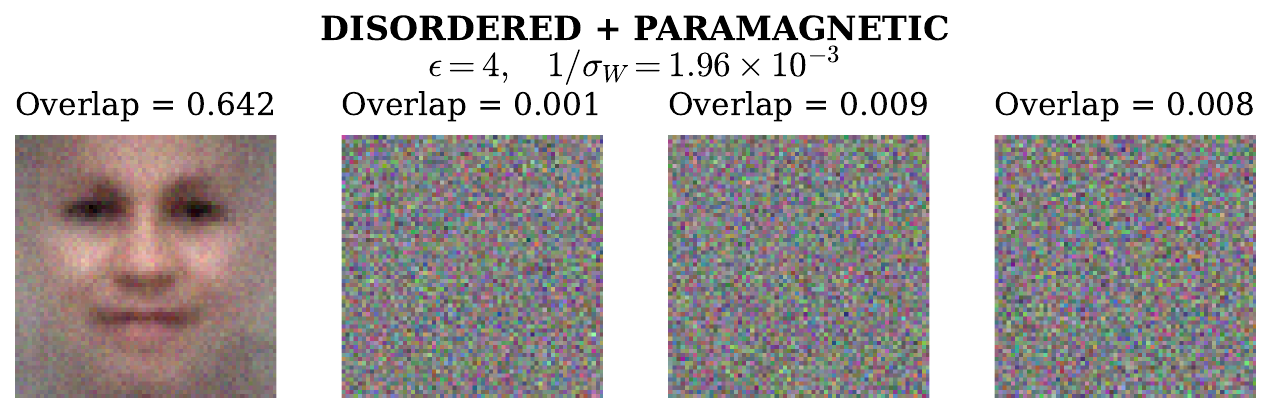}
        \includegraphics[width=\linewidth]{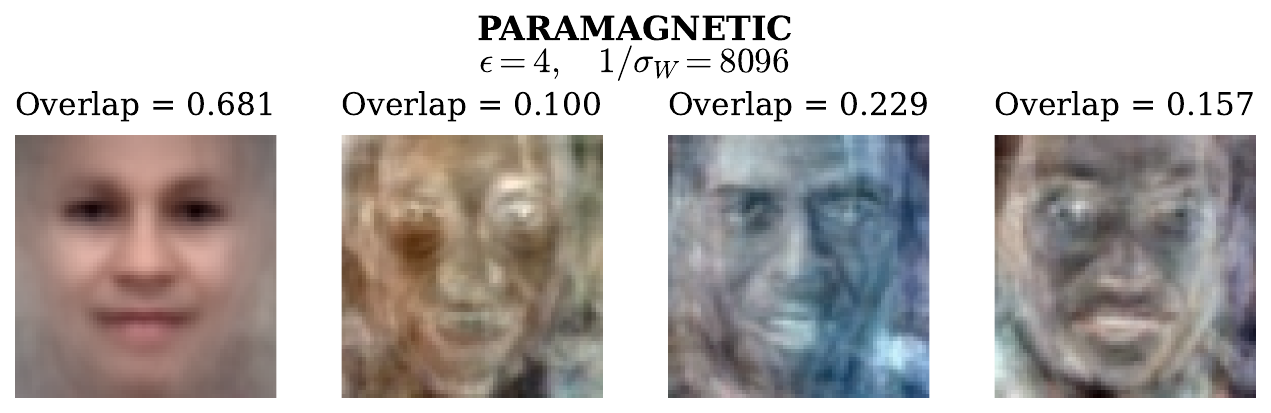}
        \includegraphics[width=\linewidth]{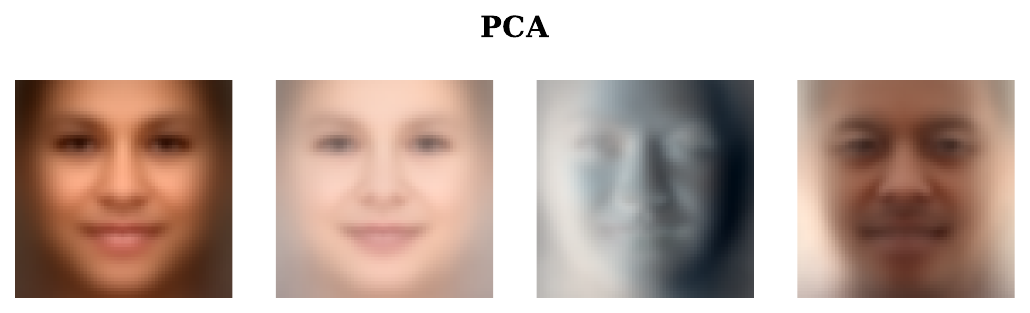}
        \caption{Features extracted from the four leading eigendirections of the trained weight matrix for each phase of the trainability diagram. From top to bottom, the panels correspond to the ferromagnetic phase, disordered phase, mixed paramagnetic-disordered phase, and paramagnetic regime. The bottom row shows the four leading principal components of the empirical data covariance matrix obtained from PCA. The numbers above each feature indicate the overlap between the singular vectors and the corresponding PCA component. In the ferromagnetic phase, the dominant modes capture nontrivial structure in the data that deviates from the principal components while remaining highly aligned across samples. The disordered phase exhibits no coherent structure and is dominated by noise. In contrast, the learned modes in the paramagnetic regime exhibit strong alignment with the leading PCA directions, indicating that optimisation primarily recovers dominant covariance modes of the data. These observations are consistent with the subspace overlap analysis and highlight qualitative differences in the representations learned across the four regimes.}
        \label{fig:real_features}
    \end{figure}
\subsection{Learned features in each phase}
\label{sec:real_features}
    
    Fig.~\ref{fig:real_features} shows the four leading learned features extracted from the different regions on the empirical trainability diagram. 
    In the ferromagnetic phase, the leading eigenmodes capture nontrivial facial features.
    In the disordered phase, the model fails to extract any meaningful structure and instead reproduces random fluctuations.

    Within the large step size region of the empirical trainability diagram, corresponding to the paramagnetic regime, two distinct types of dynamics emerge.
    Trivially, for sufficiently large step sizes, optimisation fails to converge, and the network is unable to extract meaningful structure from the data.
    However, for slightly smaller step sizes, lying between the ferromagnetic phase and the divergent regime, optimisation converges but reaches a suboptimal minimum.
    In the linear model studied in Sec.~\ref{sec:2.solvable_model}, this behaviour corresponds to the unstable convergent regime ($1 < \epsilon < 2$).
    Despite the convergence, we observe that the learned features exhibit substantial overlap with the leading principal components of the empirical data covariance matrix.
    This indicates that learning is dominated by the principal covariance directions of the data rather than by task-relevant nonlinear representations.
    For comparison, the bottom row of Fig.~\ref{fig:real_features} shows the leading principal components of the dataset.
    The labels above each panel indicate the overlap between the $i$-th learned feature and the corresponding principal component, highlighting the strong correspondence between the two.

    The overlap between the $i$-th learned feature and the corresponding PCA component is defined as
    \begin{align}
        {\rm Overlap}_i = \sexpv[\bvec{w}]{|\bvec{w}_i \cdot \bvec{v}_i|},
    \end{align}
    where $\bvec{w}_i \in \mathbb{R}^{D}$ is the $i$-th right singular vector of the trained weight matrix and $\bvec{v}_i \in \mathbb{R}^D$ is the $i$-th PCA component of the empirical data covariance matrix.
    The leading learned features in this paramagnetic regime show high overlap, while the learned features in the other phases show small overlap with the corresponding PCA components.

    To explain what is learned in this regime, recall the empirical data covariance matrix (\ref{eq:empirical_data_covariance}) and the data matrix (\ref{eq:data_matrix}) for a dataset with size $P$.
    If the data are centred, $\widehat C$ is the usual PCA covariance matrix.
    Let
    \begin{equation}
        \widehat C_{} \bvec{v}_i = \lambda_i \bvec{v}_i,
        \qquad
        \lambda_1 \geq \lambda_2 \geq \cdots \geq \lambda_{D},
    \end{equation}
    so that the vector $\bvec{v}_i$ is the $i$-th principal component of the empirical data covariance matrix.
    Now consider the first-layer weight matrix $W$. For a single sample, a gradient step has the schematic form
    \begin{equation}
        W_{ai}' = W_{ai} - \epsilon_{}  \sexpv[x\sim P]{\kappa_{a} x_i},
    \end{equation}
    where $\bvec{\kappa} = \bvec{\kappa}(W, x) \in \mathbb{R}^{N}$ is the backpropagated error at the first layer defined by substituting $l=0$ into Eq.~(\ref{eq:multi_nonlin}).

    In the pure paramagnetic regime and with SGD optimisation, the supervised component of $\bvec{\kappa}$ is effectively incoherent.
    It fluctuates strongly and does not maintain a stable alignment with the target signal. Therefore, to the leading order, one may approximate for all indices $a$ and $i$,
    \begin{align} \begin{split}
        \sexpv[x \sim P]{\kappa_a | x_i} \simeq 0, 
        \quad
        \sexpv[x \sim P]{\kappa_a \kappa_b^T | x_i} \simeq c I_{ab},
    \end{split} \end{align}
    with some constant $c >0$ and $I$ the identity. 
    Under this approximation, the average covariance of the weight increments $\delta W \equiv W' - W$ satisfies
    \begin{align}
        \sexpv[x \sim P]{{\delta W}^T \delta W}
        &= \epsilon^2_{} 
        \sexpv[x\sim P]{\bvec{x}_{} \bvec{\kappa}^T \bvec{\kappa} \bvec{x}^T} \nn\\
        &\propto \sexpv[x \sim P]{\bvec{x} \bvec{x}^T}
        = \widehat C.
    \end{align}

    Thus, although the dynamics are not learning the labels, the stochastic fluctuation of the update is still correlated with the input distribution. 
    Consequently, the right singular vectors of $W$ are expected to align with the eigenvectors of $\widehat C$ after multiple iterations, namely with the PCA directions of the data. 
    This explains why, in the paramagnetic regime, the leading directions learned by the first layer are dominated by the principal modes of the input distribution rather than task-specific representations, leading to the nonzero subspace alignment $q_k > k/D$.

    We highlight that to compute this quantity, we do not need a test set since it is evaluated by only using the trained matrices. This makes it a potentially useful tool for determining whether the model is training correctly or whether the initial hyperparameters are causing the training to be ineffective.

\section{Summary and outlook}
\label{sec:summary_outlook}
    In this work, we proposed a random matrix theoretic description of neural network optimisation dynamics based on dynamical Baik-Ben Arous-P\'ech\'e (BBP) transitions.
    Within this framework, gradient-based learning is interpreted as a competition between finite rank informative signals generated by the optimisation drift and the random spectral bulk inherited from weight initialisation and stochastic fluctuations.
    The emergence of isolated eigenvalues from the random bulk corresponds to the onset of spectral alignment between the weight matrix and the informative directions of the data.

    For a solvable linear teacher-student model trained by full-batch gradient descent, we derived the complete time-dependent trainability phase diagram analytically.
    The optimisation dynamics were shown to separate into disordered, ferromagnetic, and paramagnetic regimes, characterised respectively by the absence of alignment, stable signal extraction, and unstable optimisation dynamics.
    The BBP transition provides a precise criterion for the emergence of isolated informative eigenmodes, while the time-dependent suppression of the random bulk generates a dynamical BBP transition during training.
    We further demonstrated that similar spectral mechanisms survive in nonlinear and multilayer networks through effective rescaling of the signal and bulk contributions.
    Extending the analysis to finite datasets and stochastic optimisation revealed that finite-sample fluctuations deform the bulk spectrum and modify the transient BBP behaviour, while the stationary fluctuation spectrum near minima satisfies a fluctuation-dissipation relation involving the Hessian and gradient noise covariance. 
    %

    Several limitations remain.
    First, an exact full time-dependent calculation of the phase diagram relies on simplified teacher-student settings.
    While the numerical experiments suggest that the qualitative BBP mechanism persists in realistic architectures and datasets, the precise universality class of the transition in practical deep learning systems remains unclear.
    Second, the nonlinear and multilayer analyses are mainly restricted to the first optimisation step and the first layer, where the probability measure remains analytically tractable.
    Extending the full dynamical theory beyond the first step is considerably more challenging due to the strongly non-equilibrium and non-Gaussian nature of the evolving weight distributions.
    Third, the stochastic analysis near the stationary state relies on simplifying assumptions regarding the Hessian spectrum and gradient noise structure, particularly in the high temperature approximation.
    Understanding how these assumptions break down in strongly structured datasets and highly overparameterised architectures remains an open problem.
    %

    Interesting future directions naturally follow from the present work.
    One important extension is the study of modern adaptive optimisation algorithms such as Adam \cite{kingmaAdamMethodStochastic2014}, momentum, and related variants of SGD.
    In these algorithms, the effective update direction is dynamically rescaled, suggesting that the optimisation induces anisotropic and time-dependent deformations of both the signal and random bulk spectra.
    From the random matrix perspective, this may correspond to a dynamically evolving covariance geometry with nontrivial spectral feedback between gradient statistics and weight evolution.
    Understanding whether adaptive optimisers modify the BBP transition threshold, alter the stability of the ferromagnetic phase, or generate qualitatively different spectral universality classes would provide insight into the spectral origin of optimisation efficiency in deep learning.

    Another promising direction concerns the relation between the dynamical BBP transition and edge-of-chaos phenomena in neural networks \cite{sompolinskyChaosRandomNeural1988, pooleExponentialExpressivityDeep2016, penningtonResurrectingSigmoidDeep2017}.
    Previous studies have shown that neural networks exhibit static criticality at initialisation, where the trainability of deep networks is determined by whether signal propagation throughout the deep architecture remains stable under random initial conditions.
    In this picture, the critical phase boundary is typically fixed by architectural hyperparameters and initialisation statistics before learning begins.
    In contrast, the present work introduces a new dynamical dimension with the notion of criticality arising during optimisation itself.
    As training proceeds, gradient updates continuously reshape the competition between informative low-rank signal and random spectral disorder, dynamically driving the system across BBP transition boundaries.
    From this perspective, the BBP picture of the optimisation can be viewed as a temporal extension of the initial edge-of-chaos phase diagram.

    More broadly, the present work suggests that many aspects of neural network training may be understood through the emergence and evolution of collective spectral structures.
    Investigating how these transitions interact with heavy-tailed spectra, structured datasets, symmetry breaking, and representation learning may provide further insight into the geometric and statistical principles underlying modern deep learning systems.

\vspace*{0.2cm} 

\noindent
{\bf Acknowledgements} --  
We thank Bruno Loureiro for the discussion on gradient descent optimisation as a finite rank perturbation. CP thanks the participants of the ZiF workshop ``Random Matrix Theory for Learning and Statistical Physics", in particular Gabriele Sicuro, for the discussion on superstatistics. FD thanks École polytechnique fédérale de Lausanne (EPFL) for its hospitality during his stay, during which part of this work was conducted.

\noindent
CP is supported by the UKRI AIMLAC CDT EP/S023992/1.
DB and FD are supported by the Bando Ricerca Scientifica 2025 - Avvio alla Ricerca (D.~R.~2155/2025) of Sapienza Università di Roma, project B83C25004300005 - VESTA.
GA and BL are supported by STFC Consolidated Grant ST/X000648/1. 
GA is also supported by a Royal Society Leverhulme Trust Senior Research Fellowship.
BL is further supported by the UKRI EPSRC ExCALIBUR ExaTEPP project EP/X017168/1.

This study was conducted using the DARIAH HPC-AI cluster at CNR-NANOTEC in Lecce, funded by the ``MUR PON Ricerca e Innovazione 2014-2020" project, code PIR01\_00022 and H2IOSC Project - Humanities and cultural Heritage Italian Open Science Cloud funded by the European Union – NextGenerationEU – NRRP M4C2 - Project code IR0000029. 

\noindent
{\bf Research Data and Code Access} --
The code and data used for this manuscript will be made available in v2 of the arXiv version.

\noindent
{\bf Open Access Statement} -- For the purpose of open access, the authors have applied a Creative Commons Attribution (CC BY) licence to any Author Accepted Manuscript version arising.

\appendix

\section{Derivation of the dynamic BBP transition in linear models}
\label{appendix:A.derivation_linear}

    The evolution of the weight covariance matrix can be traced exactly in the case of linear regression with full batch gradient update.
    The weight matrix after the $t$-th iteration is given as
    \begin{align}
        W^{(t)} = W^{(0)} - \epsilon \sum_{n=0}^{t-1} K^{(n)},
    \end{align}
    where $K^{(n)}$ is the gradient averaged over the dataset at $n$-th iteration.
    For the teacher matrix $W^{\ast}$, the gradient is written in terms of the initial weight matrix and the teacher matrix 
    \begin{align}
        K^{(t)} &= - \left(W^{\ast} - W^{(t)} \right) \nonumber \\
        &= - \left(W^{\ast} - W^{(0)} + \epsilon \sum_{n=0}^{t-1} K^{(n)}  \right) \\
        &= - \left[ (1 - \epsilon)^t W^{\ast} - (1 - \epsilon)^{t} W^{(0)} \right] , \nonumber
    \end{align}
    where the last line is obtained after solving the recursive equation with initial condition $K^{(0)} = - \left(W^{\ast} - W^{(0)} \right)$.

    Substituting this equation back into the update equation, we obtain
    \begin{align} \label{eq:time_dep_linear}
        W^{(t)} = (1 - \epsilon)^t W^{(0)} + \left[1 - (1 - \epsilon)^t \right] W^{\ast},
    \end{align}
    for $t \ge 1$.
    Separating the effective bulk variance and signal strength, this equation can be written as
    \begin{align} \begin{aligned}
        W^{(t)} & =  (1 - \epsilon)^{t} W_{\perp}^{(0)} \\
        & \quad + \left[(1 - \epsilon)^{t} \bvec{w} + (1 - (1- \epsilon)^t) \bvec{u}\right] \bvec{v}^T \\
        & =  (1- \epsilon)^t W_{\perp}^{(0)} + \bvec{s}_{t} \bvec{v}^T,
    \end{aligned} \end{align}
    and for the weight covariance matrix,
    \begin{align} \begin{split}
        X^{(t)} =& (1 - \epsilon)^{2t} \sigma_W^2 \hat X_{\perp}^{(0)} 
        + \bvec{s}_t \bvec{s}_t^T.
    \end{split} \end{align}
    The expression for the effective bulk variance can be read off as
    \begin{align}
        \sigma_t^2 (\epsilon) = \left( 1 - \epsilon \right)^{2t} \sigma_W^2.
    \end{align}

    To compute the critical step size and the location of an isolated eigenvalue, consider a matrix resolvent $R_{X^{(t)}}$ of the weight covariance matrix $X^{(t)}$ after $t$ iterations,
    \begin{align} \begin{split}
        R_{X^{(t)}} (z) &= \left(zI - X^{(t)}\right)^{-1} \\
        &= \left(zI - \sigma_t^2 (\epsilon) \hat X^{(0)}_{\perp} - \bvec{s}_t {\bvec{s}_t}^T\right)^{-1}.
    \end{split} \end{align}
    Using the Sherman-Morrison formula, the critical step size is obtained by locating the pole of the resolvent matrix $R_{X^{(t)}}$, satisfying the condition,
    \begin{align} \label{eq:signal_equation}
        1 = \bvec{s}_t^T R_{\perp} (z) \bvec{s}_t,
    \end{align}
    where $R_{\perp}$ is the resolvent of the unperturbed bulk matrix $\sigma_t^2 (\epsilon) \hat X_{\perp}^{(0)}$.
    Expanding the expression, we find 
    \begin{align} \begin{split} \label{eq:linear_spike_equation}
        1 &= \left(1 - \epsilon\right)^{2t}\bvec{w}^T R_{\perp} (z) \bvec{w}\\
        &\quad + \left(1 - \left(1 - \epsilon \right)^t \right)^2 \bvec{u}^T R_{\perp} (z) \bvec{u},
    \end{split}\end{align}
    where we have used the fact that $\bvec{w}$ and $\bvec{u}$ are statistically independent.

    As the bulk resolvent $R_{\perp}$ does not contain the row vector $\bvec{w}$ by construction, the resolvent becomes
    \begin{align}
        \bvec{w}^T R_{\perp} (z) \bvec{w} 
        \ \rightarrow \ \Norm{\bvec{w}}^2 \frac{1}{N} \Tr \ 
        R_{\perp} (z) \simeq r \sigma_W^2 g (z).
    \end{align}
    Similarly, the second term on the right hand side is a standard rank-1 perturbed term.
    Since the signal vector $\bvec{u}$ is statistically uncorrelated to the resolvent $R_{\perp}$, the second term becomes
    \begin{align} \label{eq:gaussian_deterministic}
        \bvec{u}^T R_{\perp} (z) \bvec{u} \ \rightarrow \ \frac{1}{N} \Tr \ R_{\perp} (z) \simeq g (z).
    \end{align}
    Combining these terms, Eq.~(\ref{eq:linear_spike_equation}) becomes
    \begin{align} \begin{split}
        1 &= \left(\left(1 - \epsilon \right)^{2t} r \sigma_W^2 + \left( 1 - \left(1 - \epsilon \right)^t \right)^2\right) g(z)\\
        &= \theta_t^2 (\epsilon) g(z),
    \end{split} \end{align}
    where the effective signal strength $\theta_t^2(\epsilon)$ is obtained as
    \begin{align}
        \theta_t^2 (\epsilon) = \left(1 - \epsilon \right)^{2t} r \sigma_W^2 + \left( 1 - \left(1 - \epsilon \right)^t \right)^2,
    \end{align}
    and the critical step size is given as a solution to 
    \begin{align}
        \theta^2_{t}(\epsilon_c) = \frac{1}{g(\lambda_{+})},
    \end{align}
    where $\lambda_{+} = \sigma^2_t (\epsilon) (1 + \sqrt{r})^2$ is the right edge of the Marchenko-Pastur distribution.

    The location of an isolated eigenvalue $\lambda_{\rm iso}$ is determined by the functional inverse, $z(g)$, of the resolvent of the unperturbed matrix, evaluated at $g = 1/\theta^2_t(\epsilon)$ for $\epsilon > \epsilon_c$,
    \begin{align}
        \lambda_{\rm iso} = z\left(\frac{1}{\theta^2_t(\epsilon)} \right) .
    \end{align}
    For the Marchenko-Pastur distribution with scaled variance $\sigma^2_t (\epsilon)$, the inverse resolvent is given by
    \begin{align} \label{eq:inverse_resolvent}
        z(g) = \frac{1}{g} + \frac{\sigma^2_t (\epsilon)}{1 - r \sigma^2_t (\epsilon) g},
    \end{align}
    and substituting $g = 1/\theta^2_t(\epsilon)$, we obtain Eq.~(\ref{eq:iso_linear}).

    The overlap between the isolated eigenvector and the teacher direction can
    also be derived from the same resolvent formalism.
    The isolated eigenvalue satisfies the pole equation
    \begin{align}
        1 = \theta^2_t(\epsilon)\, g(\lambda_{\rm iso}) .
    \end{align} 
    The residue associated with the isolated projector gives the squared
    overlap between the maximum eigenvector of $X^{(t)}$ and the masked signal $\bvec{\hat s}_t$,
    \begin{align}
        q^2_s &\equiv  (\bvec{w}_{\rm max} \cdot \bvec{\hat s}_t)^2
        = - \frac{1}{\theta_t^4(\epsilon) g'(\lambda_{\rm iso})}.
    \end{align}
    Using the inverse-function relation
    \begin{align}
        g'(z)=\frac{1}{z'(g)} ,
    \end{align}
    together with Eq.~(\ref{eq:inverse_resolvent}),
    we obtain
    \begin{align}
        z'(g)
        =
        -\frac{1}{g^2}
        +
        \frac{r\sigma_t^4 (\epsilon) }
        {(1-r\sigma_t^2 (\epsilon) g)^2} .
    \end{align}
    Evaluating this expression at the outlier condition
    $g=1/\theta^2_t(\epsilon)$ yields
    \begin{align}
        q^2_s &= 1 
        - \frac{r\sigma_t^4 (\epsilon) }{\left(\theta^2_t(\epsilon)-r\sigma_t^2 (\epsilon) \right)^2},
    \end{align}
    and the physical overlap proportional to the teacher direction is given as
    \begin{align}
        q^2 = (\bvec{u} \cdot \bvec{\hat s})^2 q_{s}^{2}
        =\left(\frac{\theta_t^2 (\epsilon) - r \sigma_t^2 (\epsilon)}{\theta_t^2 (\epsilon)}\right) \, q_{s}^{2}.
    \end{align}

\section{\texorpdfstring{$\boldsymbol{r}$}{r} and \texorpdfstring{$\boldsymbol{\alpha}$}{a} dependence of the critical boundary}
\label{appendix:B.alpha_r}

    \begin{figure}[htp!]
        \centering
        \includegraphics[trim={0.5cm 1cm 0.5cm 1cm,},clip,width=0.45\linewidth]{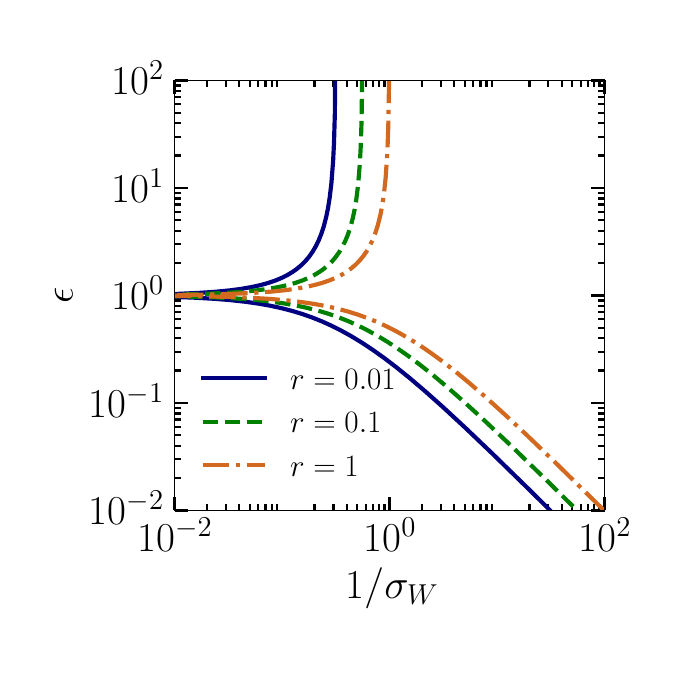}
        \includegraphics[trim={0.5cm 1cm 0.5cm 1cm,},clip,width=0.45\linewidth]{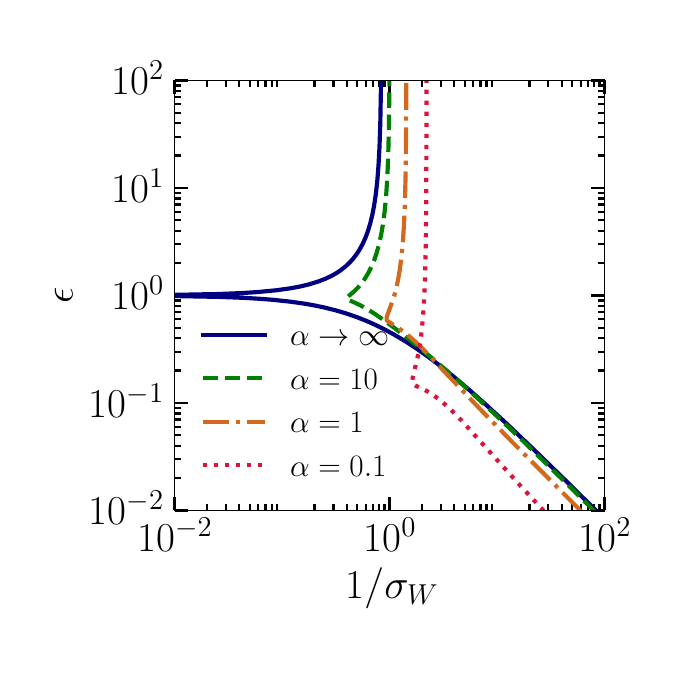}
        \caption{Trainability phase diagram of linear teacher-student model with respect to different aspect ratio $r=N/D$ (left) and  load parameter $\alpha = P/D$ (right). Depending on these ratios, the ferromagnetic phase expands or shrinks, demonstrating the interplaying effect of the dimensionality of the dataset and the network.}
        \label{fig:alpha_r}
    \end{figure}
    The dependence on the aspect ratio $r = N/D$ and the load parameter $\alpha=P/D$ of the phase diagram in the linear teacher-student case are shown in Fig.~\ref{fig:alpha_r}.
    In the gradient flow limit, studied in Sec.~\ref{sec:2.solvable_model}, where $\alpha \to \infty$, the ferromagnetic phase expands as the aspect ratio decreases.
    In the case of the proportional regime, studied in Sec.~\ref{sec:proportional_limit}, the ferromagnetic phase shrinks as the load parameter decreases, indicating that the training becomes more challenging for a smaller number of data points.

\section{Finite size scaling of \texorpdfstring{$q^2$}{q2}}
\label{appendix:C.finite_size}
    The random matrix theory predictions of isolated eigenvalues and critical step size are calculated in the large $N, D$ limit with fixed aspect ratio $r=N/D$.
    However, the simulation, as well as realistic neural networks, are carried out with a finite matrix size.
    Here, we show the finite-size scaling of the critical step size.

    \begin{figure}[htp!]
        \centering
        \includegraphics[trim={0.5cm 0.5cm 0.5cm 0.2cm,},clip,width=\linewidth]{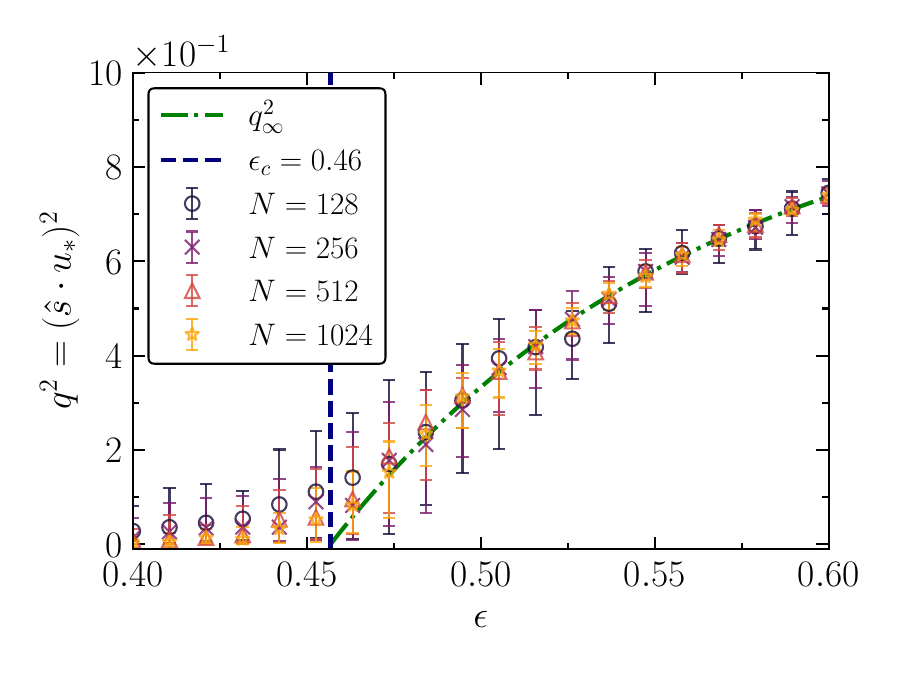}
        \caption{Finite size scaling of overlap $q^2$. The overlap approaches the theoretical prediction $q^2_{\infty}$ (green dashed dotted line) as the matrix size increases. The theoretical prediction of the critical step size $\epsilon_c$ is denoted by a blue dashed vertical line.}
        \label{fig:q2_finite_size}
    \end{figure}
    The squared overlap is evaluated over four different matrix sizes, $N = 128, 256, 512, 1024$, with fixed aspect ratio $r = 0.5$.
    Obtained overlaps are shown in Fig.~\ref{fig:q2_finite_size}, where fluctuation of the overlap decrease and the theoretical prediction is approached in the limit of large $N$ and $D$.

    \begin{figure}[htp!]
        \centering
        \includegraphics[trim={0.5cm 0.5cm 0.5cm 0.2cm,},clip,width=\linewidth]{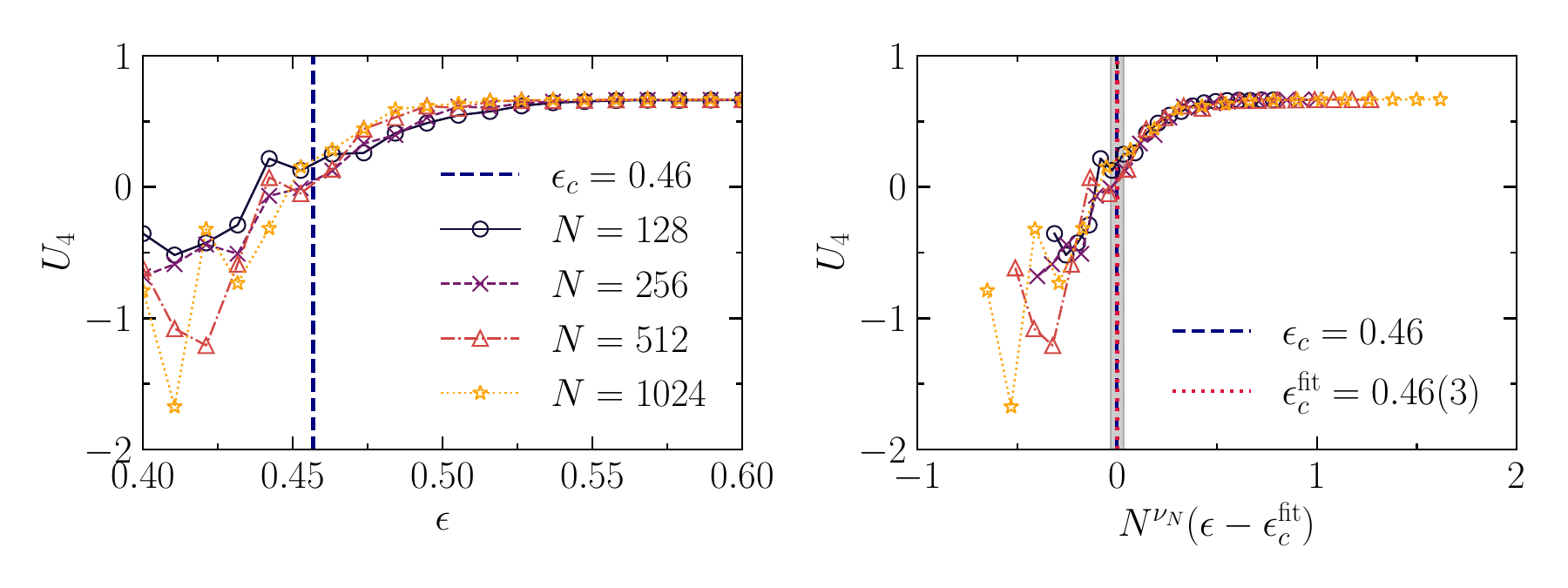}
        \caption{Binder cumulant analysis of the overlap $q^2$. (Left) the Binder cumulant $U_4$ for different matrix sizes crosses at a critical step size. (Right) rescaled Binder cumulant collapses onto the same curve for fitted critical exponent $\nu_N$ and critical step size $\epsilon_c^{\rm fit}$.}
        \label{fig:U4_finite_size}
    \end{figure}
    The numerical computation of the critical step size is obtained from the Binder cumulant analysis.
    The Binder cumulant is defined as
    \begin{align}
        U_4^{N} \equiv 1 - \frac{\sexpv{(q^2_N)^4}}{3_{} \sexpv{(q^2_N)^2}^2},
    \end{align}
    for overlaps evaluated in a different system size $N$.
    Obtained Binder cumulants are shown in Fig.~\ref{fig:U4_finite_size}.
    The Binder cumulant for different system size cross at a single point $\epsilon$, corresponding to the critical value $\epsilon_c$.
    The Binder cumulants are fitted with Pad\'e approximation of order $(n,m)=(2,2)$, and the critical step size obtained from fitting matches the theoretical prediction, as shown in the right panel of Fig.~\ref{fig:U4_finite_size}, supporting the theoretical result.

\section{Equivalence between the squared overlap \texorpdfstring{$q^2$}{q2} and self-overlap \texorpdfstring{$q^{\alpha \beta}$}{qaqb}}
\label{appendix:D.equivalence}

    Here, we show that the self-overlap $q^{\alpha \beta}$ between the student vectors is equivalent to the overlap $q^2$ between the signal direction and the student vector.
    The eigenvector $\bvec{w}$ of the student weight matrix corresponding to the maximum eigenvalue can be decomposed into the signal direction $\bvec{u}$ and other orthogonal directions $\bvec{w}_{\perp}$ as
    \begin{align}
        \bvec{w} = q_{} \bvec{u} + \sqrt{1 - q^2} \bvec{w}_{\perp},
    \end{align}
    where $q = \bvec{w} \cdot \bvec{u}$ by definition.
    Take two samples $\bvec{w}^{\alpha}$ and $\bvec{w}^{\beta}$,
    \begin{align}
        \bvec{w}^{\alpha} &= q_{} \bvec{u} + \sqrt{1 - q^2} \bvec{w}_{\perp}^{\alpha},\\
        \bvec{w}^{\beta} &= q_{} \bvec{u} + \sqrt{1 - q^2} \bvec{w}_{\perp}^{\beta},
    \end{align}
    leading to
    \begin{align}
        q^{\alpha \beta} \equiv \left|\bvec{w}^{\alpha} \cdot \bvec{w}^{\beta}\right| 
        = q^2 + \left(1 - q^{2}\right) \left|\bvec{w}^{\alpha}_{\perp} \cdot \bvec{w}^{\beta}_{\perp}\right|.
    \end{align}
    As $\bvec{w}^{\alpha}_{\perp}$ and $\bvec{w}^{\beta}_{\perp}$ are statistically uncorrelated,
    \begin{align}
        q^{\alpha \beta} \simeq q^2.
    \end{align}

\section{Time dependence of overlap \texorpdfstring{$q^2$}{q2}}
\label{appendix:E.q2_t}
    \begin{figure*}[tp!]
        \centering
        \includegraphics[trim={6cm 0.5cm 10cm 2cm,},clip,width=0.9\linewidth]{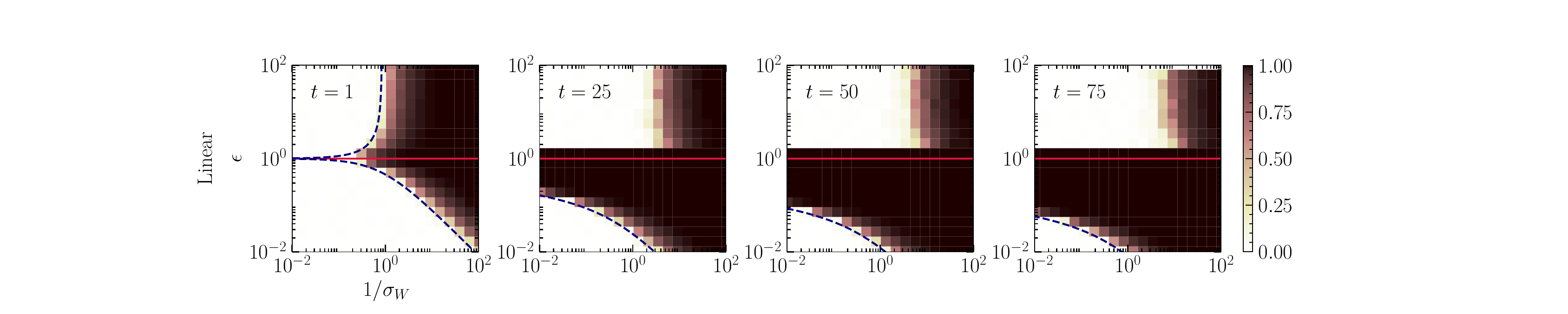}
        \includegraphics[trim={6cm 0.5cm 10cm 2cm,},clip,width=0.9\linewidth]{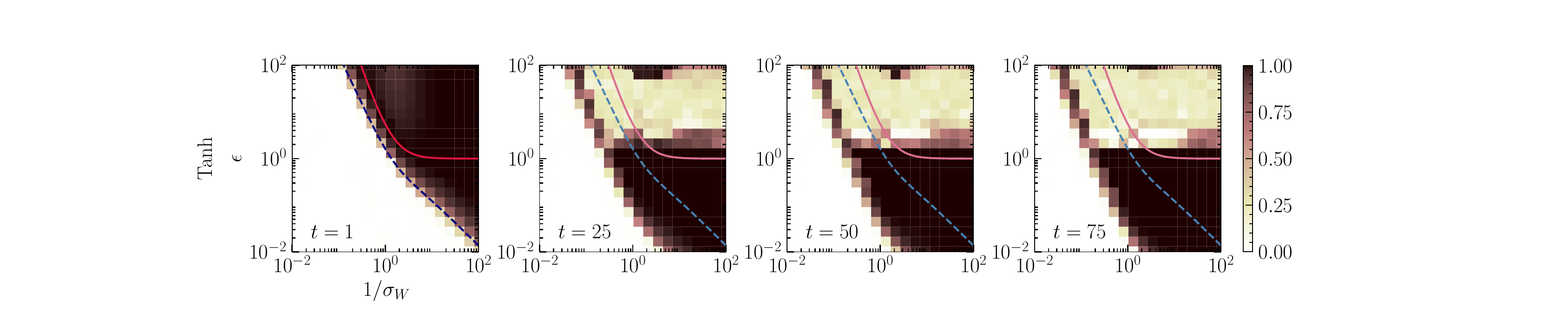}
        \includegraphics[trim={6cm 0.5cm 10cm 2cm,},clip,width=0.9\linewidth]{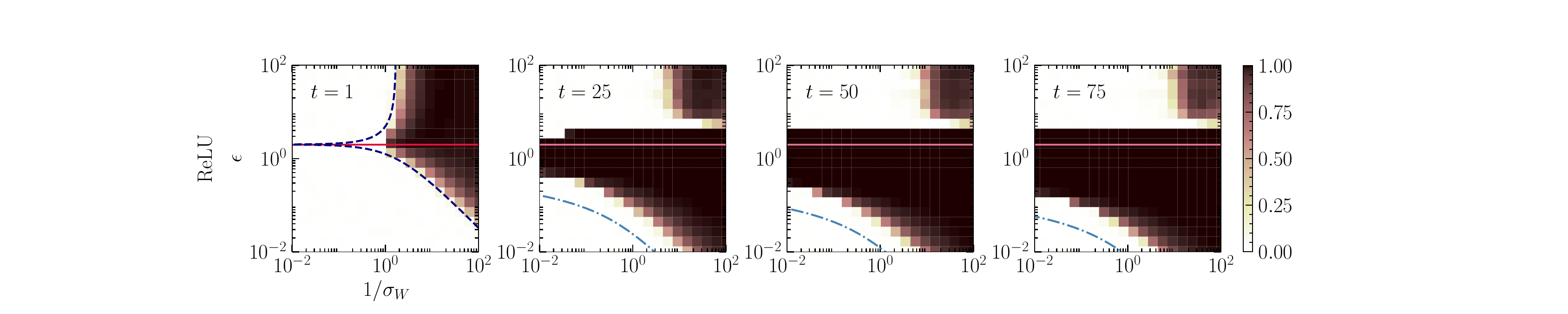}
        \caption{The square overlap $q^2$ at different number of iterations $t = (1, 25, 50, 75)$ from left to right. The overlap only persists in the ferromagnetic phase in the long training limit. The exact time-dependent phase boundaries are available for the linear case, but the single iteration phase diagram effectively remains informative also with the nonlinear activations.}
        \label{fig:q2t}
    \end{figure*}
    Time-dependent behaviour of squared overlap for different activation is shown in Fig.~\ref{fig:q2t}.
    For the linear case, the evolution of the BBP boundary is analytically obtained, while for hyperbolic tangent, the phase boundaries for the single update are overlaid.
    For ReLU activation, we use the asymptotic optimal step size $\epsilon_o=2$ obtained in the long-training limit and overlay the critical boundary $\epsilon_c$ from the linear theory to illustrate the qualitative dynamics of the phase diagram.

    Even though the exact time-dependent critical step size is unavailable for nonlinear models, a similar diminishing critical step size and increasing ferromagnetic phase are observed.
    This is due to the effect of bulk rescaling, below the paramagnetic boundary.
    Moreover, the initial overlap in the paramagnetic regime decreases as the training proceeds across different activation functions, where the optimisation does not converge to the correct target.

\section{Effective bulk variance and signal strength of nonlinear models}
\label{appendix:F.variance_and_signal_nonlinear}

    With nonlinear activation functions, the rescaling factors $\mu_{1,a}$ and $\mu_{2,a}$ defined in Eq.~(\ref{eq:non_lin_rescale}) modifies the effective bulk variance and signal strength.
    To obtain the expressions for the effective bulk variance and signal strength, we follow the procedure introduced in Appendix~\ref{appendix:A.derivation_linear}.
    Using the rank-1 decomposition of the teacher matrix, Eq.~(\ref{eq:teacher_w}), the single step update Eq.~(\ref{eq:ts_decomposition}) is decomposed into
    \begin{align}\begin{split} \label{eq:nonlinear_update_equation}
        W_{ai}' &= \left( 1 - \epsilon \mu_{2,a}\right) {W_{\perp}}_{ai} \\
        &\quad + \left[\left( 1 - \epsilon \mu_{2,a} \right) w_{a} + \epsilon \mu_{1,a} u_a\right] v_i^T \\
        &= \left(1 - \epsilon \mu_{2,a} \right) {W_{\perp}}_{ai}
        + s_a v_i^t,
    \end{split}\end{align}
    where we have defined the effective signal vector
    \begin{align}
        \bvec{s} = \left( \bvec{I} - \epsilon \bvec{\mu}_{2} \right) \odot \bvec{w} + \epsilon \bvec{\mu}_{1} \odot \bvec{u},
    \end{align}
    and we denote the elementwise product with $\odot$.

    If the signal vector $\bvec{u} \in \mathbb{R}^N$ contains a macroscopic condensation, say $\bar u$, the self-averaging assumption is no longer valid.
    For simplicity, we consider the representative case in which the signal vector is aligned with one of the canonical coordinate $\bvec{e}$, which corresponds to the maximum value of the condensate, $\bar u = 1$.
    \begin{align}
        \lim_{N \to \infty} \max |u_i|^2 \to \bar u = 1.
    \end{align}
    In general cases, standard measure of the condensation can be used, such as occupation number \cite{rosHighdimensionalRandomLandscapes2025}.

    Decompose the rescaling factors into self-averaging bulk ($u \to 0$) and condensing components ($u \to 1$),
    \begin{align}
    \begin{aligned}
        \bar \mu &\equiv \lim_{N \to \infty} \frac{1}{N} \sum_{a=1}^{N} \mu_a (u\to0),
        \\
        \tilde \mu &\equiv \bvec{\mu} \cdot \bvec{e} - \bar \mu,
    \end{aligned}
    \end{align}
    where $\bvec{e}$ is the canonical coordinate in which the signal vector condenses into.
    Explicitly,
    \begin{align}\begin{split}
        \bar \mu_1 &= \sexpv[h]{\phi'(0)\phi'(h)}, \\
        \tilde \mu_1 &= \sexpv[h,z]{\left( \phi'(z) - \phi'(0)\right)\phi'(h)},
    \end{split}\end{align}
    and
    \begin{align}\begin{split}
        \bar \mu_2 &= \sexpv[h]{\phi'(h)^2 + \left( \phi (h) - \phi(0))\phi''(h)\right)},\\
        \tilde \mu_2 &= -\sexpv[h,z]{(\phi(z) - \phi(0))\phi''(h)}.
    \end{split}\end{align}
    The arguments $h$ and $z$ are Gaussian random numbers with variance $\sigma_W^2$ and 1, respectively.
    Then, the rescaling factors $\bvec{\mu}$ are decomposed into the bulk contribution and signal direction as
    \begin{align}\begin{split}
        \bvec{\mu}_1 &= \bar \mu_1 \bvec{I} + \tilde \mu_1 \bvec{e},\\
        \bvec{\mu}_2 &= \bar \mu_2 \bvec{I} + \tilde \mu_2 \bvec{e}.
    \end{split}\end{align}
    The random bulk term of Eq.~(\ref{eq:nonlinear_update_equation}) is
    \begin{align} \begin{split}
        X_{\perp}' &= \left(\bvec{I} - \epsilon \bvec{\mu}_2 \right) \odot X_{\perp} \odot \left(\bvec{I} - \epsilon \bvec{\mu}_2 \right) \\
        &= \sigma^2 (\epsilon) \odot \hat X_{\perp},
    \end{split} \end{align}
    where the effective variance is defined as
    \begin{align}
        \sigma_{ab}^2 (\epsilon) = \left(1 - \epsilon \mu_{2,a} \right) \left(1 - \epsilon \mu_{2,b} \right) \sigma_W^2.
    \end{align}
    However, when computing the bulk spectral density, the signal aligned direction in $\bvec{\mu}_2$ does not contribute in the large $N, D$ limit, as it is an $\mathcal{O}(1)$ contribution, and the other components self-average.
    The effective variance after assuming the self-averaging property becomes 
    \begin{align} \label{eq:nonlinear_effective_bulk_variance}
        \sigma^2 (\epsilon) \simeq \left(1 - \epsilon \bar \mu_2 \right)^2.
    \end{align}
    The optimal step size can be approximated as
    \begin{align}
        \epsilon_o \simeq \frac{1}{\bar \mu_2}.
    \end{align}

    The effective signal strength is defined through Eq.~(\ref{eq:signal_equation}).
    The effective signal vector is decomposed as
    \begin{align}
        \bvec{s} 
        &= \left( 1 - \epsilon \bar \mu_2 \right) \bvec{w} - \epsilon \tilde \mu_2 \left(\bvec{e} \odot \bvec{w} \right) + \epsilon \left(\bar \mu_1 + \tilde \mu_1 \right) \bvec{e}.
    \end{align}
    The following terms contribute to the signal,
    \begin{align}\begin{split}
        \bvec{w}^T R_{\perp} \bvec{w} &\simeq r \sigma_W^2 g, \\
        \left( \bvec{e} \odot \bvec{w} \right)^T R_{\perp}\left( \bvec{e} \odot \bvec{w} \right) &\simeq r \sigma_W^2 g,\\
        \left( \bvec{e} \odot \bvec{w} \right)^T R_{\perp} \bvec{w}
        &\simeq r \sigma_W^2 g,\\
        \bvec{e}^T R_{\perp} \bvec{w} &\simeq 0,\\
        \bvec{e}^T R_{\perp} \bvec{e} &\simeq g,
    \end{split} \end{align}
    where we have used Gaussian deterministic equivalence.
    Substituting these relations into Eq.~(\ref{eq:signal_equation}) and organising the terms according to
    \begin{align}
        1 = \theta^2(\epsilon) g(z),
    \end{align}
    gives the effective signal strength, 
    \begin{align} \label{eq:nonlinear_effective_signal_strength}
        \theta^2 (\epsilon) = \left(1 - \left( \bar \mu_2 + \tilde \mu_2 \right) \epsilon\right)^2 r \sigma_W^2 + \epsilon^2 \left(\bar \mu_1 + \tilde \mu_1 \right)^2.
    \end{align}
    The critical step size $\epsilon_c$ is defined as the solution to Eq.~(\ref{eq:eps_c_linear}), with effective variance and signal strength given by Eq.~(\ref{eq:nonlinear_effective_bulk_variance}) and (\ref{eq:nonlinear_effective_signal_strength}).
    Explicitly, the critical step size $\epsilon_c$ is the solution to the quadratic equation
    \begin{align}
        c_2 \epsilon^2_c + c_1 \epsilon_c + c_0 = 0,
    \end{align}
    with coefficients
    \begin{align}
        c_2 &= \tilde \mu_2 \left( \tilde \mu_2 + 2 \bar \mu_2 \right) r \sigma_W^2 - \bar \mu_2^2 \sqrt{r} \sigma_W^2 + \left(\bar \mu_1 + \tilde \mu_1 \right)^2, \nn\\
        c_1 &= 2 \left(\bar \mu_2 \sqrt{r} - \tilde \mu_2 r \right) \sigma_W^2,\\
        c_0 &= - \sqrt{r} \sigma_W^2. \nn
    \end{align}

    For hyperbolic tangent activation, or in general for any odd activation functions, $\bvec{\mu}_2$ simplifies as the expectation value over a Gaussian measure vanishes,
    \begin{align}
        \sexpv[z]{\tanh(u_a z)} \sim 0,
    \end{align}
    and $\bvec{\mu}_2$ does not depend on the signal direction $\bvec{e}$, leading to 
    \begin{align}
        \tilde \mu_2 = 0.
    \end{align}
    In this case, the expression for the critical step size $\epsilon_c$ simplifies to
    \begin{align}
        \epsilon_c = \left(\bar \mu_2 \pm \left(\bar \mu_1 + \tilde \mu_1 \right) \frac{1}{r^{1/4} \sigma_W} \right)^{-1}.
    \end{align}

    \begin{figure}[thp!]
        \centering
        \includegraphics[width=\linewidth]{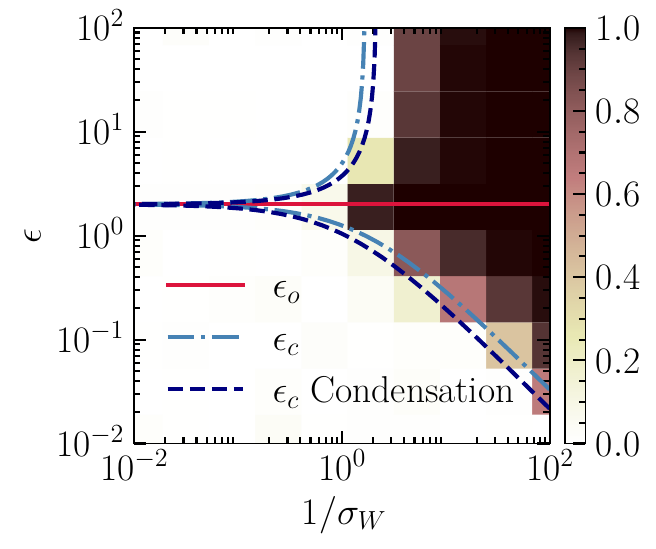}
        \caption{Correction to the phase diagram in the case of condensing signal vector $\bvec{u} = \bvec{e}$. The optimal boundary does not change as it does not depend on the signal vector, but the BBP transition boundary is slightly modified. The colour map is the simulated value of $q^2$ overlap with $\bvec{u} = \bvec{e}_1$.}
        \label{fig:condense}
    \end{figure}
    For ReLU activation, expressions for $\bvec{\mu}_1$ and $\bvec{\mu}_2$ can be obtained analytically.
    \begin{align}
        \mu_{1,a} &= \sexpv[\bvec{h},z]{\Theta(u_a z) \Theta({h_{\perp}}_a)} \nn\\
        &= \int_{0}^{\infty} \frac{dz}{\sqrt{2 \pi}} e^{-\frac{1}{2} z^2}
        \int_{0}^{\infty} \frac{d{h_{\perp}}_a}{\sqrt{2\pi \sigma_W^2}} e^{-\frac{1}{2\sigma_W^2} {h_{\perp}}_a^2} \nn\\
        &= \frac{1}{4},
    \end{align}
    and
    \begin{align}
        \mu_{2,a} &= 
        \mathbb{E}_{\bvec{h},z}\Big[\Theta({h_{\perp}}_a)^2 \nn\\
        &\quad + \left({h_{\perp}}_a \Theta({h_{\perp}}_a) - u_a z \Theta(u_a z) \right) \delta({h_{\perp}}_a)\Big]\nn\\
        &= \frac{1}{2} - u_a \frac{1}{\sqrt{2\pi \sigma_W^2}}\int_{0}^{\infty} \frac{dz}{\sqrt{2\pi}} z e^{-\frac{1}{2} z^2}\nn\\
        &= \frac{1}{2} - \frac{1}{2 \pi \sigma_W} u_a.
    \end{align}
    The four scaling parameters read as
    \begin{align}
        \bar \mu_1 = \frac{1}{4},
        \quad
        \tilde \mu_1 = 0,
        \quad
        \bar \mu_2 = \frac{1}{2},
        \quad
        \tilde \mu_2 = -\frac{1}{2\pi \sigma_W},
    \end{align}
    Compared to the self-averaging case (\ref{eq:relu_rescaling_sa}).
    Comparison between the phase boundaries obtained from the self-averaging assumption and condensing signal vector is shown in Fig.~\ref{fig:condense}.
    For the simulation, we used $N=100$, $\alpha=200$ and
    \begin{align}
        \bvec{u} = \bvec{e}_1 = \begin{pmatrix}1 \\ 0 \\ \vdots \\ 0\end{pmatrix} \in \mathbb{R}^N.
    \end{align}

\section{Rescaling factors of multilayer nonlinear networks}
\label{appendix:G.mult_nonlin}
    
    For nonlinear models with multiple layers, the gradient of the loss function with respect to the $l$-th layer weight matrix is given by Eq.~(\ref{eq:multi_nonlin}). Here we derive Eq.~(\ref{eq:multi_non_lin_rescale}).
    The expectation value of the first layer gradient is calculated by applying Wick-Isserlis's theorem,
    \begin{align} \begin{split} \label{eq:mult_nonlin_steins}
        \sexpv[x]{\frac{\partial \ell}{\partial W_{ai}^{0}}}
        &= \sexpv[x]{J_{ab}^{1} \phi_{b}^{L} {\phi_{a}^{1}}' x_i - H_{a}^{1} {\phi_{a}^{1}}' x_i} \\
        &= \sexpv[x]{\partial_{x_i} \left(J_{ab}^{1} \phi_{b}^{L}
        {\phi_{a}^{1}}' - H_{a}^{1} {\phi_{a}^{1}}' \right)}.
    \end{split} \end{align}
    Each of $\partial_{x_i} J$, $\partial_{x_i} \phi$, and $\partial_{x_i}
    \phi'$ will return a factor of $W$, and for the teacher-student setting,
    $\partial_{x_i} H$ will return a factor of $W^{\ast}$,
    \begin{align}
        \partial_{x_i} \phi_{a}^{\ell} &= {\phi_{a}^{\ell}}' Z_{ab}^{\ell - 1}
        \partial_{x_i} \phi_{b}^{\ell - 1} = T_{ab}^{\ell, 2} {\phi_{b}^{1}}' W_{bi}, \nn\\
        \quad \partial_{x_i} \phi_{a}^{1} &= {\phi_{a}^{1}}' W_{ai}, \nn\\
        \partial_{x_i} {\phi_{a}^{\ell}}' &= {\phi_{a}^{\ell}}'' Z_{ab} \partial_{x_i} \phi_{b}^{\ell - 1} = {\phi_{a}^{\ell}}'' Z_{ab}^{\ell -1} T_{bc}^{\ell-1,2} {\phi_{c}^{1}}' W_{ci}, \nn \\
        \partial_{x_i} T_{ab}^{\ell, \ell-1} &= \partial_{x_i} {\phi_{a}^{\ell}}' Z_{ab}^{\ell-1} = {\phi_{a}^{\ell}}'' Z_{ab}^{\ell-1} Z_{ad}^{\ell -1} T_{dc}^{\ell-1,2} {\phi_{c}^{1}}' W_{ci} , \nn\\
        \partial_{x_i} J_{ab}^{1}&= J_{fb}\Big( (\partial_{x_i} T_{fd}^{L,L-1}) T_{da}^{L-1,1} + \cdots \nn \\
        &\quad 
        + T_{fd}^{L,2}\partial_{x_i}T_{da}^{2,1}\Big) 
        = J_{fb} \mathcal{M}_{afc} {\phi_{c}^{1}}' W_{ci}, 
    \label{eq:mult_nonlin_derivative}
     \end{align}
    where $\mathcal{M}$ is a function of the transfer matrices and the activation function derivatives, defined implicitly above.

    Using the set of equations (\ref{eq:mult_nonlin_derivative}), the first
    term in the Eq.~(\ref{eq:mult_nonlin_steins}) becomes
    \begin{align} 
       \partial_{x_i} \left( J_{ab}^{1} \phi_b^{L} {\phi_{a}^{1}}' \right)
      & = \Big[ J_{fb} \mathcal{M}_{afc} {\phi_{c}^{1}}' \phi_{b}^{L} {\phi_{a}^{1}}' 
      + J_{ab}^{1} T_{bc}^{L,2} {\phi_{c}^{1}}' {\phi_{a}^{1}}' \nn \\
      & \quad + \delta_{ac} J_{ab}^{1} \phi_{b}^{L} {\phi_{c}^{1}}'' \Big] W_{ci} \nn \\
      &= g_{ac} W_{ci} .
     \end{align}
    For the second term in the teacher-student setting,
    \begin{align} 
        \partial_{x_i} H_{a}^{l} & = \mathcal{J}_{db} \partial_{x_i}
        \left(\phi_{b}^{\ast \, L} T_{da}^{L, l} \right) \nn\\
        & = \mathcal{J}_{db} T_{bc}^{\ast \, L,2} T_{da}^{L, l} {\phi_{c}^{\ast \, 1}}' W_{ci}^{\ast} \nn\\
        &\quad + \mathcal{J}_{db} \phi_{b}^{\ast \, L} \mathcal{M}_{adc} {\phi_{c}^{1}}' W_{ci} \, 
        \partial_{x_i} \left( H_{a}^{1} {\phi_{a}^{1}}' \right) \nn\\
        &= \mathcal{J}_{db} T_{bc}^{\ast \, L,2} T_{da}^{L,1} {\phi_{c}^{\ast \, 1}}' W_{ci}^{\ast} \\
        &\quad+ \left( \mathcal{J}_{db} \phi_{b}^{\ast \, L} \mathcal{M}_{adc}
        {\phi_{c}^{1}}' + \delta_{ac} H_{b} T_{bc}^{L,1} {\phi_{c}^{1}}' \right) W_{ci} , \nn
     \end{align}
    where $T^{\ast}$ is $T$ equivalent for the teacher network.
    Putting everything together, the rescaling factors $\mu_1$ and $\mu_2$ are
    given by Eq.~(\ref{eq:multi_non_lin_rescale}).

\section{Finite sample size effect}
\label{appendix:finite_sample_size}
    In the proportional training regime and stochastic gradient descent limit, the finite sample size effect introduces stochastic noise in the training trajectory.
    Specifically, the randomness is introduced when the covariance matrix of the dataset is replaced with the empirical covariance matrix in the gradient drift.
    Recalling the finite dataset update (\ref{eq:prop_update}), the weight covariance matrix after a single update is given by
    \begin{align}
        X' = \widetilde W_{\perp} \widetilde W_{\perp}^T + \bvec{s}_{} \bvec{s}^T\,.
    \end{align}
    Here, we denote quantities masked by the empirical covariance with a tilde, and normalised objects with a hat.

\subsection{Bulk resolvent}
    The bulk part $\widetilde X_{\perp} = \widetilde W_{\perp} \widetilde W_{\perp}^T$ is written explicitly 
    \begin{align} \begin{split}
        \widetilde X_{\perp} &= \widetilde W_{\perp} \widetilde W_{\perp}^T\\
        &= W\left(I - \epsilon \widehat C \right) \left(I - \bvec{\hat v} \bvec{\hat v}^T \right)
        \left(I - \epsilon \widehat C \right)^T W^T\\
        &\simeq W \left( I - \epsilon \widehat C\right)^2 W^T,
    \end{split} \end{align}
    where the term proportional to $\bvec{\hat v} \bvec{\hat v}^T$ in the second line is neglected as it is a rank-1 correction, which does not alter the bulk spectrum.
    This is a standard correlated Wishart matrix with temporal correlation \cite{pottersFirstCourseRandom}, where the resolvent is
    \begin{align} \label{eq:cov_resolvent}
        g(z) = \frac{1}{z} + \frac{1}{z}\int \frac{\sigma_W^2 ( 1 - \epsilon x)^2 g(z)}{1 - r (1 - \epsilon x)^2 g(z)} \rho_{{\rm MP}_{1/\alpha}}(x) dx,
    \end{align}
    or in terms of inverse resolvent,
    \begin{align}
        z(g) = \frac{1}{g} + \int \frac{\sigma_W^2 (1 - \epsilon x)^2}{1 - r \sigma_W^2 (1 - \epsilon x)^2 g} \rho_{{\rm MP}_{1/\alpha}}(x) dx.
    \end{align}

\subsection{Masked rank-1 signal}
    Denote the bulk matrix resolvent as
    \begin{align}
        R_{\perp} (z) = \left(z I - \widetilde X_{\perp} \right)^{-1}.
    \end{align}
    Then, using the matrix determinant lemma,
    \begin{align}
        \det \left(z I - X'\right) = \det \left(z I - \widetilde X_{\perp} \right) \left(1 - \bvec{s}^T R_{\perp} (z) \bvec{s}\right),
    \end{align}
    the signal equation corresponding to Eq.~(\ref{eq:bbp_condition}) becomes
    \begin{align}
        1 = \bvec{s}^T R_{\perp} (z_{\rm iso}) \bvec{s}.
    \end{align}
    Using the definition of the masked signal $\bvec{s} = \bvec{\tilde w} + \epsilon \vartheta \bvec{u}$, we find
    \begin{align} \label{eq:cov_spike}
        1 = \bvec{\tilde w}^T R_{\perp}(z) \bvec{\tilde w} + \epsilon^2 \vartheta^2 \bvec{u}^T R_{\perp} (z) \bvec{u}\,.
    \end{align}
    The second term on the right hand side is a standard rank-1 perturbed term.
    Using the relation~(\ref{eq:gaussian_deterministic}), the second term becomes
    \begin{align}
        \epsilon^2 \vartheta^2 \bvec{u}^T R_{\perp} (z) \bvec{u}
        \rightarrow
        \epsilon^2 \vartheta^2 g(z),
    \end{align}
    by Gaussian deterministic equivalence, and we have used the relation $g = g_{\perp} + \mathcal{O}(N^{-1})$.

    On the other hand, the vector $\bvec{\tilde w}$ is not independent from the matrix resolvent $R_{\perp}$, so the Gaussian deterministic equivalence does not hold.
    One can obtain the expression for $\bvec{\tilde w}^T R_{\perp} (z) \bvec{\tilde w}$, by using the cavity method.
    Denote the matrix resolvent for the whole matrix $\tilde X$ as
    \begin{align}
        R (z) = \left( z I - \tilde X\right)^{-1}.
    \end{align}
    Using the Sherman-Morrison formula, the orthogonal proportion of the resolvent $R_{\perp}$ and the original resolvent are related by
    \begin{align}
        \bvec{\tilde w}^T R_{\perp} (z) \bvec{\tilde w} = \frac{\bvec{\tilde w}^T R (z) \bvec{\tilde w}}{1 + \bvec{\tilde w}^T R(z) \bvec{\tilde w}}.
    \end{align}
    Transforming into the eigenbasis of the empirical covariance matrix,
    \begin{align}
        \left(I - \epsilon \widehat C\right)^2 = O^T D O,
    \end{align}
    where $D = {\rm diag}(a_1, a_2, \cdots)$, and $a_i = \sigma_W^2 (1 - \epsilon x_i)^2$, with $x_i \sim \rho_{{\rm MP}_{1/\alpha}}$, we can write
    \begin{align} \label{eq:full_appendix}
        \bvec{\tilde w}^T R (z) \bvec{\tilde w} = \bvec{\hat v} D W^T
        R(z) W D \bvec{\hat v}^T,
    \end{align}
    as the rotation $O$ does not affect the spectral properties.

    As $D$ is a diagonal matrix, it is sufficient to calculate the diagonal elements of $W^T R (z) W$.
    Using cavity method, the $i$-th diagonal component $(W^T R(z) W)_{ii}$ is
    \begin{align} \label{eq:sm_appendix}
        \bvec{w}_i^T R(z) \bvec{w}_i = \frac{\bvec{w}_i^T R_{/i}(z) \bvec{w}_i}{1 - a_i \bvec{w}_i^T R_{/i}(z) \bvec{w}_i},
    \end{align}
    where $\bvec{w}_i \in \mathbb{R}^N$ is the $i$-th column of $W$ and $R_{/i} (z)$ is the matrix resolvent with $i$-th column removed.
    Utilising the fact that $\bvec{w}_i$ is statistically independent from $R_{/i}(z)$, Gaussian deterministic equivalence gives
    \begin{align}
        \bvec{w}_i^T R_{/i} (z) \bvec{w}_i \simeq r \sigma_W^2 g(z).
    \end{align}
    Substituting it into Eqs.~(\ref{eq:full_appendix}) and (\ref{eq:sm_appendix}) yields
    \begin{align}
        \bvec{\tilde w}^T R (z) \bvec{\tilde w} = \sum_{i=1}^D \left( \bvec{\hat v} D \right)^2_i \frac{r \sigma_W^2 g(z)}{1 - a_i r \sigma_W^2 g(z)}.
    \end{align}
    From the definition of $\hat v$,
    \begin{align} 
        \left( \hat v D\right)_i^2 = \frac{a_i x_i^2 v_i^2}{\sum_{j=1}^D x_j^2 v_j^2}
        \simeq \frac{1}{D} \frac{a_i x_i^2}{m_2},
     \end{align}
    where we have defined the second moment of the matrix $\widehat C$ as
    \begin{align}
        \frac{1}{D} \sum_{j=1}^{D} x_j^2 \simeq m_2
        \equiv \int x^2 \rho_{{\rm MP}_{1/\alpha}}(x) dx.
    \end{align}
    Putting all things back together,
    \begin{align}
        \bvec{\tilde w}^T R(z) \bvec{\tilde w} &= \sum_{i=1}^{D} \frac{a_i x_i^2}{D m_2}
        \frac{r \sigma_W^2 g(z)}{1 - a_i r\sigma_W^2 g(z)}\\
        &\simeq \frac{1}{m_2} \int \frac{(1 - \epsilon x)^2 x^2 r \sigma_W^2 g(z)}{1 - (1 - \epsilon x)^2 r \sigma_W^2 g(z)} \rho_{{\rm MP}_{1/\alpha}}(x) dx. \nn
    \end{align}
    Substituting this equation with Eq.~(\ref{eq:cov_resolvent}) into Eq.~(\ref{eq:cov_spike}) defines Eq.~(\ref{eq:prop_spike}).

\section{Non-Gaussian data distribution}
\label{appendix:I.non-gaussian}
    The effect of a non-Gaussian data distribution at the stationary limit can be analysed in the framework of Sec.~\ref{sec:4.stochastic}.
    For a quadratic loss function, the gradient is
    \begin{align} \begin{split}
        \ell = \frac{1}{2} \sum_{a=1}^{N} \left( \sum_{i=1}^{D} W_{ai}^{\ast} x_i
        - \sum_{i=1}^{D} W_{ai} x_i \right)^2 ,
    \end{split} \end{align}
    The gradient and Hessian are given by
    \begin{align} \begin{split}
        & \frac{\partial \ell}{\partial W_{ai}} = \left( \sum_{j=1}^{D} W_{aj}^{\ast} x_{j} -
        \sum_{j=1}^{D} W_{aj} x_j \right) x_i, \\
        & \frac{\partial^2 \ell}{\partial W_{ai} \partial W_{bj}} = \delta_{ab} x_i x_j.
    \end{split} \end{align}
    Denoting the $k$-th order moment of the data distribution as
    \begin{align}
        m_{k} \equiv \sexpv[x]{\bvec{x}^k},
    \end{align}
    the expectation value of the gradient, Hessian, and the Fisher information matrix are written as
    \begin{align} 
        K_{ai} &= \sum_{j=1}^{D} \left(W_{aj}^{\ast} - W_{aj} \right)
        \sexpv[x]{x_j x_i} = m_2 \Delta_{ai}, \\
        H_{ai, bj} &= \delta_{ab} \sexpv[x]{x_i x_j} = m_2 \delta_{ab} \delta_{ij}, \\
        F_{ai, bi} &= \sum_{j,k=1}^{D} \left(W^{\ast}_{aj} - W_{aj} \right)
        \left(W^{\ast}_{bk} - W_{bk} \right) \sexpv[x]{ x_i^2 x_j x_k} \nn\\
        &= \sum_{j, k=1}^{D} \Delta_{aj}
        \Delta_{bk} f_{ijk},
     \end{align}
    where $f_{ijk}$ is a function of moments of the data distribution,
    \begin{align} \begin{split}
        f_{ijk} & = (1 - \delta_{ij})(1 - \delta_{ik})(1 - \delta_{jk}) m_2 m_1^2 \\
        &\quad + \left[\delta_{ij}(1 - \delta_{ik}) + \delta_{ik}(1 - \delta_{ij})\right] m_3 m_1 \\
        &\quad + \delta_{jk}(1 - \delta_{ij})m_2^2 + \delta_{ij} \delta_{ik} m_4 .
    \end{split} \end{align}

    For a centred Gaussian data distribution with unit variance, the Lyapunov equation simplifies to
    \begin{align}
      \Sigma_{ab} = 0 ,
    \end{align}
    leading to the trivial solution $X = W^{\ast} {W^{\ast}}^T$, which is consistent with the
    result obtained in Sec.~\ref{sec:2.solvable_model}.
    For non-centred distributions, such as Poisson or power law distributions, the
    odd moments are nonzero, leading to nontrivial solutions of the Lyapunov
    equation,
    \begin{align} \label{eq:non_gauss_data}
    X_{ab} = \left(W^{\ast} {W^{\ast}}^T\right)_{ab} + \frac{TS_{ab}}{2 m_2 - T m_4} ,
    \end{align}
    where $S_{ab}$ is an off-diagonal contribution to the weight covariance matrix, which is given by
    \begin{align} \begin{aligned}
    S_{ab} = \Sigma_{ai, bk} \Big( (1 - \delta_{ij})(1 - \delta_{ik}) (1 - \delta_{jk}) m_2 m_1^2 & \\
    +  \left( \delta_{ij}(1 - \delta_{ik}) + \delta_{ik}(1 - \delta_{ij})\right) m_3 m_1
    \Big) , &
    \end{aligned} \end{align}
    and the indices $i,j,k$ are summed over.
    The additional term appearing in Eq.~(\ref{eq:non_gauss_data}) is the correction from the non-Gaussian data distribution.

\section{Deep architectures in the case of real data}
\label{appendix:deep_real}
   \begin{figure}[tp!]
        \centering
        {\includegraphics[trim={0cm 0.2cm 0cm 0cm},clip,width=\linewidth]{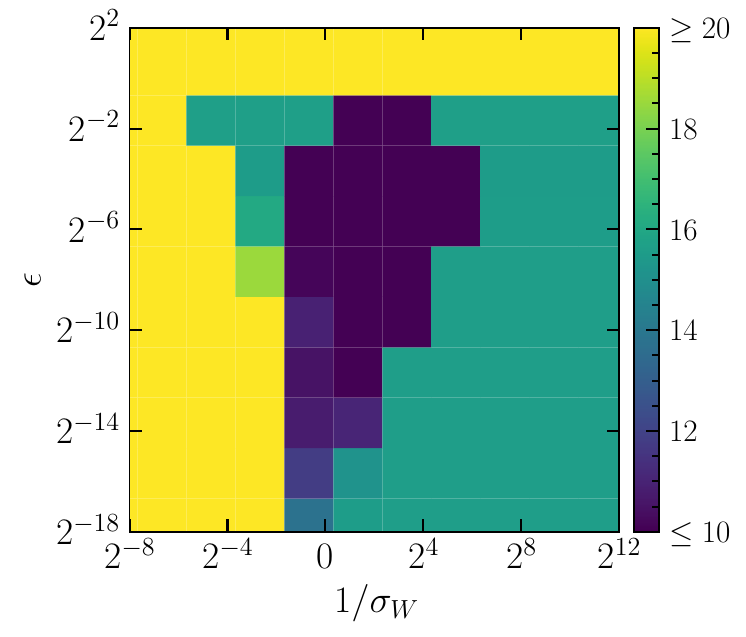}}
        \caption{The final test loss of a three-layer ReLU network after 300 SGD epochs trained on the UTKFace age regression task. The empirical trainability phase diagram shows different dynamical regimes of training as a function of the step size and initial variance $\sigma_W^2$. The darker region corresponds to the lower final loss, and the brighter region corresponds to the higher final loss.}
        \label{fig:real_loss_BIG}
    \end{figure}
    The final test loss for the three-hidden-layer ReLU network after 300 SGD epochs trained on the UTKFace age regression task is shown in Fig.~\ref{fig:real_loss_BIG}.
    Compared to the shallow network shown in Fig.~\ref{fig:real_loss}, the three-hidden-layer architecture exhibits a qualitatively similar trainability phase diagram.
    In the lower and right regions of the phase diagram, the learning signal is too weak relative to the initial random bulk, preventing spectral alignment and leading to a disordered phase.
    At large step sizes, two paramagnetic regimes are observed.
    A paramagnetic regime in which optimisation converges to a trivial PCA representation, and a divergent regime.
    Successful learning occurs only in the intermediate region of the phase diagram, where the balance between signal and disorder allows spectral alignment to emerge and the optimisation converges to the correct solution.

    \begin{figure}[tp!]
        \centering
        \includegraphics[width=\linewidth]{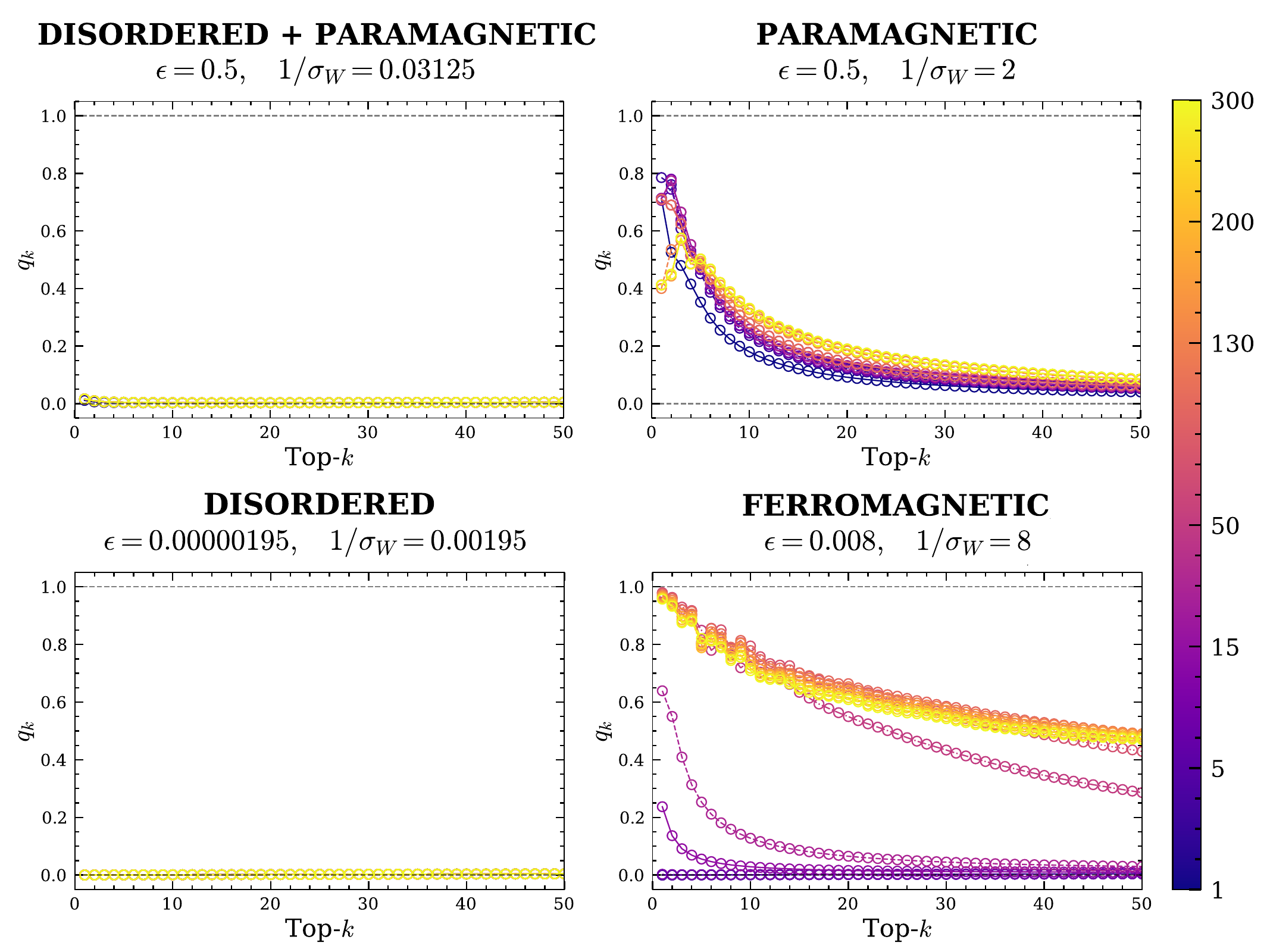}
        \caption{The top-$k$ subspace overlap computed for the first layer of the three-layer ReLU network. The computation was performed after 300 SGD epochs. (Top left) mixed paramagnetic-disordered phase, (top right) paramagnetic regime, (bottom left) disordered phase, and (bottom right) ferromagnetic phase. The darker colour corresponds to the earlier epoch in the training and the brighter colour to the later. As in the shallow network shown in Sec.~\ref{sec:5.realistic}, similar dynamical phases are classified.}
        \label{fig:real_overlaps_BIG}
    \end{figure}
    \begin{figure}[thp!]
        \centering
        \includegraphics[width=\linewidth]{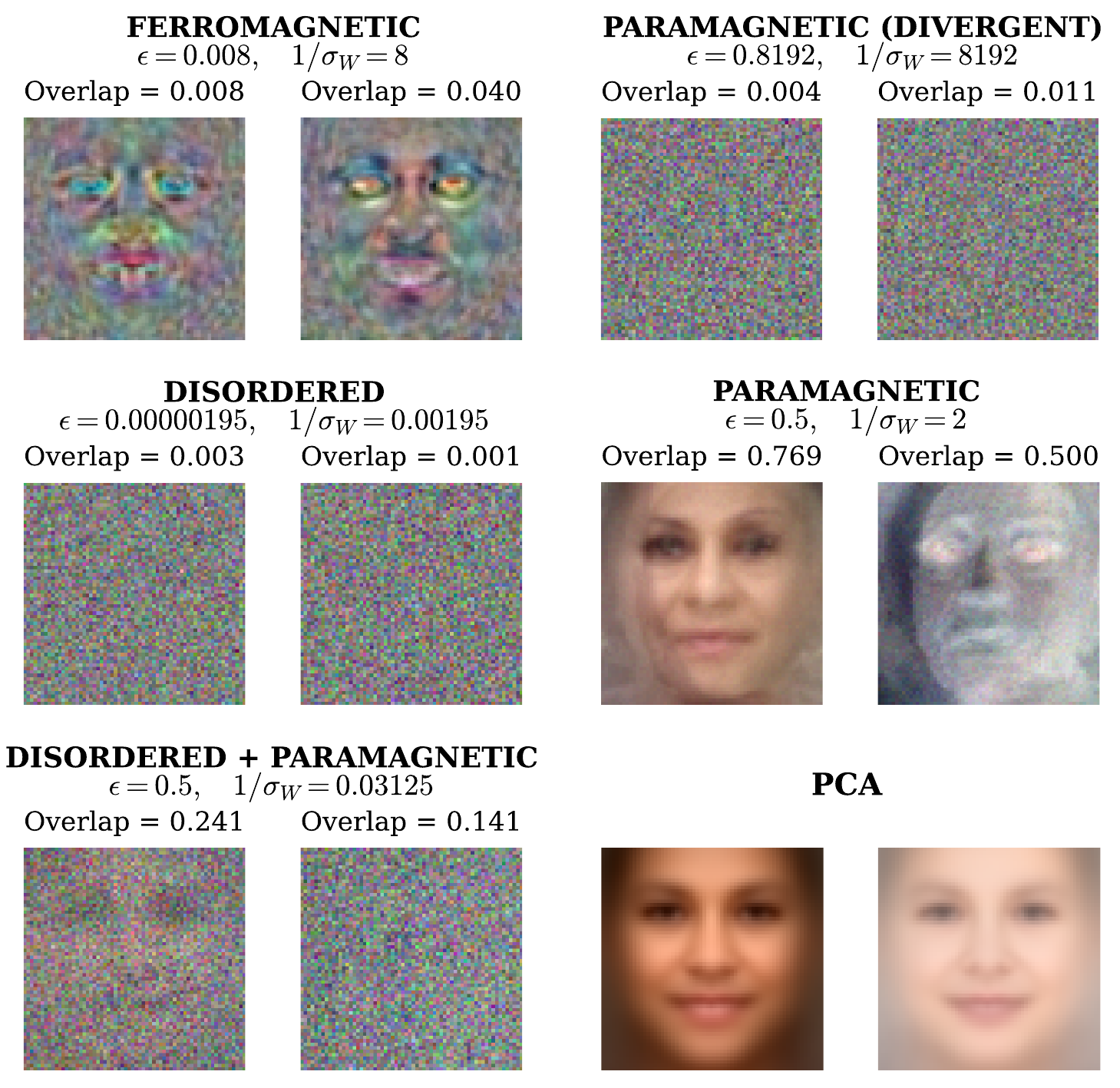}
        \caption{Learned features after the training in different phases. The leading two features of (left column from top to bottom) ferromagnetic, disordered, mixed paramagnetic-disordered phases, and (right column from top to bottom) divergent paramagnetic, PCA paramagnetic, and corresponding PCA components are shown.}
        \label{fig:real_features_deep}
    \end{figure}
    Fig.~\ref{fig:real_overlaps_BIG} shows the top-$k$ self-overlaps of the first-layer weight matrix for representative points in the phase diagram.
    From the top left and proceeding clockwise, the panels correspond to the mixed paramagnetic-disordered, paramagnetic, ferromagnetic, and disordered phases.
    The overall behaviour is qualitatively identical to that of the shallow network.
    In the disordered phase, the overlap remains close to the random-subspace expectation, indicating the absence of coherent spectral alignment.
    In the paramagnetic regime, alignment is dominated by a small number of leading directions, reflecting the emergence of a PCA representation.
    By contrast, the ferromagnetic phase exhibits substantial and progressively growing subspace alignment, signalling the formation of informative low-dimensional representations.
    These observations demonstrate that the dynamical BBP transition and the associated spectral organisation persist in deeper architectures.

    The learned features extracted from the leading two eigendirections of the first-layer weight matrix are shown in Fig.~\ref{fig:real_features_deep}.
    The phenomenology is qualitatively identical to that of the shallow network.
    The ferromagnetic phase develops informative nonlinear representations, whereas the convergent paramagnetic regime primarily recovers the dominant principal components of the dataset rather than task-relevant features.
    In the disordered and mixed phases, no coherent representation emerges, while the divergent paramagnetic regime fails to converge.

    As discussed in Sec.~\ref{sec:3.extensions}, the spectral framework developed in this work applies directly only to the first layer of a deep network.
    For internal layers, left and right rotational and permutational symmetries obstruct the construction of a unique self-overlap order parameter, making a complete spectral description substantially more challenging.
    Extending the theory to these hidden representations remains an important direction for future work.
    Despite this limitation, the results presented here demonstrate that the first-layer spectrum already captures the essential trainability and representation-learning phenomena.
    The persistence of the trainability phase diagram, spectral alignment, and learned feature structure in deeper architectures provides strong evidence that the dynamical BBP framework remains relevant beyond the shallow-network setting.

\section{Details of the architectures and numerical settings}
\label{appendix:real_data}
    For the simulation presented in Sec.~\ref{sec:5.realistic} and Appendix~\ref{appendix:deep_real}, a fully connected dense neural network with ReLU activation is used.
    Details of the architectures are shown in Table~\ref{tab:architecture}.
    The UTKFace dataset contains 2,0000 samples of $64 \times 64$ pixel facial images with 3 colour channels.
    Each image is flattened into 12288 dimensional vector and fed into the networks.
    \begin{table}[htp!]
        \centering
        \begin{tabular}{lr}
            \toprule
            One hidden layer & Sec.~\ref{sec:5.realistic} \\
            \hline\hline
            \textit{Activation} & \textit{Layer width} \\
            ReLU & 12288 \\
            ReLU & 256 \\
            Linear & 1 \\
            \hline
            \toprule
            Three hidden layers & App.~\ref{appendix:deep_real} \\
            \hline\hline
            \textit{Activation} & \textit{Layer width} \\
            ReLU & 12288 \\
            ReLU & 256 \\
            ReLU & 64 \\
            ReLU & 16 \\
            Linear & 1 \\
            \hline
        \end{tabular}
       \caption{
       Architectures used in the numerical experiments. 
       The single hidden layer ReLU network is used in Sec.~\ref{sec:5.realistic} 
       and the three hidden layers ReLU network is used
       in Appendix~\ref{appendix:deep_real}.
       }
       \label{tab:architecture}
    \end{table}

    The networks are trained with stochastic gradient descent with a batch size 128, and the spectrum of the weight covariance matrix is monitored during training.
    The results shown in Sec.~\ref{sec:5.realistic} and Appendix~\ref{appendix:deep_real} are taken after 300 epochs of training.
    The gradient clipping was used with the gradient norm threshold of $10^{12}$.
    %


\providecommand{\href}[2]{#2}\begingroup\raggedright\endgroup

\end{document}